\documentclass[
aps,prb,reprint,superscriptaddress
]{revtex4-2}

\usepackage{amsmath}
\usepackage{amsfonts}
\usepackage{txfonts}
\usepackage{mathtools}
\usepackage{comment}
\usepackage{graphicx}
\usepackage{dcolumn}
\usepackage{bm}
\usepackage{xspace}
\usepackage{physics}
\usepackage{siunitx}
\usepackage{nicematrix}
\usepackage{diagbox}
\usepackage{here}
\usepackage{color}%
\usepackage[normalem]{ulem}
\usepackage{hyperref}
\hypersetup{
 colorlinks=true,%
 linkcolor=blue,
 citecolor=blue,
}

\DeclareSIUnit\angstrom{\text{Å}}

\DeclareMathOperator{\Pf}{Pf}
\newcommand{\Z}{\mathbb{Z}}

\newcommand{\kinv}{\bm{k}_{\mathrm{inv}}}

\definecolor{myorange}{rgb}{0.7,0.5,0.0}
\definecolor{mygreen}{rgb}{0.1,0.6,0.2}
\definecolor{purple}{rgb}{0.75,0.0,1.0}

\newcommand{\AppliedPhys}{Department of Applied Physics, University of Tokyo, Bunkyo, Tokyo 113-8656, Japan}
\newcommand{\RIKEN}{RIKEN Center for Emergent Matter Science, Wako, Saitama 351-0198, Japan}

\begin{document}

\title{
Three-dimensional spinless Euler insulators with rotational symmetry
}

\author{Manabu Sato} \affiliation{\AppliedPhys} 
\author{Shingo Kobayashi} \affiliation{\RIKEN}
\author{Motoaki Hirayama} \affiliation{\AppliedPhys} \affiliation{\RIKEN}
\author{Akira Furusaki} \affiliation{\RIKEN}

\date{\today}

\begin{abstract}
The Euler class is a $\mathbb{Z}$-valued topological invariant that characterizes a pair of real bands in a two-dimensional Brillouin zone.
One of the symmetries that permits its definition is $\mathcal{C}_{2z}\mathcal{T}$, where $\mathcal{C}_{2z}$ denotes a twofold rotation about the $z$ axis and $\mathcal{T}$ denotes time-reversal symmetry.
Here, we study three-dimensional spinless insulators characterized by the Euler class, focusing on the case where additional $\mathcal{C}_{4z}$ or $\mathcal{C}_{6z}$ rotational symmetry is present, and investigate the relationship between the Euler class of the occupied bands and their rotation eigenvalues.
We first consider two-dimensional systems and clarify the transformation rules for the real Berry connection and curvature under point group operations, using the corresponding sewing matrices.
Applying these rules to $\mathcal{C}_{4z}$ and $\mathcal{C}_{6z}$ operations, we obtain explicit formulas that relate the Euler class to the rotation eigenvalues at high-symmetry points. 
We then extend our analysis to three-dimensional systems, focusing on the difference in the Euler class between the two $C_{2z}T$-invariant planes.
We derive analytic expressions that relate the difference in the Euler class to two types of representation-protected invariants and analyze their phase transitions.
We further construct tight-binding models and perform numerical calculations to support our analysis and elucidate the bulk-boundary correspondence.
\end{abstract}


\maketitle

\section{Introduction}

Topological insulators~\cite{RevModPhys.83.1057,RevModPhys.82.3045} have recently attracted considerable attention within the condensed matter physics community, emerging as a vibrant and rapidly evolving field. 
One particularly striking aspect of these insulators is the existence of robust metallic surface states, which arise from the nontrivial topology of the bulk electronic structure. 
Protected by time-reversal symmetry, these surface states exhibit notable resilience against disorder and other symmetry-preserving perturbations, offering a promising platform for exploring exotic quantum phenomena and potential applications in spintronics and quantum computation.

To systematically classify these topological phases, researchers have developed powerful theoretical frameworks based on onsite and/or crystal symmetry.
For example, the K-theory approach~\cite{PhysRevB.78.195125,10.1063/1.3149495,Ryu_2010}, which builds on the notion of stable equivalence, provides a robust method for classifying topological phases by focusing on properties that remain invariant under the addition of extra trivial bands.
Moreover, topological quantum chemistry~\cite{Bradlyn2017,PhysRevB.97.035138,PhysRevB.97.035139,annurev:/content/journals/10.1146/annurev-conmatphys-041720-124134} and symmetry-based indicators~\cite{PhysRevX.7.041069,Po2017,PhysRevX.8.031070,Po_2020} have provided concrete criteria for identifying various topological phases.
These approaches enable the efficient classification of a wide range of topological phases by relying solely on symmetry eigenvalues at high-symmetry points in the Brillouin zone (BZ), which greatly streamlines the identification process and facilitates high-throughput searches for candidate materials~\cite{Tang2019, Zhang2019,Vergniory2019,doi:10.1126/sciadv.aau8725,PhysRevB.100.195108,doi:10.1126/science.abg9094}.

According to the K-theoretical classification, stable topological insulator phases in class AI (spinless fermions) do not exist in one, two, or three spatial dimensions. 
Nonetheless, outside the scope of K-theory, there exist topological invariants characterizing spinless insulators with crystalline symmetry.
Notably, these phases are considered unstable because they can only be defined for systems that satisfy specific conditions regarding the number or representations of the occupied bands.
In three-dimensional (3D) spinless systems, the Fu model~\cite{PhysRevLett.106.106802} is a  prominent example of such unstable topological phases.
The Fu model exhibits gapless surface states with a quadratic band touching, which has been experimentally observed in a photonic crystal~\cite{Kim2022}.
This model has fourfold rotational symmetry $\mathcal{C}_{4z}$ and time-reversal symmetry $\mathcal{T}$, and its basis consists of $p_x$ and $p_y$ orbitals, which transform in the two-dimensional (2D) representation of $\mathcal{C}_{4z}$ and $\mathcal{T}$.
The nontrivial topology of the Fu model is characterized by a $\mathbb{Z}_2$ invariant that can be defined only when the occupied bands consist entirely of pairs belonging to the 2D representation of $\mathcal{C}_{4z}$ and $\mathcal{T}$.
Thus, the Fu model is classified as a so-called representation-protected topological phase~\cite{PhysRevB.102.115117,PhysRevX.10.031001,PhysRevB.108.155137,PhysRevB.110.L100508}.
Other examples in this category include halved mirror chirality~\cite{PhysRevLett.113.116403}, which is defined for the 2D representations of the $C_{nv}$ groups for $n=3,4,6$.

Besides the $\mathbb{Z}_2$ invariant, the Fu model is also characterized by the Euler class~\cite{PhysRevB.104.195114}. 
The Euler class is a $\mathbb{Z}$-valued invariant that characterizes a pair of real bands in a 2D system and is defined for both spinless and spinful systems~\cite{PhysRevX.9.021013,Ahn_2019,PhysRevB.102.115135,hatcherVBKT}. 
$\mathcal{PT}$ symmetry in spinless systems and $\mathcal{C}_{2z}\mathcal{T}$ symmetry in both spinless and spinful systems, each of which squares to $+1$, guarantee the reality condition~\cite{PhysRevB.91.161105,PhysRevLett.118.056401,PhysRevLett.118.156401} and allow the definition of the Euler class, where $\mathcal{C}_{2z}$ and $\mathcal{P}$ represent twofold rotation around the $z$-axis and inversion symmetry, respectively.
Remarkably, phase transitions associated with the change in the Euler class are accompanied by a non-Abelian braiding of band nodes~\cite{PhysRevX.9.021013,doi:10.1126/science.aau8740}.
Such non-Abelian braiding processes have received growing attention and have been explored in a wide range of systems.
Theoretical studies have proposed real materials that exhibit the braiding process in the electronic~\cite{Bouhon2020,PhysRevResearch.3.L042017,PhysRevB.105.L081117,PhysRevLett.133.093404,lee2024eulerbandtopologyspinorbit} or phonon~\cite{Park2021,Peng2022,PhysRevB.105.085115} bands.
Moreover, studies on the braiding process have been extended to artificial quantum systems~\cite{PhysRevLett.125.053601,PhysRevB.103.205303,Guo2021,Jiang2021,Jiang2021.2,Park2021.ACS.Photo,PhysRevB.105.214108,Zhao2022,Qiu2023,JIANG20241653,PhysRevB.109.165125,doi:10.1126/science.adf9621,doi:10.1126/sciadv.ads5081,karle2024anomalousmultigaptopologicalphases}, and some of these phenomena have been experimentally observed.

Despite extensive research endeavors on the Euler class, there remain unexplored aspects in 3D spinless systems.
To date, no material realization of 3D Euler insulators has been discovered.
Here, a 3D Euler insulator refers to a band insulator where the Euler classes defined on two $C_{2z}T$-invariant planes in the 3D BZ are different.
The same holds for the $\mathbb{Z}_2$ invariants that characterize the Fu model or the halved mirror chirality.
The primary obstacle arises from the fact that determining these invariants involves computing Wilson loop spectra~\cite{PhysRevB.102.115117,PhysRevB.93.205104,PhysRevLett.121.106403,PhysRevB.100.195135,PhysRevB.84.075119,PhysRevB.89.155114,PhysRevX.6.021008,PhysRevLett.120.266401,PhysRevB.99.045140} and that the general relationships among these invariants remain elusive.

In this work, we investigate the relationship between the Euler class and the rotational eigenvalues of the system.
Our focus is on 3D spinless insulators with two occupied bands that respect time-reversal symmetry $\mathcal{T}$ and either fourfold ($\mathcal{C}_{4z}$) or sixfold ($\mathcal{C}_{6z}$) rotational symmetry.
Moreover, we extend our analysis to include systems with additional mirror symmetry.
Based on the relationship between the Euler class and rotational symmetry, we derive general formulas that relate the Euler class to representation-protected topological invariants.
Guided by these relations, we identify novel topological phases and construct tight-binding models that realize them.

This paper is organized as follows.
Our analysis begins in Sec.~\ref{sec:Transformation_rule} with 2D systems.
Using the sewing matrices, we describe how the real Berry curvature, which is used to define the Euler class, transforms under point group operations.
We clarify that the transformation laws differ depending on whether the determinant of the sewing matrices in the real-gauge is $+1$ or $-1$.
By applying our general framework to the twofold rotation $\mathcal{C}_{2z}$, we confirm that it reproduces the results of previous studies on the Euler class and the second Stiefel-Whitney class.

Next, in Sec.~\ref{sec:Euler_C4z}, we examine $\mathcal{C}_{4z}$ symmetry as an example of a point group operation.
We derive a general formula that relates the Euler class to the $\mathcal{C}_{4z}$ and  $\mathcal{C}_{2z}$ eigenvalues of the occupied bands.
Although this formula is not universally applicable for determining the Euler class, examining the conditions for its applicability naturally leads to the notion of representation-protected topological phases.
We then establish a relationship between the Euler class and the $\mathbb{Z}_2$ invariant defined by Fu~\cite{PhysRevLett.106.106802}.
Notably, possible combinations of the values of these two invariants depend on the $\mathcal{C}_{2z}$ eigenvalue at the X point.
Based on the results above, our analysis is extended to 3D systems.
We focus on the differences between the topological invariants defined on the $k_z = 0$ and $k_z = \pi$ planes and derive an explicit formula connecting them.
Additionally, we investigate a previously unexplored 3D topological phase that is predicted to exist by our general formula.
This phase is characterized by a nontrivial Euler class and the trivial $\mathbb{Z}_2$ invariant.
We elucidate the bulk-boundary correspondence of this phase through the construction and numerical calculation of a 3D tight-binding model.
Subsequently, we extend our framework to systems with $\mathcal{C}_{4v}$ symmetry, incorporating halved mirror chirality into our formula.
The derived relationship provides a unified picture of the phase transition processes for each topological invariant.
Furthermore, the degenerate points of the occupied bulk bands are located at nearly the same positions as those of the gapless surface bands.

Finally, in Sec.~\ref{sec:Euler_C6z}, we turn to $\mathcal{C}_{6z}$ symmetry as an example of a point group operation.
Although we follow the same computational procedure as in $\mathcal{C}_{4z}$-symmetric systems, several conclusions differ in the case of 3D systems. 
We derive a general expression that relates the Euler class to the $\mathcal{C}_{6z}$, $\mathcal{C}_{3z}$, and $\mathcal{C}_{2z}$ eigenvalues of the occupied bands.
After examining the conditions under which the Euler class can be determined without ambiguity, we establish a relationship between the Euler class and the $\mathbb{Z}_2$ invariant, which incorporates the $\mathcal{C}_{2z}$ eigenvalues at the $\Gamma$ and M points. 
We then focus on the differences between the topological invariants defined on the two $C_{2z} T$-invariant planes in 3D systems.
The relationship between them differs significantly from that in $\mathcal{C}_{4z}$-symmetric systems.
Notably, the relative sign of the Euler classes defined on the two $C_{2z} T$-invariant planes plays a crucial role in relation to the $\mathbb{Z}_2$ invariant.
Building on this insight, we construct a tight-binding model that realizes a topological phase in which the Euler classes on the $k_z = 0$ and $k_z = \pi$ planes have opposite signs.
We demonstrate this sign discrepancy through numerical calculations of the topological surface states.
Finally, we incorporate mirror symmetry into our discussion.
We derive a formula connecting the Euler class, $\mathbb{Z}_2$ invariants, and halved mirror chirality.
The relation between the $\mathbb{Z}_2$ invariants and the halved mirror chirality takes the same form as in $\mathcal{C}_{4v}$-symmetric systems.

In the Appendices, we present additional models and generalize the results for the representation-protected topological invariants. 
In Appendix~\ref{sec:2DTBs_e2_nu4}, we introduce $\mathcal{C}_{4z}$-symmetric 2D tight-binding models that realize phases where either the Euler class or the $\mathbb{Z}_2$ invariant is trivial, while the other is nontrivial. 
In Appendix~\ref{sec:nu4_chi4_Nbands}, we generalize the relationship between the two types of representation-protected invariants in $\mathcal{C}_{4v}$-symmetric systems to the general case with an arbitrary number of occupied bands. 
In Appendix~\ref{sec:C6TB_surf_PT}, we present the linking nodal line structure for a $\mathcal{PT}$-symmetric tight-binding model and discuss its connection to the Euler class computed on the $C_{2z}T$-invariant planes.
In Appendix~\ref{sec:C6TB_surf_different_cell}, we present the surface states of the tight-binding model discussed in the main text for a different choice of unit cell.
Finally, in Appendix~\ref{sec:nu6_chi6_Nbands}, we present a generalization of the relationship for the $\mathcal{C}_{6v}$-symmetric case, analogous to that presented in Appendix~\ref{sec:nu4_chi4_Nbands} for the $\mathcal{C}_{4v}$-symmetric case.

In this paper, we assume that spin-orbit coupling is negligibly weak and treat electrons as spinless fermions.
We denote symmetry operations acting on electronic states by calligraphic letters and those in the $k$-space by italic letters.

\section{Transformation rule for the real Berry curvature under point group symmetries}
\label{sec:Transformation_rule}

We start our discussion with 2D systems defined in the $xy$-plane and possessing $\mathcal{C}_{2z}\mathcal{T}$ symmetry.
Since the $\mathcal{C}_{2z}\mathcal{T}$ operator satisfies $(\mathcal{C}_{2z}\mathcal{T})^2 = +1$, which holds in both spinless and spinful systems, we can adopt a real-gauge for the wavefunctions $\ket{u_n(\bm{k})}$ so that they satisfy
\begin{equation}
\mathcal{C}_{2z}\mathcal{T} \ket{\tilde{u}_n(\bm{k})} = \ket{\tilde{u}_n(\bm{k})},
\label{eq:real-gauge}
\end{equation}
where $n$ is the band index.
Hereafter, we use the tilde to denote the real-gauge.
When the number of occupied bands is two, the gapped Hamiltonian is topologically classified by a $\Z$-valued invariant called the Euler class~\cite{PhysRevX.9.021013,PhysRevB.102.115135,hatcherVBKT}.
The Euler class $e_2$ is defined by the flux integral over the BZ as
\begin{equation}
    e_2 = \dfrac{1}{2\pi} \int_{\mathrm{BZ}} d\bm{S} \cdot \tilde{\bm{F}}_{12} ,
\end{equation}
where $\tilde{\bm{F}}$ denotes the real Berry curvature $\tilde{\bm{F}}_{mn}(\bm{k}) = \nabla_{\bm{k}} \cross \tilde{\bm{A}}_{mn}(\bm{k})$ and $\tilde{\bm{A}}$ the real Berry connection $\tilde{\bm{A}}_{mn}(\bm{k}) = \braket{\tilde{u}_m(\bm{k})}{\nabla_{\bm{k}}\tilde{u}_n(\bm{k})}$ ($m, n \in \{1, 2\}$).
Since $\tilde{\bm{A}}$ and $\tilde{\bm{F}}$ are antisymmetric, $e_2$ can also be expressed as 
\begin{equation}
    e_2 = \dfrac{1}{4\pi i} \int_{\mathrm{BZ}} d\bm{S} \cdot \Tr\qty[\sigma_2 \tilde{\bm{F}}]
\end{equation}
using the second Pauli matrix $\sigma_2$.
Note that the Euler class is well-defined only for orientable real states.
Hence, we restrict gauge transformations to matrices in the SO(2) group.

Let the Hamiltonian $H(\bm{k})$ be symmetric under point group operation $\mathcal{R}$, i.e., $\mathcal{R} H(\bm{k}) \mathcal{R}^\dagger = H(R\bm{k})$, where the $2 \times 2$ matrix $R$ represents the point group operation in the $k$-space.
As noted at the end of the Introduction, symmetry operations acting on electronic states are denoted by calligraphic letters, while those in $k$-space are denoted by italic letters.
The sewing matrix, which is defined as
\begin{equation}
    \qty[\tilde{\mathcal{B}}_R(\bm{k})]_{mn} = \mel{\tilde{u}_m(R\bm{k})}{\mathcal{R}}{\tilde{u}_n(\bm{k})},
\end{equation}
relates the real Berry connection at the wave vector $\bm{k}$ to that at $R\bm{k}$ through
\begin{align}
    \label{eq:A_transform_R}
    & \qty[\tilde{A}_\mu(\bm{k})]_{mn} \nonumber\\
    =& \sum_{p,q} \qty[\tilde{\mathcal{B}}_R(\bm{k})]_{pm}^* \bra{\tilde{u}_p(R\bm{k})} \mathcal{R} \partial_\mu \qty[\mathcal{R}^\dagger \ket{\tilde{u}_q(R\bm{k})} \qty[\tilde{\mathcal{B}}_R(\bm{k})]_{qn}] \nonumber\\
    =& \sum_\nu R^{\mathrm{T}}_{\mu\nu} \qty[\tilde{\mathcal{B}}_R^\dagger(\bm{k}) \tilde{A}_\nu(R\bm{k}) \tilde{\mathcal{B}}_R(\bm{k})]_{mn} + \qty[\tilde{\mathcal{B}}_R^\dagger(\bm{k}) \partial_\mu \tilde{\mathcal{B}}_R(\bm{k})]_{mn},
\end{align}
where $\mu, \nu \in \{x, y\}$ and $\partial_\mu$ denotes the derivative with respect to $k_\mu$.
In the real-gauge, the sewing matrix can be represented as a real orthogonal matrix, whose determinant is restricted to either $+1$ or $-1$.
When $\det \qty[\tilde{\mathcal{B}}_R(\bm{k})] = +1$, it can be expressed as $\tilde{\mathcal{B}}_R(\bm{k}) = \exp[-i \phi_R(\bm{k}) \sigma_2]$, with $\phi_R(\bm{k})$ being a real function defined in the BZ.
By substituting this expression into Eq.~\eqref{eq:A_transform_R}, we obtain
\begin{equation}
    \label{eq:A_symmR_det+1}
    \tilde{A}_\mu(\bm{k}) = \sum_\nu  R^{\mathrm{T}}_{\mu\nu} \tilde{A}_\nu(R\bm{k}) - i \sigma_2 \partial_{\mu} \phi_R(\bm{k}).
\end{equation}
As a consequence, the real Berry curvature satisfies
\begin{align}
    \label{eq:F_symmR_det+1}
    \tilde{F}_z(\bm{k}) &= \partial_{k_x} \qty(\sum_\nu  R^{\mathrm{T}}_{y\nu} \tilde{A}_\nu(R\bm{k}) - i \sigma_2 \partial_{k_y} \phi_R(\bm{k})) - (k_x \leftrightarrow k_y) \nonumber\\ 
    &= \sum_{\mu, \nu} R^{\mathrm{T}}_{x\mu}  R^{\mathrm{T}}_{y\nu} \qty(\partial_\mu \tilde{A}_\nu(R\bm{k}) - \partial_\nu \tilde{A}_\mu(R\bm{k})).
\end{align}
On the other hand, when $\det \qty[\tilde{\mathcal{B}}_R(\bm{k})] = -1$, it can be expressed as $\tilde{\mathcal{B}}_R(\bm{k}) = \sigma_3 \exp[-i \phi_R(\bm{k}) \sigma_2]$, where $\sigma_3$ is the third Pauli matrix.
By substituting this expression into Eq.~\eqref{eq:A_transform_R}, we obtain
\begin{equation}
    \label{eq:A_symmR_det-1}
    \tilde{A}_\mu(\bm{k}) = - \sum_\nu  R^{\mathrm{T}}_{\mu\nu} \tilde{A}_\nu(R\bm{k}) - i \sigma_2 \partial_{\mu} \phi_R(\bm{k}),
\end{equation}
which leads to
\begin{align}
    \label{eq:F_symmR_det-1}
    \tilde{F}_z(\bm{k}) = - \sum_{\mu, \nu} R^{\mathrm{T}}_{x\mu}  R^{\mathrm{T}}_{y\nu} \qty(\partial_\mu \tilde{A}_\nu(R\bm{k}) - \partial_\nu \tilde{A}_\mu(R\bm{k})).
\end{align}
By respectively comparing Eq.~\eqref{eq:A_symmR_det+1} with Eq.~\eqref{eq:A_symmR_det-1} and Eq.~\eqref{eq:F_symmR_det+1} with Eq.~\eqref{eq:F_symmR_det-1}, we find that the sign of each transformation rule depends on the symmetry of the occupied bands through the form of $\tilde{\mathcal{B}}_R(\bm{k})$.

\begin{figure}[htp]
\includegraphics[width=8.5cm]{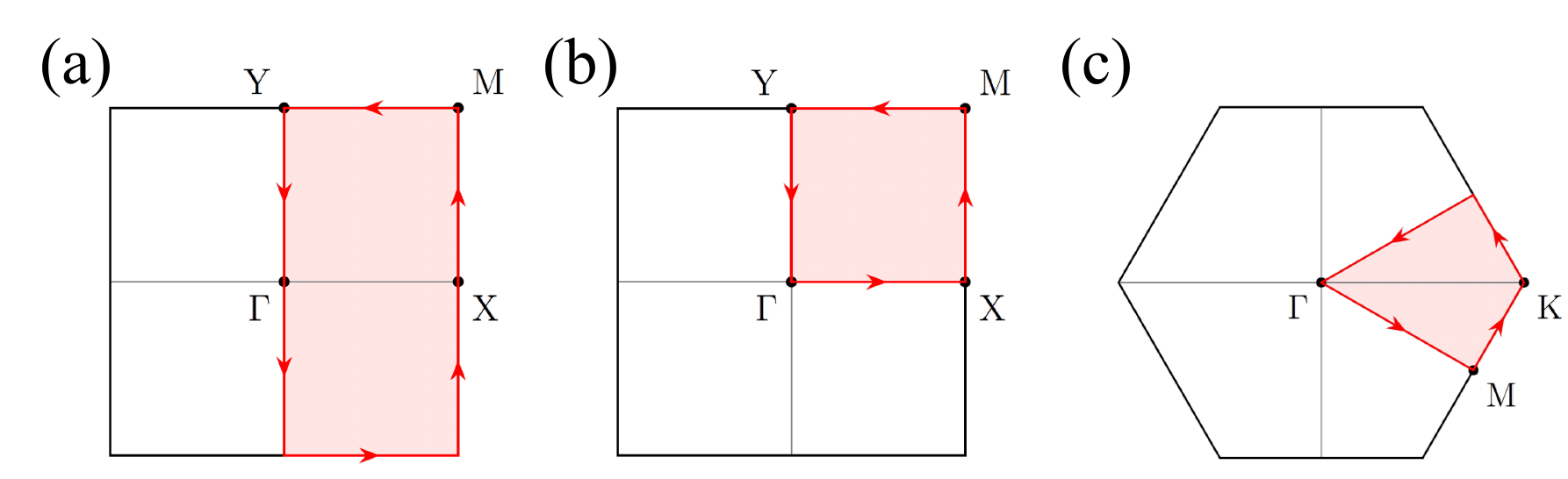}
\caption{
Integration domains and paths for calculating the Euler class in (a) $\mathcal{C}_{2z}$-, (b) $\mathcal{C}_{4z}$-, and (c) $\mathcal{C}_{6z}$-symmetric systems.
The red-shaded regions in (a), (b), and (c) indicate the hBZ, qBZ, and sBZ, respectively, while the red lines show the integration paths, with arrows indicating their directions.
High-symmetry points in the BZs are also shown.
}
\label{fig:BZ}
\end{figure}

As an example of the point group operation $\mathcal{R}$, we consider the twofold rotation $\mathcal{C}_{2z}$.
In this case, the transformation matrix in the $k$-space is given by $R = -\sigma_0$, where $\sigma_0$ denotes the $2 \times 2$ identity matrix.  
When $\tilde{\mathcal{B}}_{C_{2z}}(\bm{k}) = \exp[-i \phi_{C_{2z}}(\bm{k}) \sigma_2]$,  
Eq.~\eqref{eq:A_symmR_det+1} becomes  
\begin{equation}
    \label{eq:A_C2z_det+1}
    \tilde{A}_\mu(\bm{k}) = - \tilde{A}_\mu(-\bm{k}) - i \sigma_2 \partial_{\mu} \phi_{C_{2z}}(\bm{k}),
\end{equation}
and Eq.~\eqref{eq:F_symmR_det+1} simplifies to
\begin{equation}
    \tilde{F}_z(\bm{k}) = \tilde{F}_z(-\bm{k}).
\end{equation}
Since $\tilde{F}_z(\bm{k})$ is an even function of $\bm{k}$, the Euler class $e_2$ is given by
\begin{align}
    \dfrac{1}{2} e_2 &= \dfrac{1}{4\pi i} \int_{\mathrm{hBZ}} d\bm{S} \cdot \Tr\qty[\sigma_2 \tilde{\bm{F}}] \nonumber\\
    &= \dfrac{1}{4\pi i} \int_{-\pi}^\pi dk_y \; \Tr\qty[\sigma_2 \tilde{A}_y(\pi, k_y)] \nonumber\\
    &\phantom{{}={}} - \dfrac{1}{4\pi i} \int_{-\pi}^\pi dk_y \; \Tr\qty[\sigma_2 \tilde{A}_y(0, k_y)],
\end{align}
where hBZ denotes a half of the BZ shown in Fig.~\ref{fig:BZ}(a).
From Eq.~\eqref{eq:A_C2z_det+1}, the first term in the last line can be evaluated as
\begin{align}
    &\int_{-\pi}^\pi dk_y \; \Tr\qty[\sigma_2 \tilde{A}_y(\pi, k_y)] \nonumber\\
    =& \int_{0}^\pi dk_y \; \Tr\qty[\sigma_2 \qty(\tilde{A}_y(\pi, k_y) + \tilde{A}_y(\pi, -k_y))] \nonumber\\
    =& \int_{0}^\pi dk_y \; (-2i) \partial_{k_y} \phi_{C_{2z}}(\pi, k_y) \nonumber\\
    =& -2i \qty(\phi_{C_{2z}}(\pi, \pi) - \phi_{C_{2z}}(\pi, 0)).
\end{align}
Applying the same computation to the second term, we obtain
\begin{align}
    \dfrac{1}{2} e_2 &= - \dfrac{1}{2\pi} \qty(\phi_{C_{2z}}(\pi, \pi) - \phi_{C_{2z}}(\pi, 0)) \nonumber\\
    &\phantom{{}={}} + \dfrac{1}{2\pi} \qty(\phi_{C_{2z}}(0, \pi) - \phi_{C_{2z}}(0, 0)) \mod 1.
 \end{align}
Then, it follows that
\begin{equation}
    \label{eq:e2_C2z_det+1}
    (-1)^{e_2} = e^{-i\phi_{C_{2z}}(\Gamma)} e^{-i\phi_{C_{2z}}(\mathrm{M})} e^{+i\phi_{C_{2z}}(\mathrm{X})} e^{+i\phi_{C_{2z}}(\mathrm{Y})},
\end{equation}
where $\Gamma = (0, 0)$, $\mathrm{M} = (\pi, \pi)$, $\mathrm{X} = (\pi, 0)$, and $\mathrm{Y} = (0, \pi)$.
The eigenvalues of $\tilde{\mathcal{B}}_{C_{2z}}(\bm{k}) = \exp[-i \phi_{C_{2z}}(\bm{k}) \sigma_2]$ are $e^{+i \phi_{C_{2z}}(\bm{k})}$ and $e^{-i \phi_{C_{2z}}(\bm{k})}$.
For $C_{2z}$-invariant $\bm{k}$ points, such as $\Gamma$, M, X, and Y, these eigenvalues correspond to the $\mathcal{C}_{2z}$ eigenvalues of the occupied bands at those points.
Therefore, Eq.~\eqref{eq:e2_C2z_det+1} establishes the connection between $e_2$ and the $\mathcal{C}_{2z}$ eigenvalues of the occupied bands.
Here we must note an important point: Equation~\eqref{eq:e2_C2z_det+1} contains only one eigenvalue from the complex conjugate pair $\qty{e^{+i \phi_{C_{2z}}(\bm{k})}, e^{-i \phi_{C_{2z}}(\bm{k})}}$.
Thus, in order to obtain $e_2$ by using Eq.~\eqref{eq:e2_C2z_det+1}, we generally need to determine which of the $\mathcal{C}_{2z}$ eigenvalues corresponds to $e^{+i \phi_{\mathcal{C}_{2z}}(\bm{k})}$ and which corresponds to $e^{-i \phi_{\mathcal{C}_{2z}}(\bm{k})}$.  
However, this issue does not arise when considering the $\mathcal{C}_{2z}$ operation in spinless systems, since the eigenvalues are real and necessarily satisfy $e^{+i \phi_{\mathcal{C}_{2z}}(\bm{k})} = e^{-i \phi_{\mathcal{C}_{2z}}(\bm{k})}$.
Consequently, using Eq.~\eqref{eq:e2_C2z_det+1}, we can determine $e_2$ modulo 2 from the $\mathcal{C}_{2z}$ eigenvalues.

On the other hand, when $\tilde{\mathcal{B}}_{C_{2z}}(\bm{k}) = \sigma_3 \exp[-i \phi_{C_{2z}}(\bm{k}) \sigma_2]$, Eq.~\eqref{eq:A_symmR_det-1} becomes
\begin{equation}
    \label{eq:A_C2z_det-1}
    \tilde{A}_\mu(\bm{k}) = \tilde{A}_\mu(-\bm{k}) - i \sigma_2 \partial_{\mu} \phi_{C_{2z}}(\bm{k}),
\end{equation}
and Eq.~\eqref{eq:F_symmR_det-1} simplifies to
\begin{equation}
    \tilde{F}_z(\bm{k}) = - \tilde{F}_z(-\bm{k}).
\end{equation}
Thus, we obtain $e_2 = 0$.

The above results are consistent with the formula~\cite{PhysRevLett.121.106403} for the second Stiefel-Whitney class $w_2$, which is a $\mathbb{Z}_2$ invariant equal to $e_2$ modulo 2:
\begin{equation}
    (-1)^{w_2} = \prod_{i=1}^4 (-1)^{\lfloor N_{\mathrm{occ}}^-(\Gamma_i)/2 \rfloor},
\end{equation}
where $\{\Gamma_i\}$ are the $C_{2z}$-invariant points, $N_{\mathrm{occ}}^{-}(\Gamma_i)$ is the number of occupied bands with negative $\mathcal{C}_{2z}$ eigenvalues at $\Gamma_i$, and $\lfloor\cdot\rfloor$ represents the floor function. 

\begin{table}[tb]
\caption{
Summary of formulas expressing the Euler class in $\mathcal{C}_{nz}$-symmetric systems.
The second column indicates whether the determinant of the sewing matrix for the $\mathcal{C}_{nz}$ operator is $+1$ or $-1$ when computed in the real-gauge.
The third column shows the formulas that relate the Euler class $e_2$ to the rotation eigenvalues of the system, with references to equations in the main text.
}
\label{tab:summary_e2}
\begin{tabular}{ccc}
\hline\hline
\noalign{\vskip 1.5pt}
$n$ & $\det \qty[\tilde{\mathcal{B}}_{C_{nz}}]$ & Formulas for the Euler class $e_2$ \\
\noalign{\vskip 1.5pt}
\hline 
\noalign{\vskip 1.2pt}
2 & $+1$ & $(-1)^{e_2} = e^{-i\phi_{C_{2z}}(\Gamma)} e^{-i\phi_{C_{2z}}(\mathrm{M})} e^{+i\phi_{C_{2z}}(\mathrm{X})} e^{+i\phi_{C_{2z}}(\mathrm{Y})}$\, [Eq.~\eqref{eq:e2_C2z_det+1}] \\
\noalign{\vskip 1.2pt}
2 & $-1$ & $e_2 = 0$ \\
\noalign{\vskip 1.2pt}
4 & $+1$ & $i^{e_2} = e^{+i\phi_{C_{4z}}(\Gamma)} e^{+i\phi_{C_{4z}}(\mathrm{M})} e^{-i\phi_{C_{2z}}(\mathrm{X})}$\, [Eq.~\eqref{eq:e2_C4z_det+1}] \\
\noalign{\vskip 1.2pt}
4 & $-1$ & $e_2 = 0$ \\
\noalign{\vskip 1.2pt}
6 & $+1$ & $e^{i \pi e_2/3} = e^{+i\phi_{C_{6z}}(\Gamma)} e^{-i\phi_{C_{2z}}(\mathrm{M})} e^{+i\phi_{C_{3z}}(\mathrm{K})}$\, [Eq.~\eqref{eq:e2_C6z_det+1}] \\
\noalign{\vskip 1.2pt}
6 & $-1$ & $e_2 = 0$ \\
\hline\hline
\end{tabular} 
\end{table}

In the following sections, we apply this framework to the $\mathcal{C}_{4z}$ and $\mathcal{C}_{6z}$ operations to derive explicit formulas for the Euler class. 
The results are summarized in Table~\ref{tab:summary_e2}.

\section{Euler class in $\mathcal{C}_{4z}$-symmetric systems}
\label{sec:Euler_C4z}

In this section, we consider the fourfold rotation $\mathcal{C}_{4z}$ in spinless systems, whose transformation matrix in the 2D $k$-space is given by $R = -i \sigma_2$.

We begin by discussing 2D systems.
We first derive a formula that relates the Euler class $e_2$ to the rotational eigenvalues of the occupied bands.
Subsequently, we introduce another topological invariant protected by $\mathcal{C}_{4z}$ and $\mathcal{T}$ symmetries, denoted by $\nu_4$, and establish an explicit relation between $e_2$ and $\nu_4$ by expressing $\nu_4$ in the real-gauge.
We then extend our analysis to 3D systems.
We examine the differences between each type of topological invariant characterizing the $C_{2z} T$-invariant $k_z = 0, \pi$ planes, based on both analytical calculations and numerical results for a tight-binding model.
Moreover, we explore the connection with an additional topological invariant that arises in the presence of mirror symmetry.
Finally, we provide a unified perspective on the phase transition processes for these topological invariants.

\subsection{Euler class in 2D systems}
\label{subsec:Euler_C4z_2D}
First, we consider the case where $\tilde{\mathcal{B}}_{C_{4z}}(\bm{k}) = \sigma_3 \exp[-i \phi_{C_{4z}}(\bm{k}) \sigma_2]$.  
Under this assumption, Eqs.~\eqref{eq:A_symmR_det-1} and \eqref{eq:F_symmR_det-1} are rewritten as  
\begin{equation}
    \label{eq:A_C4z_det-1}
    \left\{ \,
        \begin{aligned}
        & \tilde{A}_x(\bm{k}) = - \tilde{A}_y(C_{4z}\bm{k}) - i \sigma_2 \partial_{k_x} \phi_{C_{4z}}(\bm{k}) \\
        & \tilde{A}_y(\bm{k}) = \tilde{A}_x(C_{4z}\bm{k}) - i \sigma_2 \partial_{k_y} \phi_{C_{4z}}(\bm{k}) 
        \end{aligned}
    \right.
\end{equation}
and  
\begin{equation}
    \tilde{F}_z(\bm{k}) = - \tilde{F}_z(C_{4z}\bm{k}),
\end{equation}
respectively.
Consequently, $e_2 = 0$ follows.

In contrast, when $\tilde{\mathcal{B}}_{C_{4z}}(\bm{k}) = \exp[-i \phi_{C_{4z}}(\bm{k}) \sigma_2]$, Eqs.~\eqref{eq:A_symmR_det+1} and \eqref{eq:F_symmR_det+1} transform into
\begin{equation}
    \label{eq:A_C4z_det+1}
    \left\{ \,
        \begin{aligned}
        & \tilde{A}_x(\bm{k}) = \tilde{A}_y(C_{4z}\bm{k}) - i \sigma_2 \partial_{k_x} \phi_{C_{4z}}(\bm{k}) \\
        & \tilde{A}_y(\bm{k}) = - \tilde{A}_x(C_{4z}\bm{k}) - i \sigma_2 \partial_{k_y} \phi_{C_{4z}}(\bm{k}) 
        \end{aligned}
    \right.
\end{equation}
and
\begin{equation}
    \tilde{F}_z(\bm{k}) = \tilde{F}_z(C_{4z}\bm{k}),
\end{equation}
respectively.
Thus, the Euler class $e_2$ is given by
\begin{align}
    \label{eq:e2_C4z_Stokes}
    \dfrac{1}{4} e_2 &= \dfrac{1}{4\pi i} \int_{\mathrm{qBZ}} d\bm{S} \cdot \Tr\qty[\sigma_2 \tilde{\bm{F}}] \nonumber\\
    &= \dfrac{1}{4\pi i} \int_{0}^\pi dk_x \; \Tr\qty[\sigma_2 \tilde{A}_x(k_x, 0)] \nonumber\\
    &\phantom{{}={}} + \dfrac{1}{4\pi i} \int_{0}^\pi dk_y \; \Tr\qty[\sigma_2 \tilde{A}_y(\pi, k_y)] \nonumber\\
    &\phantom{{}={}} - \dfrac{1}{4\pi i} \int_{0}^\pi dk_x \; \Tr\qty[\sigma_2 \tilde{A}_x(k_x, \pi)] \nonumber\\
    &\phantom{{}={}} - \dfrac{1}{4\pi i} \int_{0}^\pi dk_y \; \Tr\qty[\sigma_2 \tilde{A}_y(0, k_y)],
\end{align}
where qBZ denotes a quarter of the BZ shown in Fig.~\ref{fig:BZ}(b).
As Eq.~\eqref{eq:A_C4z_det+1} leads to
\begin{align}
    &\int_{0}^\pi dk_x \; \Tr\qty[\sigma_2 \tilde{A}_x(k_x, 0)] - \int_{0}^\pi dk_y \; \Tr\qty[\sigma_2 \tilde{A}_y(0, k_y)] \nonumber\\
    =& \int_{0}^\pi dk_x \; \Tr\qty[\sigma_2 \qty{\tilde{A}_y(0, k_x) - i \sigma_2 \partial_{k_x} \phi_{C_{4z}}(k_x, 0)}] \nonumber\\
    \phantom{{}={}}& - \int_{0}^\pi dk_y \; \Tr\qty[\sigma_2 \tilde{A}_y(0, k_y)] \nonumber\\
    =& -2i \qty(\phi_{C_{4z}}(\pi, 0) - \phi_{C_{4z}}(0, 0)),
\end{align}
Eq.~\eqref{eq:e2_C4z_Stokes} simplifies to
\begin{align}
    \dfrac{1}{4} e_2 &= - \dfrac{1}{2\pi} \qty(\phi_{C_{4z}}(\pi, 0) - \phi_{C_{4z}}(0, 0)) \nonumber\\
    &\phantom{{}={}} + \dfrac{1}{2\pi} \qty(\phi_{C_{4z}}(\pi, \pi) - \phi_{C_{4z}}(0, \pi)) \mod 1.
\end{align}
Thus, we find that
\begin{align}
    \label{eq:e2_C4z_det+1}
    i^{e_2} 
    &= e^{+i\phi_{C_{4z}}(\Gamma)} e^{+i\phi_{C_{4z}}(\mathrm{M})} e^{-i[\phi_{C_{4z}}(\mathrm{X}) + \phi_{C_{4z}}(\mathrm{Y})]} \nonumber\\
    &= e^{+i\phi_{C_{4z}}(\Gamma)} e^{+i\phi_{C_{4z}}(\mathrm{M})} e^{-i\phi_{C_{2z}}(\mathrm{X})},
\end{align}
where the last equality follows from
\begin{align}
    \tilde{\mathcal{B}}_{C_{2z}}(\mathrm{X}) &= \tilde{\mathcal{B}}_{C_{4z}}(\mathrm{Y}) \tilde{\mathcal{B}}_{C_{4z}}(\mathrm{X}) \nonumber\\
    &= \exp\{-i [\phi_{C_{4z}}(\mathrm{X}) + \phi_{C_{4z}}(\mathrm{Y})] \sigma_2\}.
\end{align}
Equation~\eqref{eq:e2_C4z_det+1} relates the value of $e_2$ to the product of the $\mathcal{C}_{4z}$ eigenvalues at the $\Gamma$ and M points and the $\mathcal{C}_{2z}$ eigenvalue at the X point.
Note that the choice between the X and Y points in the last line of Eq.~\eqref{eq:e2_C4z_det+1} is arbitrary due to $\mathcal{C}_{4z}$ symmetry.
The relationship between $e_2$ and the rotational eigenvalues shown in Eq.~\eqref{eq:e2_C4z_det+1} is similar to that for the Chern number~\cite{PhysRevB.86.115112}.
However, unlike in the case of the Chern number, Eq.~\eqref{eq:e2_C4z_det+1} contains only one eigenvalue from each complex conjugate pair.
Since the $\mathcal{C}_{2z}$ eigenvalues in spinless systems are always real, the problem reduces to whether the $\mathcal{C}_{4z}$ eigenvalues are real or not.
If the $\mathcal{C}_{4z}$ eigenvalues are real, Eq.~\eqref{eq:e2_C4z_det+1} allows us to determine the value of $e_2$ modulo 4 from the rotational eigenvalues.
On the other hand, if the $\mathcal{C}_{4z}$ eigenvalues are $\pm i$, Eq.~\eqref{eq:e2_C4z_det+1} cannot be used to determine $e_2$ because the value of $e_2$ depends on whether $e^{+i\phi_{C_{4z}}}$ is taken to be $+i$ or $-i$.
In this case, we can instead derive another expression for $e_2$.
Taking the real part of Eq.~\eqref{eq:e2_C4z_det+1}, we obtain
\begin{align}
    \label{eq:e2_C4z_det+1_re}
    \cos\qty(\dfrac{\pi}{2}e_2) &= e^{-i \phi_{C_{2z}}(\mathrm{X})} \Bigl\{ \cos [\phi_{C_{4z}}(\Gamma)] \cos [\phi_{C_{4z}}(\mathrm{M})] \nonumber\\
    &\phantom{{}={}} - \sin [\phi_{C_{4z}}(\Gamma)] \sin [\phi_{C_{4z}}(\mathrm{M})] \Bigr\}.
\end{align}
For the $\mathcal{C}_{4z}$ eigenvalues at the $C_{4z}$-invariant momenta $\kinv$, one finds $e^{+ i\phi_{C_{4z}}(\kinv)} = \pm i$, implying $\cos [\phi_{C_{4z}}(\kinv)] = 0$ and thus eliminating the first term of Eq.~\eqref{eq:e2_C4z_det+1_re}.
Moreover, in this case, $\tilde{\mathcal{B}}_{C_{4z}}(\kinv)$ is an antisymmetric matrix with $\pm \sin [\phi_{C_{4z}}(\kinv)]$ as its off-diagonal elements, which gives $\sin [\phi_{C_{4z}}(\kinv)] = - \Pf \qty[\tilde{\mathcal{B}}_{C_{4z}}(\kinv)]$.  
Therefore, we find
\begin{equation}
    \label{eq:e2_Pf_GM}
    \cos\qty(\dfrac{\pi}{2}e_2) = - e^{-i \phi_{C_{2z}}(\mathrm{X})} \Pf \qty[\tilde{\mathcal{B}}_{C_{4z}}(\Gamma)] \Pf \qty[\tilde{\mathcal{B}}_{C_{4z}}(\mathrm{M})].
\end{equation}
This provides another expression for $e_2$, valid only when the $\mathcal{C}_{4z}$ eigenvalues are complex.

\subsection{Relationship between $e_2$ and $\nu_4$}
\label{subsec:relation_e2_nu4}

A basis consisting of two orbitals with angular momentum $l$ forms a 2D irreducible representation of $\mathcal{C}_{nz}$ under $\mathcal{T}$ symmetry when $2l \neq 0 \mod n$: $\mathcal{C}_{nz}=\exp(\frac{2\pi il}{n}\sigma_2)$.
We refer to such a basis as an orbital doublet.
For $n = 4$, a basis with $l = 1$ forms an orbital doublet, which corresponds to the $\{p_x, p_y\}$ or $\{d_{zx}, d_{yz}\}$ orbitals.
For such orbital doublets in $\mathcal{C}_{4z}$-symmetric systems, we can define the $\mathbb{Z}_2$ invariant $\nu_{4}$~\cite{PhysRevLett.106.106802}, which is given by
\begin{equation}
    \label{eq:nu4_def}
    (-1)^{\nu_{4}} = \exp\qty(i \int_{\Gamma}^{\mathrm{M}} d\bm{k} \cdot \bm{\mathcal{A}}(\bm{k})) \dfrac{\Pf \qty[w_4(\mathrm{M})]}{\Pf \qty[w_4(\Gamma)]},
\end{equation}
where $\bm{\mathcal{A}}(\bm{k})$ is the U(1) Berry connection
\begin{equation}
    \bm{\mathcal{A}}(\bm{k}) = -i \sum_{n}^{\mathrm{occ.}} \braket{u_n(\bm{k})}{\nabla_{\bm{k}}u_n(\bm{k})},
\end{equation}
and $w_4(\kinv)$ is an antisymmetric matrix defined by $\qty[w_4(\kinv)]_{mn} = \mel{u_m(\kinv)}{\mathcal{C}_{4z}\mathcal{T}}{u_n(\kinv)}$ at the $C_{4z}T$-invariant momenta $\kinv \in \{\Gamma, \mathrm{M}\}$, where energy levels are twofold degenerate due to the relation $(\mathcal{C}_{4z}\mathcal{T})^2 = -1$.
We use a calligraphic letter to denote the U(1) Berry connection, in order to distinguish it from the real Berry connection, which is written in an italic font.

The $\mathcal{C}_{nz}$ eigenvalues of two orbitals with angular momentum $l$ are given by $\exp(\pm i2\pi l/n)$, which are complex for orbital doublets.
Thus, the condition on band representations for defining $\nu_{4}$ is equivalent to the condition under which the expression for $e_2$ in Eq.~\eqref{eq:e2_Pf_GM} is valid.
To establish the relationship between $e_2$ and $\nu_4$, we express $\nu_4$ in the real-gauge.
When we adopt the real-gauge, the U(1) Berry connection $\mathcal{A}(\bm{k})$ vanishes, and $w_4(\kinv)$ can be expressed in terms of the sewing matrix of the $\mathcal{C}_{4z}$ operation as follows:
\begin{align}
    \qty[\tilde{w}_4(\kinv)]_{mn} &= \mel{\tilde{u}_m(\kinv)}{\mathcal{C}_{4z}\mathcal{C}_{2z}\mathcal{C}_{2z}\mathcal{T}}{\tilde{u}_n(\kinv)} \nonumber\\
    &= \mel{\tilde{u}_m(\kinv)}{\mathcal{C}_{4z}\mathcal{C}_{2z}}{\tilde{u}_n(\kinv)} \nonumber\\
    &= - \qty[\tilde{\mathcal{B}}_{C_{4z}}(\kinv)]_{mn},
\end{align}
where we utilized the fact that the 2D representation under consideration shares the same symmetry as the $\{p_x, p_y\}$ orbitals and acquires a minus sign under the $\mathcal{C}_{2z}$ operation.
Therefore, we obtain 
\begin{equation}
    \label{eq:nu4_Pf_GM}
    (-1)^{\nu_{4}} = \dfrac{\Pf \qty[\tilde{\mathcal{B}}_{C_{4z}}(\mathrm{M})]}{\Pf \qty[\tilde{\mathcal{B}}_{C_{4z}}(\Gamma)]}.
\end{equation}
We note that $\Pf \qty[\tilde{\mathcal{B}}_{C_{4z}}(\kinv)]$ is gauge-invariant since we consider only gauge transformations within the SO(2) group as discussed in Sec.~\ref{sec:Transformation_rule}.
Noting that $\Pf \qty[\tilde{\mathcal{B}}_{C_{4z}}(\kinv)] = \pm 1$, we obtain  
\begin{equation}
    \label{eq:relation_e2_nu4}
    \cos\qty(\dfrac{\pi}{2}e_2) = - e^{-i \phi_{C_{2z}}(\mathrm{X})} (-1)^{\nu_{4}}
\end{equation}
from Eqs.~\eqref{eq:e2_Pf_GM} and \eqref{eq:nu4_Pf_GM}.
This establishes a general relationship between the two topological invariants, $e_2$ and $\nu_4$.

Although computing $e_2$ and $\nu_4$ previously required evaluating the Wilson loop spectra along different paths~\cite{PhysRevB.93.205104,PhysRevLett.121.106403,PhysRevX.9.021013,PhysRevB.99.235125,PhysRevB.100.195135,PhysRevLett.123.036401}, Eq.~\eqref{eq:relation_e2_nu4} provides a more direct and efficient way to determine $e_2$ from $\nu_4$ and the $\mathcal{C}_{2z}$ eigenvalue at the X point.
Table~\ref{tab:e2_nu4_values} presents the possible combinations of $e_2$ and $\nu_4$ values allowed by Eq.~\eqref{eq:relation_e2_nu4}.
For example, it is known that the Fu model consists of $p$ orbitals with the $\mathcal{C}_{2z}$ eigenvalue of $-1$, and its $C_{2z}T$-invariant planes realize the phases $(e_2, \nu_4) = (0,0)$ and $(2,1)$~\cite{PhysRevLett.106.106802,PhysRevB.104.195114}.
Table~\ref{tab:e2_nu4_values} includes both phases, ensuring consistency between our analysis and the numerical results reported in previous studies.
In cases where the $\mathcal{C}_{2z}$ eigenvalue at the X point is $+1$, one can find topological phases where either $e_2$ or $\nu_4$ is trivial while the other is nontrivial.
Tight-binding models that realize these phases are presented in Appendix~\ref{sec:2DTBs_e2_nu4}.

\begin{table}[thp]
    \centering
    \caption{Possible combinations of $e_2$ and $\nu_{4}$ in 2D systems where both invariants can be defined. These combinations satisfy the relation in Eq.~\eqref{eq:relation_e2_nu4}, with the open (filled) circles indicating cases where the $\mathcal{C}_{2z}$ eigenvalue at the X point is $-1$ ($+1$).
    }
    \label{tab:e2_nu4_values}
    \begin{tabular}{|l||l|l|l|l|l|l|l|l|l|}
        \hline
        \diagbox[width=4em,height=2em]{$\nu_{4}$}{$e_2$} & 0 & 1 & 2 & 3 & 4 & 5 & 6 & 7 & $\cdots$ \\ \hline\hline
        0 & $\circ$ &   & $\bullet$ &   & $\circ$ &   & $\bullet$ &  & \\ \hline
        1 & $\bullet$ &   & $\circ$ &   & $\bullet$ &   & $\circ$ &   & \\ \hline
    \end{tabular}
\end{table}

\subsection{Euler class in 3D systems}
\label{subsec:Euler_C4z_3D}

Next, we extend our discussion to 3D systems.  
Since the $k_z = 0$ and $k_z = \pi$ planes are $C_{2z}T$-invariant, the Euler class $e_2$ can be defined on these planes.  
Furthermore, these planes contain $C_{4z}T$-invariant points, and thus, when the occupied bands belong to the 2D representation, the $\mathbb{Z}_2$ invariant $\nu_4$ can also be defined.  
Therefore, the discussion on 2D systems presented above can be directly applied to the $k_z = 0, \pi$ planes in the 3D BZ.
Note that although the transformation matrix $R$ in the $k$-space extends to a $3 \times 3$ matrix, this extension is irrelevant to our discussion, as we focus solely on transformations within the $xy$-plane.

Let $\bar{k}_z \in \{0, \pi\}$, and denote the Euler class on the $k_z = \bar{k}_z$ plane by $e_2(\bar{k}_z)$.  
Similarly, we denote the $\mathbb{Z}_2$ invariant $\nu_4$ on the $k_z = 0$ plane, as defined in Eq.~\eqref{eq:nu4_def}, by $\nu_4(0)$.  
For the $k_z = \pi$ plane, $\nu_4(\pi)$ is defined analogously to Eq.~\eqref{eq:nu4_def}, using the $\mathrm{Z} = (0, 0, \pi)$ and $\mathrm{A} = (\pi, \pi, \pi)$ points instead of $\Gamma$ and M points.  
Then, the 3D topological phases are characterized by their differences $\bar{e}_2 = e_2(\pi) - e_2(0)$ and $\bar{\nu}_4 = \nu_4(\pi) - \nu_4(0) \mod 2$ ~\cite{PhysRevLett.106.106802,PhysRevB.90.165114,PhysRevB.99.235125,PhysRevB.104.195114}.

To see why 3D systems are characterized by the difference $\bar{e}_2$, it is instructive to consider $\mathcal{PT}$-symmetric systems.
In $\mathcal{PT}$-symmetric spinless systems, where the real-gauge can be chosen throughout the entire BZ, the Euler class can be defined on any closed surface that has no band degeneracy point.
It is known that any closed surface with a nontrivial Euler class necessarily contains linked nodal lines~\cite{PhysRevLett.121.106403,bouhon2022multigaptopologicalconversioneuler}.
Thus, assuming the absence of nodes on the $C_{2z} T$-invariant planes ($k_z=0,\pi$), band structures are characterized by the Euler class on the closed surface that encloses all nodes in the region $0 < k_z < \pi$.
Such a surface can be continuously deformed into a closed surface consisting of the $k_z = 0$ and $k_z = \pi$ planes together with the BZ boundary at $0 < k_z < \pi$.
By taking into account the cancellation due to the periodicity of the BZ, $e_2$ on the closed surface lying within $0 < k_z < \pi$ reduces to the difference between $e_2(0)$ and $e_2(\pi)$.
When inversion symmetry is broken while preserving $\mathcal{C}_{2z} \mathcal{T}$ symmetry, the linked nodal lines become gapped, but the difference $\bar{e}_2$ remains well-defined, thereby characterizing the topology of $\mathcal{C}_{2z} \mathcal{T}$-invariant 3D insulators.
Note that the sign of $\bar{e}_2$ is not gauge-invariant since the signs of $e_2(\bar{k}_z)$ flip under an O(2) transformation with a determinant of $-1$.
Moreover, the calculation of $\bar{e}_2$ requires consideration of the relative sign between $e_2(0)$ and $e_2(\pi)$.
In the following, we comment on the issue of the relative sign whenever it becomes relevant.

Following the discussion in the previous subsection, we can also derive the relationship between $\bar{e}_2$ and $\bar{\nu}_4$.
As indicated in Table~\ref{tab:e2_nu4_values}, for bands where $\nu_4$ can be defined, $e_2$ only takes even integer values.
This implies that
\begin{equation}
    \cos\qty(\dfrac{\pi}{2}\bar{e}_2) = \cos\qty(\dfrac{\pi}{2}e_2(\pi)) \cos\qty(\dfrac{\pi}{2}e_2(0)).
\end{equation}
Thus, from Eq.~\eqref{eq:e2_Pf_GM}, we obtain
\begin{align}
    \label{eq:e2bar_Pf_GMZA}
    &\cos\qty(\dfrac{\pi}{2}\bar{e}_2) \nonumber\\
    =& \Pf \qty[\tilde{\mathcal{B}}_{C_{4z}}(\mathrm{Z})] \Pf \qty[\tilde{\mathcal{B}}_{C_{4z}}(\mathrm{A})] \Pf \qty[\tilde{\mathcal{B}}_{C_{4z}}(\Gamma)] \Pf \qty[\tilde{\mathcal{B}}_{C_{4z}}(\mathrm{M})],
\end{align}
where we used the fact that the $\mathcal{C}_{2z}$ eigenvalues at the X point and at $\mathrm{R} = (\pi, 0, \pi)$ point are identical in $\mathcal{C}_{2z}$-symmetric insulators.
In addition, from Eq.~\eqref{eq:nu4_Pf_GM}, we obtain  
\begin{equation}
    \label{eq:nu4bar_Pf_GMZA}
    (-1)^{\bar{\nu}_4} = \dfrac{\Pf \qty[\tilde{\mathcal{B}}_{C_{4z}}(\mathrm{A})]}{\Pf \qty[\tilde{\mathcal{B}}_{C_{4z}}(\mathrm{Z})]} \dfrac{\Pf \qty[\tilde{\mathcal{B}}_{C_{4z}}(\Gamma)]}{\Pf \qty[\tilde{\mathcal{B}}_{C_{4z}}(\mathrm{M})]}.
\end{equation}
Consequently, we establish the relationship between $\bar{e}_2$ and $\bar{\nu}_4$ as  
\begin{equation}
    \label{eq:relation_e2bar_nu4bar}
    \cos\qty(\dfrac{\pi}{2}\bar{e}_2) = (-1)^{\bar{\nu}_4}.
\end{equation}
This result is consistent with the topological phase of the Fu model, characterized by $(\bar{e}_2, \bar{\nu}_4) = (2, 1)$~\cite{PhysRevLett.106.106802,PhysRevB.104.195114}.
We note that the left hand side of Eq.~\eqref{eq:relation_e2bar_nu4bar} is invariant not only under the simultaneous sign reversal $(e_2(0), e_2(\pi)) \to (-e_2(0), -e_2(\pi))$ but also under individual sign reversals, and therefore the issue of the relative sign does not arise in its application.

\subsection{Tight-binding model}

When $\bar{e}_2 \neq 0$ or $\bar{\nu}_4 = 1$, the system is classified as a strong topological insulator and is expected to host gapless surface states on the (001) surface respecting $\mathcal{C}_{4z}$ symmetry.
For instance, the Fu model, which realizes $(\bar{e}_2, \bar{\nu}_4) = (2, 1)$, exhibits nontrivial bulk topology in both invariants, which results in gapless surface bands on the (001) surface with a quadratic band touching at the $\bar{\mathrm{M}}$ point.

\begin{figure}[htp]
\includegraphics[width=8.5cm]{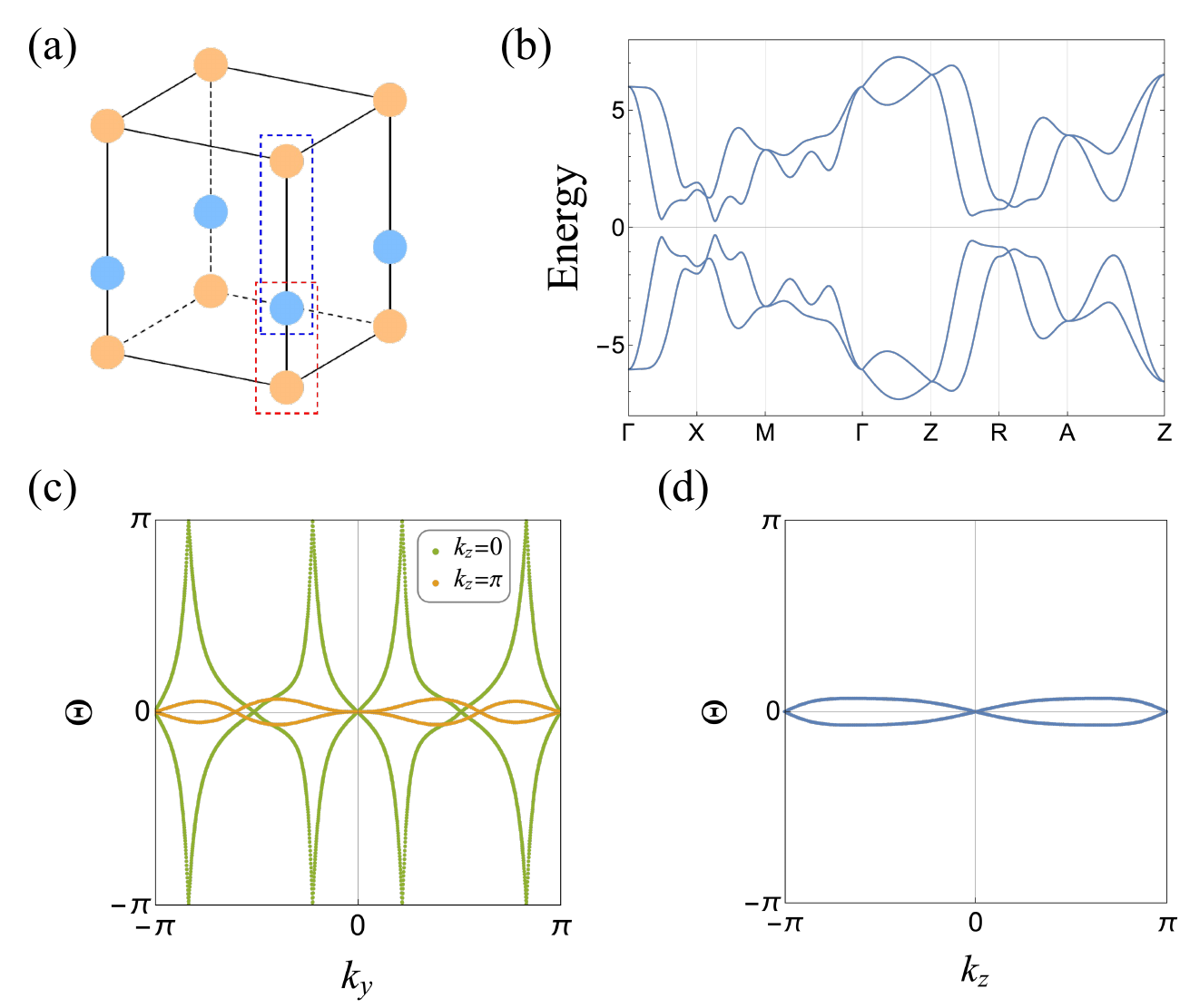}
\caption{
Lattice and topological properties of the Hamiltonian in Eq.~\eqref{eq:Hamil_3D_e2bar=4_nu4bar=0}, where the parameters are chosen as $(M, t_1, t_2, t_3, t_4, t_5, t_6, t_7, t_8, t_9, t_{1}', t_{2}', t_{3}', t_4') = (1.5, 0.5, 0.4, 0.1, 0.6, -0.8, 0.2, 0.3, 0.8, 0.3, -0.3, 0.15, 0.8, 0.4)$.
(a) Tetragonal lattice structure with two atoms per unit cell.
The red and blue dashed lines indicate different possible choices for defining the unit cell.
(b) Bulk band structure along high-symmetry lines.
(c) Wilson loop spectra obtained by integrating along the $k_x$ direction while keeping $k_y$ fixed within each $C_{2z} T$-invariant plane.
(d) Wilson loop spectrum obtained by integrating along a polygonal path connecting $(k_x, k_y) = (-\pi, \pi)$, $(0,0)$, and $(\pi, \pi)$ with fixed $k_z$.
}
\label{fig:3DTB_C4_bulk}
\end{figure}

Interestingly, Eq.~\eqref{eq:relation_e2bar_nu4bar} allows for a novel 3D topological phase with $(\bar{e}_2, \bar{\nu}_4) = (4, 0)$.
In this case, the presence of gapless surface states is suggested by $\bar{e}_2 = 4$; however, $\bar{\nu}_4 = 0$ appears to imply their absence.  
To investigate how this bulk topology manifests in surface states, we study the following 3D tight-binding model:
\begin{align}
    \label{eq:Hamil_3D_e2bar=4_nu4bar=0}
    &H(k_x, k_y, k_z) \nonumber\\
    =& [ M + 2t_1(\cos k_x + \cos k_y)  \nonumber\\
    \phantom{{}={}}& + 4t_2 \cos k_x \cos k_y + 2t_1' \cos k_z ] \Gamma_{3,0} \nonumber\\
    \phantom{{}={}}& + [2t_3 (\cos k_x - \cos k_y) + 2t_4 (\cos 2k_x - \cos 2k_y)] \Gamma_{3,3} \nonumber\\
    \phantom{{}={}}& + [ t_5 + 2t_6 (\cos k_x + \cos k_y) \nonumber\\
    \phantom{{}={}}& + 2 (t_7 + 2t_2' \cos k_z) (\cos 2k_x + \cos 2k_y)] \Gamma_{1,0} \nonumber\\
    \phantom{{}={}}& + 2t_8 (\cos k_x + \cos k_y) \Gamma_{2,2} + 4t_9 \sin k_x \sin k_y \Gamma_{3,1} \nonumber\\
    \phantom{{}={}}& + 2t_3' \sin k_z \Gamma_{2,0} + 2t_4' \sin k_z \Gamma_{1,2},
\end{align}
where the $\Gamma$ matrices are the direct products of the Pauli matrices $\sigma_i$ ($i=1,2,3$) and the $2\times2$ identity matrix $\sigma_0$, i.e.,  $\Gamma_{i,j} = \sigma_i \otimes \sigma_j$.
This Hamiltonian has fourfold rotation symmetry
\begin{equation}
    \mathcal{C}_{4z} H(k_x, k_y, k_z) \mathcal{C}_{4z}^\dagger = H(-k_y, k_x, k_z), \quad \mathcal{C}_{4z} = - i \Gamma_{0,2}
\end{equation}
and time-reversal symmetry
\begin{equation}
    \mathcal{T} H(k_x, k_y, k_z) \mathcal{T}^\dagger = H(-k_x, -k_y, -k_z), \quad \mathcal{T} = \Gamma_{0,0} K,
\end{equation}
where $K$ is the complex conjugation operator.
This model describes the $\{p_x, p_y\}$ orbitals on two sites aligned along the $z$-axis in a tetragonal lattice [see Fig.~\ref{fig:3DTB_C4_bulk}(a)], so that the matrix representations of the symmetry operations are identical to those in the Fu model~\cite{PhysRevLett.106.106802,PhysRevB.104.195114}.
Thus, the bands serve as the basis for the real 2D representation.
Under these conditions, both $e_2$ and $\nu_4$ can be defined for the gapped phases of this model.

Figure~\ref{fig:3DTB_C4_bulk}(b) shows the bulk band structure of this model, where all $C_{4z}T$-invariant points exhibit twofold degeneracy.
We now compute the bulk topological invariants using the Wilson loop method~\cite{PhysRevB.84.075119,PhysRevB.89.155114,PhysRevB.100.195135,PhysRevB.102.115117,PhysRevLett.121.106403,PhysRevX.6.021008}.
Figure~\ref{fig:3DTB_C4_bulk}(c) presents the Wilson loop spectra obtained by integrating along the $k_x$ direction while keeping $k_y$ fixed within each $C_{2z} T$-invariant plane. 
The Euler class $e_2(\bar{k}_z)$ is determined from the winding number of the Wilson loop spectrum in the $k_z = \bar{k}_z$ plane~\cite{PhysRevLett.121.106403,PhysRevX.9.021013,PhysRevB.99.235125,PhysRevB.100.195135}.
In the $k_z = 0$ plane, the Wilson loop spectrum consists of curves with winding numbers of $+4$ and $-4$, showing that $e_2(0) = 4$.  
On the other hand, the Wilson loop spectrum in the $k_z = \pi$ plane exhibits no winding, indicating that $e_2(\pi) = 0$.
Figure~\ref{fig:3DTB_C4_bulk}(d) illustrates the Wilson loop spectrum obtained by integrating along a polygonal path connecting $(k_x, k_y) = (-\pi, \pi)$, $(0,0)$, and $(\pi, \pi)$ with fixed $k_z$.
The $\mathbb{Z}_2$ invariant $\nu_4(\bar{k}_z)$ is equal to the parity of half the number of eigenvalues $\Theta(\bar{k}_z) = \pm\pi$~\cite{PhysRevB.93.205104}.  
Thus, we obtain $\nu_4(0) = \nu_4(\pi) = 0$.
The topological indices of the $k_z = 0, \pi$ planes computed by the Wilson loop method are both listed in Table~\ref{tab:e2_nu4_values}, supporting our discussion based on rotational symmetry.

\begin{figure}[htp]
\includegraphics[width=8.5cm]{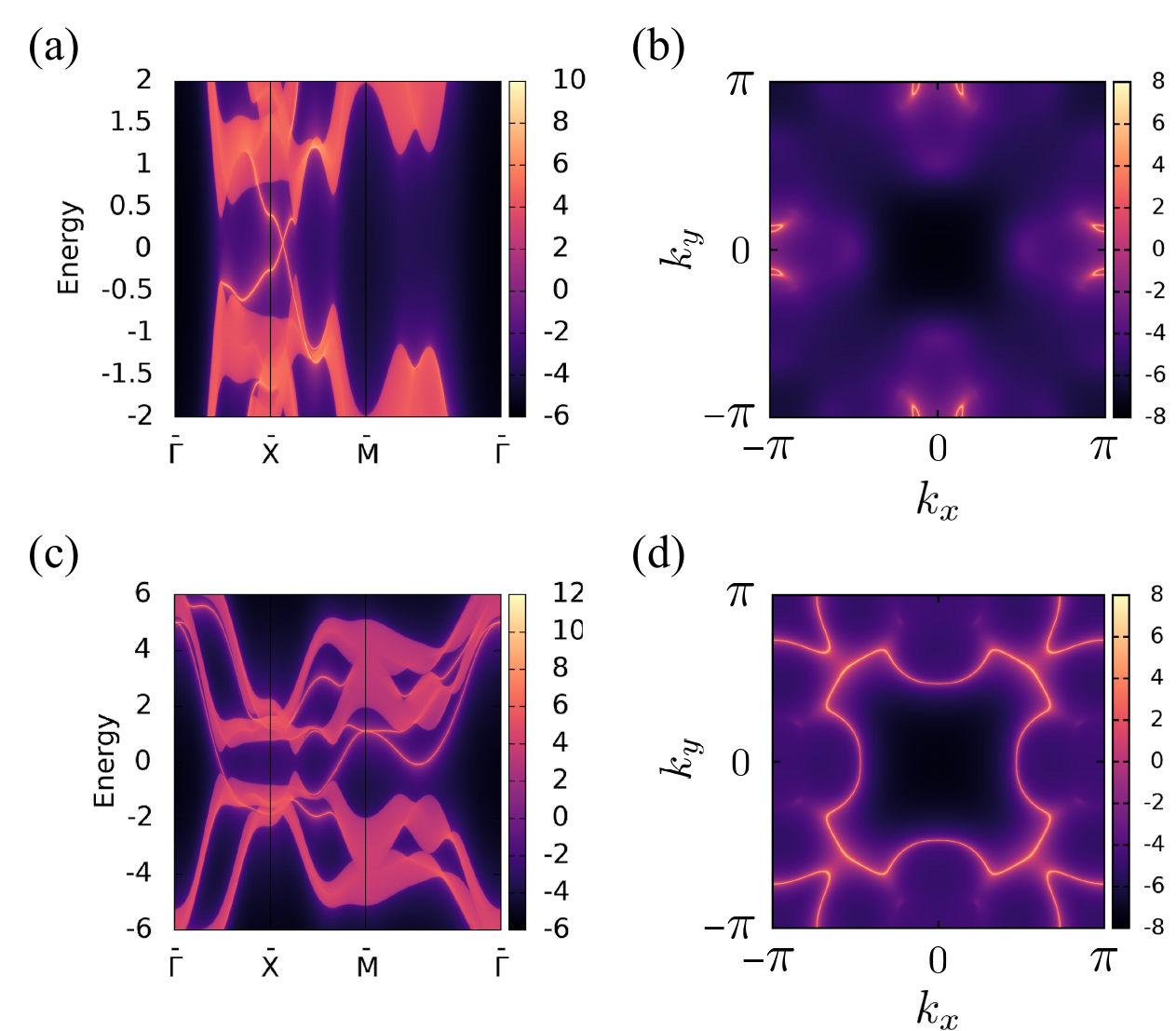}
\caption{
Surface states on the (001) surface of the Hamiltonian in Eq.~\eqref{eq:Hamil_3D_e2bar=4_nu4bar=0}, obtained using the Green's function method~\cite{PhysRevB.28.4397, MPLopezSancho_1984, MPLopezSancho_1985} implemented in WannierTools~\cite{WU2017}.
The parameters are the same as those in Fig.~\ref{fig:3DTB_C4_bulk}.
(a) Surface band structure along the high-symmetry lines in the surface BZ.
(b) Fermi lines for the (001) surface.
(c) and (d) show the same quantities as (a) and (b), respectively, but computed with a different unit cell.
}
\label{fig:3DTB_C4_surf}
\end{figure}

Having established that the model given by Eq.~\eqref{eq:Hamil_3D_e2bar=4_nu4bar=0} indeed realizes $\bar{e}_2 = 4$ and $\bar{\nu}_4 = 0$, we now investigate its surface states on the $C_{2z}T$-symmetric (001) surface.   
As shown in Fig.~\ref{fig:3DTB_C4_surf}(a), gapless surface states emerge, featuring a band-crossing point with linear dispersion on the $\bar{\mathrm{X}}$–$\bar{\mathrm{M}}$ line.
The Fermi lines (surfaces) shown in Fig.~\ref{fig:3DTB_C4_surf}(b) further reveal that four such surface Dirac cones are present within the surface BZ, arranged in a $C_{4z}$-symmetric configuration.
In general, twofold degeneracies in the $C_{2z}T$-invariant plane are topologically protected by the $\Z$-valued winding number~\cite{PhysRevB.104.195114}.
In this connection, the value of $\bar{e}_2$ is anticipated to coincide with the total winding number of the surface bands~\cite{PhysRevB.104.195114}.
The surface states shown in Figs.~\ref{fig:3DTB_C4_surf}(a) and (b) are consistent with $\bar{e}_2 = 4$.
Meanwhile, it has been anticipated that robust surface states appear only when $\bar{\nu}_4 = 1$.
For example, the Fu model characterized by $\bar{\nu}_4 = 1$ is known to host a quadratic band touching at the $\bar{\mathrm{M}}$ point on the (001) surface.
However, our model does not align with this expectation.

To address this apparent discrepancy, we reconsider the role of $\bar{\nu}_4$ in determining surface states.
Considering that $\nu_4$ is a $\mathbb{Z}_2$ topological invariant and is defined using the information at the $C_{4z}T$-invariant points, as shown in Eqs.~\eqref{eq:nu4_def} and \eqref{eq:nu4_Pf_GM}, it is natural to expect that $\bar{\nu}_4$ should determine the parity of the number of quadratic band touchings appearing at the $C_{4z}T$-invariant points in the surface BZ.
To validate this expectation, we recompute the surface states using a different unit cell [see Fig.~\ref{fig:3DTB_C4_bulk}(a)].
In general, changes in the definition of the unit cell alter the surface termination, which in turn affects the surface band structure.
For instance, in the Fu model, adopting a different unit cell shifts the quadratic band touching from the $\bar{\mathrm{M}}$ point to the $\bar{\Gamma}$ point.
In our model, the use of a different unit cell causes the surface Dirac cones to disappear, and instead, quadratic band touchings emerge at both the $\bar{\Gamma}$ and $\bar{\mathrm{M}}$ points, as illustrated in Figs.~\ref{fig:3DTB_C4_surf}(c) and (d).
Since a quadratic band touching carries a winding number of 2~\cite{PhysRevB.104.195114}, the surface states remain consistent with $\bar{e}_2 = 4$.
Moreover, the presence of two quadratic band touchings at $C_{4z}T$-invariant points further substantiates the bulk-boundary correspondence associated with $\bar{\nu}_4 = 0$.
Before closing this subsection, we note that when $\mathcal{C}_{4z}$ symmetry is reduced to $\mathcal{C}_{2z}$, $\nu_4$ is no longer defined, and each of the quadratic band touchings on the surface splits into two Dirac cones.
Nevertheless, the nontrivial Euler class and the total winding number of the surface bands remain preserved.

\subsection{Relationship among $\bar{e}_2$, $\bar{\nu}_4$, and the halved mirror chirality}

\begin{figure}[htp]
\includegraphics[width=8.5cm]{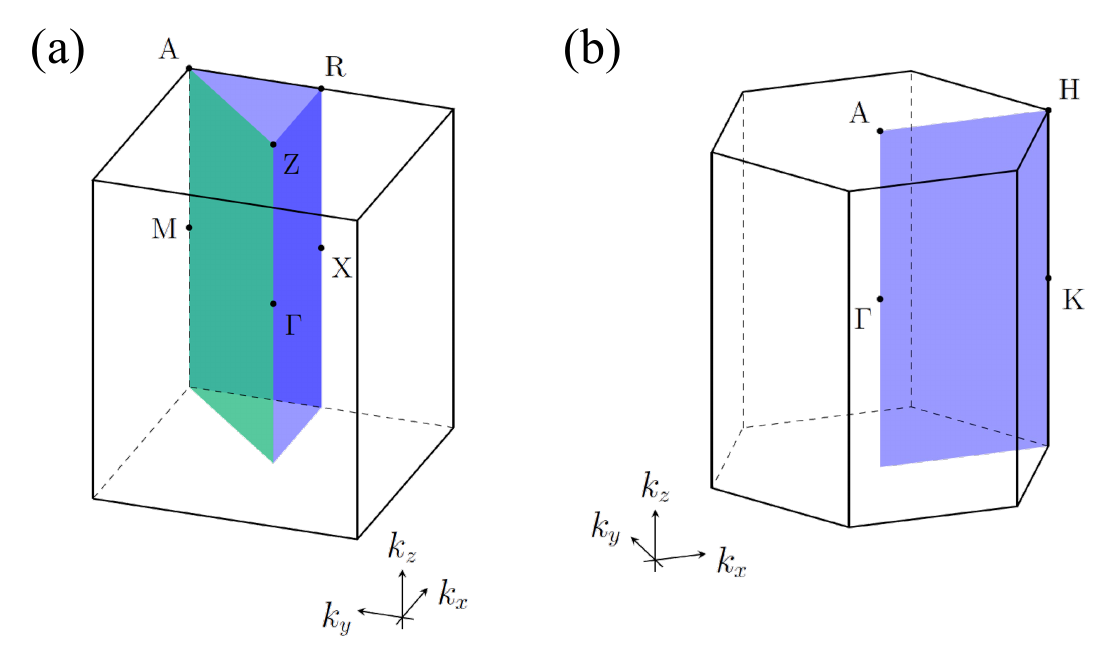}
\caption{
Half-mirror planes (HMPs) in the 3D BZs.
(a) HMPs in $\mathcal{C}_{4v}$-symmetric systems.
The green and blue planes indicate $\mathrm{HMP}_{4,1}$ and $\mathrm{HMP}_{4,2}$, respectively.
(b) HMPs in $\mathcal{C}_{6v}$-symmetric systems.
The blue plane indicates $\mathrm{HMP}_{6,1}$.
}
\label{fig:HMP}
\end{figure}

Next, we discuss the case where vertical mirror symmetry is present.
In $\mathcal{C}_{nv}$-symmetric systems ($n=3,4,6$), mirror-invariant planes bounded by two distinct $C_{mz}$-invariant lines ($m > 2$) and the $k_z = \pm\pi$ planes are referred to as half-mirror planes~\cite{PhysRevLett.113.116403}.
We denote the $i$-th half-mirror plane in a $\mathcal{C}_{nv}$-symmetric system as $\mathrm{HMP}_{n,i}$ (see Fig.~\ref{fig:HMP} for the cases with $n = 4, 6$).
When all occupied bands form orbital doublets, another $\Z$-valued topological invariant, known as the halved mirror chirality, can be defined by the integral over the $\mathrm{HMP}_{n,i}$ as~\cite{PhysRevLett.113.116403}
\begin{equation}
    \label{eq:chi_def}
    \chi_{n,i} = \dfrac{1}{2\pi} \int_{\mathrm{HMP}_{n,i}} dt_{n,i} dk_z \; [\mathcal{F}_{\mathrm{e}}(t_{n,i}, k_z) - \mathcal{F}_{\mathrm{o}}(t_{n,i}, k_z)],
\end{equation}
where $\mathcal{F}_{\mathrm{e}}(t, k_z)$ and $\mathcal{F}_{\mathrm{o}}(t, k_z)$ are the Berry curvature of occupied doublet bands in the mirror-even and mirror-odd subspaces, respectively, defined by
\begin{equation}
    \mathcal{F}_{\eta}(t, k_z) = \partial_t \mathcal{A}^\eta_z(t, k_z) - \partial_z \mathcal{A}^\eta_t(t, k_z)
\end{equation}
with the U(1) Berry connection in each mirror subspace
\begin{equation}
    \mathcal{A}^\eta_\mu(t, k_z) = -i \sum_{n}^{\mathrm{occ.}} \braket{u^{\eta}_n(t, k_z)}{\partial_{\mu} u^{\eta}_n(t, k_z)}
\end{equation}
for $\eta \in \qty{\mathrm{e}, \mathrm{o}}$.
Here, $(t, k_z)$ parametrizes the momenta on a half-mirror plane.
As previously mentioned, the U(1) Berry connection and curvature, obtained by summing over the occupied band indices, are denoted by calligraphic letters.
In the case of $\mathcal{C}_{4v}$ symmetry, there are two half-mirror planes:
\begin{align}
    \mathrm{HMP}_{4,1} = \left\{
        \qty(\pi (1-t_{4,1}), \pi (1-t_{4,1}), k_z) 
        \;\middle|\;
        \begin{array}{l}
            0 \leq t_{4,1} \leq 1, \\
            -\pi \leq k_z \leq \pi
        \end{array}
    \right\}
\end{align}
and
\begin{align}
    &\mathrm{HMP}_{4,2} \nonumber\\
    =& \left\{(2\pi t_{4,2}, 0, k_z) 
        \;\middle|\;
        \begin{array}{l}
            0 \leq t_{4,2} \leq 1/2, \\
            -\pi \leq k_z \leq \pi
        \end{array}
    \right\} \nonumber \\
    & \quad \cup 
    \left\{ (\pi, 2\pi (t_{4,2} - 1/2), k_z) 
        \;\middle|\;
        \begin{array}{l}
            1/2 \leq t_{4,2} \leq 1, \\
            -\pi \leq k_z \leq \pi
        \end{array}
    \right\},
\end{align}
as shown in Fig.~\ref{fig:HMP}(a).
These planes are referred to as $\mathrm{HMP}_4$ and $\mathrm{HMP}_5$, respectively, in Ref.~\cite{PhysRevLett.113.116403}.

Here we consider systems with two occupied bands and derive the relationship among $\bar{e}_2$, $\bar{\nu}_4$, and $\chi_{4,i}$.
Note that the occupied bands are uniquely specified by their mirror eigenvalue $\eta$ alone, since here we consider a case where there is one occupied band with mirror eigenvalue $+1$ and one with $-1$.
We adopt a gauge that satisfies
\begin{equation}
    \label{eq:real-gauge_3D}
    \mathcal{C}_{2z}\mathcal{T} \ket{\tilde{u}_n(k_x, k_y, k_z)} = \ket{\tilde{u}_n(k_x, k_y, -k_z)},
\end{equation}
which corresponds to the real-gauge in Eq.~\eqref{eq:real-gauge} on the $k_z = 0,\pi$ planes.
Due to the $\mathcal{C}_{2z} \mathcal{T}$ symmetry, the relation $\mathcal{F}_\eta(t_{n,i}, k_z) = \mathcal{F}_\eta(t_{n,i}, -k_z)$ holds for both $\eta \in \qty{\mathrm{e}, \mathrm{o}}$.
Thus, defining
\begin{equation}
    B_\eta^{(n,i)} = \int_{\mathrm{HMP}_{n,i}} dt_{n,i} dk_z \; \mathcal{F}_{\eta}(t_{n,i}, k_z),
\end{equation}
we obtain
\begin{align}
    \dfrac{1}{2} B_\eta^{(n,i)} &= \int_{
        \substack{
            0 \leq t_{n,i} \leq 1 \\
            0 \leq k_z \leq \pi
        }
        } dt_{n,i} dk_z \; \mathcal{F}_{\eta}(t_{n,i}, k_z) \nonumber\\
    &= \int_{0}^\pi dk_z \; \tilde{\mathcal{A}}^\eta_z(1, k_z) - \int_{0}^\pi dk_z \; \tilde{\mathcal{A}}^\eta_z(0, k_z) \nonumber\\
    &\phantom{{}={}} + \int_0^1 dt_{n,i} \; \tilde{\mathcal{A}}^\eta_t(t_{n,i}, 0) - \int_0^1 dt_{n,i} \; \tilde{\mathcal{A}}^\eta_t(t_{n,i}, \pi).
\end{align}
Since the gauge choice in Eq.~\eqref{eq:real-gauge_3D} corresponds to the real-gauge on the $C_{2z}T$-invariant $k_z = 0, \pi$ planes, we obtain
\begin{align}
    \label{eq:B_eta_Stokes_real-gauge_2bands}
    \dfrac{1}{2} B_\eta^{(n,i)} &= \int_{0}^\pi dk_z \; \tilde{\mathcal{A}}^\eta_z(1, k_z) - \int_{0}^\pi dk_z \; \tilde{\mathcal{A}}^\eta_z(0, k_z).
\end{align}
Now, let $\bar{t} \in \qty{0, 1}$.
According to the relation $\mathcal{M}_{4,i} \mathcal{C}_{4z} \mathcal{M}_{4,i}^{-1} = \mathcal{C}_{4z}^{-1}$ for the mirror operations $\mathcal{M}_{4,i}$ that leave $\mathrm{HMP}_{4,i}$ invariant, the state given by
\begin{align}
    \label{eq:C4v_u_-eta}
    \ket{\tilde{u}^{-\eta}(\bar{t}, k_z)} &= e^{i\varphi_\eta(\bar{t}, k_z)} \dfrac{1}{2} \qty(\mathcal{C}_{4z} - \mathcal{C}_{4z}^{-1}) \ket{\tilde{u}^\eta(\bar{t}, k_z)} \nonumber\\
    &= e^{i\varphi_\eta(\bar{t}, k_z)} \mathcal{C}_{4z} \ket{\tilde{u}^\eta(\bar{t}, k_z)}
\end{align}
represents the energy eigenstate with the same energy as $\ket{\tilde{u}^\eta(\bar{t}, k_z)}$ but with the opposite mirror eigenvalue~\footnote{We also denote the mirror eigenvalues $\pm 1$ by $\eta$ with a slight abuse of notation.}, where $e^{i\varphi_\eta(\bar{t}, k_z)}$ accounts for the U(1) phase ambiguity.
In other words, at $C_{4z}$-invariant momenta specified by $t=\bar{t}$, $\ket{\tilde{u}^\eta(\bar{t}, k_z)}$ and $\ket{\tilde{u}^{-\eta}(\bar{t}, k_z)}$ form the 2D representation of the point group $\mathcal{C}_{4v}$.
From Eq.~\eqref{eq:C4v_u_-eta}, the Berry connection in the mirror subspace $-\eta$ satisfies
\begin{equation}
    \label{eq:A_eta_C4v}
    \tilde{\mathcal{A}}^{-\eta}_z(\bar{t}, k_z) = \tilde{\mathcal{A}}^{\eta}_z(\bar{t}, k_z) + \partial_{k_z} \varphi_\eta(\bar{t}, k_z).
\end{equation}
Based on the above discussion, the halved mirror chirality $\chi_{4,i}$ can be expressed as
\begin{align}
    \dfrac{1}{2} \chi_{4,i} &= \dfrac{1}{2\pi} \dfrac{1}{2} \qty(B_{\mathrm{e}}^{(4,i)} - B_{\mathrm{o}}^{(4,i)}) \nonumber\\
    =& \dfrac{1}{2\pi} [\varphi_{\mathrm{e}}(1, 0) - \varphi_{\mathrm{e}}(1, \pi) + \varphi_{\mathrm{e}}(0, \pi) - \varphi_{\mathrm{e}}(0, 0)] \mod 1,
\end{align}
which results in
\begin{align}
    \label{eq:chi_C4v}
    (-1)^{\chi_{4,i}} &= e^{-i\varphi_{\mathrm{e}}(0, 0)} e^{i\varphi_{\mathrm{e}}(1, 0)} e^{i\varphi_{\mathrm{e}}(0, \pi)} e^{-i\varphi_{\mathrm{e}}(1, \pi)}.
\end{align}
The right-hand side can be evaluated as
\begin{equation}
    e^{-i\varphi_{\mathrm{e}}(\bar{t}, k_z)} = \mel{\tilde{u}^{\mathrm{o}}(\bar{t}, k_z)}{\mathcal{C}_{4z}}{\tilde{u}^{\mathrm{e}}(\bar{t}, k_z)},
\end{equation}
which follows from Eq.~\eqref{eq:C4v_u_-eta}.
As discussed in Sec.~\ref{subsec:Euler_C4z_2D}, for the real 2D representation under $\mathcal{C}_{4z}$ symmetry, the sewing matrix of the $\mathcal{C}_{4z}$ operation at the $C_{4z}T$-invariant momenta is antisymmetric.
Thus, at $k_z = \bar{k}_z \in \{0, \pi\}$, the matrix element $\mel{\tilde{u}^{\mathrm{o}}(\bar{t}, \bar{k}_z)}{\mathcal{C}_{4z}}{\tilde{u}^{\mathrm{e}}(\bar{t}, \bar{k}_z)}$ represents an off-diagonal element of the $2 \times 2$ antisymmetric matrix $\tilde{\mathcal{B}}_{C_{4z}}(\bar{t}, \bar{k}_z)$, leading to
\begin{equation}
    \mel{\tilde{u}^{\mathrm{o}}(\bar{t}, \bar{k}_z)}{\mathcal{C}_{4z}}{\tilde{u}^{\mathrm{e}}(\bar{t}, \bar{k}_z)} = \pm \Pf \qty[\tilde{\mathcal{B}}_{C_{4z}}(\bar{t}, \bar{k}_z)]
\end{equation}
where the sign $\pm$ depends on the ordering of the basis used in the matrix representation.
As a consequence of the above discussion, we obtain
\begin{equation}
    (-1)^{\chi_{4,i}} = \dfrac{\Pf \qty[\tilde{\mathcal{B}}_{C_{4z}}(0, 0)]}{\Pf \qty[\tilde{\mathcal{B}}_{C_{4z}}(1, 0)]} \dfrac{\Pf \qty[\tilde{\mathcal{B}}_{C_{4z}}(1, \pi)]}{\Pf \qty[\tilde{\mathcal{B}}_{C_{4z}}(0, \pi)]}.
\end{equation}
Specifically,
\begin{equation}
    \label{eq:chi41_Pf_GMZA}
    (-1)^{\chi_{4,1}} = \dfrac{\Pf \qty[\tilde{\mathcal{B}}_{C_{4z}}(\mathrm{M})]}{\Pf \qty[\tilde{\mathcal{B}}_{C_{4z}}(\Gamma)]} \dfrac{\Pf \qty[\tilde{\mathcal{B}}_{C_{4z}}(\mathrm{Z})]}{\Pf \qty[\tilde{\mathcal{B}}_{C_{4z}}(\mathrm{A})]}
\end{equation}
and
\begin{equation}
    \label{eq:chi42_Pf_GMZA}
    (-1)^{\chi_{4,2}} = \dfrac{\Pf \qty[\tilde{\mathcal{B}}_{C_{4z}}(\Gamma)]}{\Pf \qty[\tilde{\mathcal{B}}_{C_{4z}}(\mathrm{M})]} \dfrac{\Pf \qty[\tilde{\mathcal{B}}_{C_{4z}}(\mathrm{A})]}{\Pf \qty[\tilde{\mathcal{B}}_{C_{4z}}(\mathrm{Z})]}
\end{equation}
hold.
It is worth making the following remark.
From Eqs.~\eqref{eq:chi41_Pf_GMZA} and \eqref{eq:chi42_Pf_GMZA}, we obtain $\chi_{4,1} + \chi_{4,2} = 0 \mod 2$, whereas it is known that $\chi_{4,1} + \chi_{4,2} = 1 \mod 2$ can occur in Weyl semimetal phases~\cite{PhysRevLett.113.116403}.
This apparent contradiction arises because our derivation is based on the real-gauge.
When a Weyl point exists on the $C_{2z}T$-invariant plane, a line of singularity intersects the HMP [see Fig.~\ref{fig:C4_phase_transition}(b) and its caption], which invalidates the calculation.
Our calculation is therefore valid only in insulating phases and does not contradict the possibility of $\chi_{4,1} + \chi_{4,2} = 1 \mod 2$ in Weyl semimetal phases.

By combining the results from Eqs.~\eqref{eq:e2bar_Pf_GMZA}, \eqref{eq:nu4bar_Pf_GMZA}, \eqref{eq:chi41_Pf_GMZA}, and \eqref{eq:chi42_Pf_GMZA}, and noting that $\Pf \qty[\tilde{\mathcal{B}}_{C_{4z}}(\kinv)] = \pm 1$ for the $C_{4z}T$-invariant momenta $\kinv$, we obtain
\begin{equation}
    \label{eq:relation_e2bar_nu4bar_chi}
    \cos\qty(\dfrac{\pi}{2}\bar{e}_2) = (-1)^{\bar{\nu}_4} = (-1)^{\chi_{4,1}} = (-1)^{\chi_{4,2}}
\end{equation}
as the general relationship among the three types of topological invariants, $\bar{e}_2$, $\bar{\nu}_4$, and $\chi_{4,i}$.
The relationship between $\bar{\nu}_4$ and $\chi_{4,i}$ remains valid for an arbitrary number of bands in the doublet representation, which will be proven in Appendix~\ref{sec:nu4_chi4_Nbands}.

\subsection{Phase transitions}
\label{subsec:C4z_phase_transitions}
It is of great importance to describe how topological invariants change.
Previous studies have discussed topological phase transitions associated with $e_2$ and $\chi_{n,i}$ separately~\cite{PhysRevLett.113.116403,PhysRevX.9.021013}.
In this subsection, we verify the consistency of these phase transition processes by using the relationship in Eq.~\eqref{eq:relation_e2bar_nu4bar_chi} and elucidate the connection between the band-touching points (i.e., nodal points) in the occupied bands and those of the gapless surface bands.

The Euler class $e_2$ of the occupied bands is related to the total winding number $N_{\mathrm{t}}$ of their nodal points located in the $C_{2z}T$-invariant plane through the relation~\cite{PhysRevX.9.021013,Mathai2017}
\begin{equation}
    \label{eq:e2_Nt}
    e_2 = - \frac{1}{2} N_{\mathrm{t}}.
\end{equation}
Importantly, nodal points (i.e., Weyl points) between an occupied band and an unoccupied band create branch cuts for the occupied bands, and when a nodal point between the occupied bands crosses these cuts, the sign of its winding number reverses~\cite{PhysRevX.9.021013,doi:10.1126/science.aau8740}.
Thus, the calculation of $N_{\mathrm{t}}$ must take into account branch cuts in the $C_{2z}T$-invariant plane.
Indeed, the value of $e_2$ changes when Weyl points are pair-created between the occupied and unoccupied bands, follow trajectories that enclose a nodal point within the occupied bands, and eventually undergo pair-annihilation.
As for the halved mirror chirality, it has been shown that $\chi_{4,i}$ changes when Weyl points are transferred between two distinct HMPs during the pair creation and annihilation process~\cite{PhysRevLett.113.116403}.

\begin{figure}[htp]
\includegraphics[width=8.5cm]{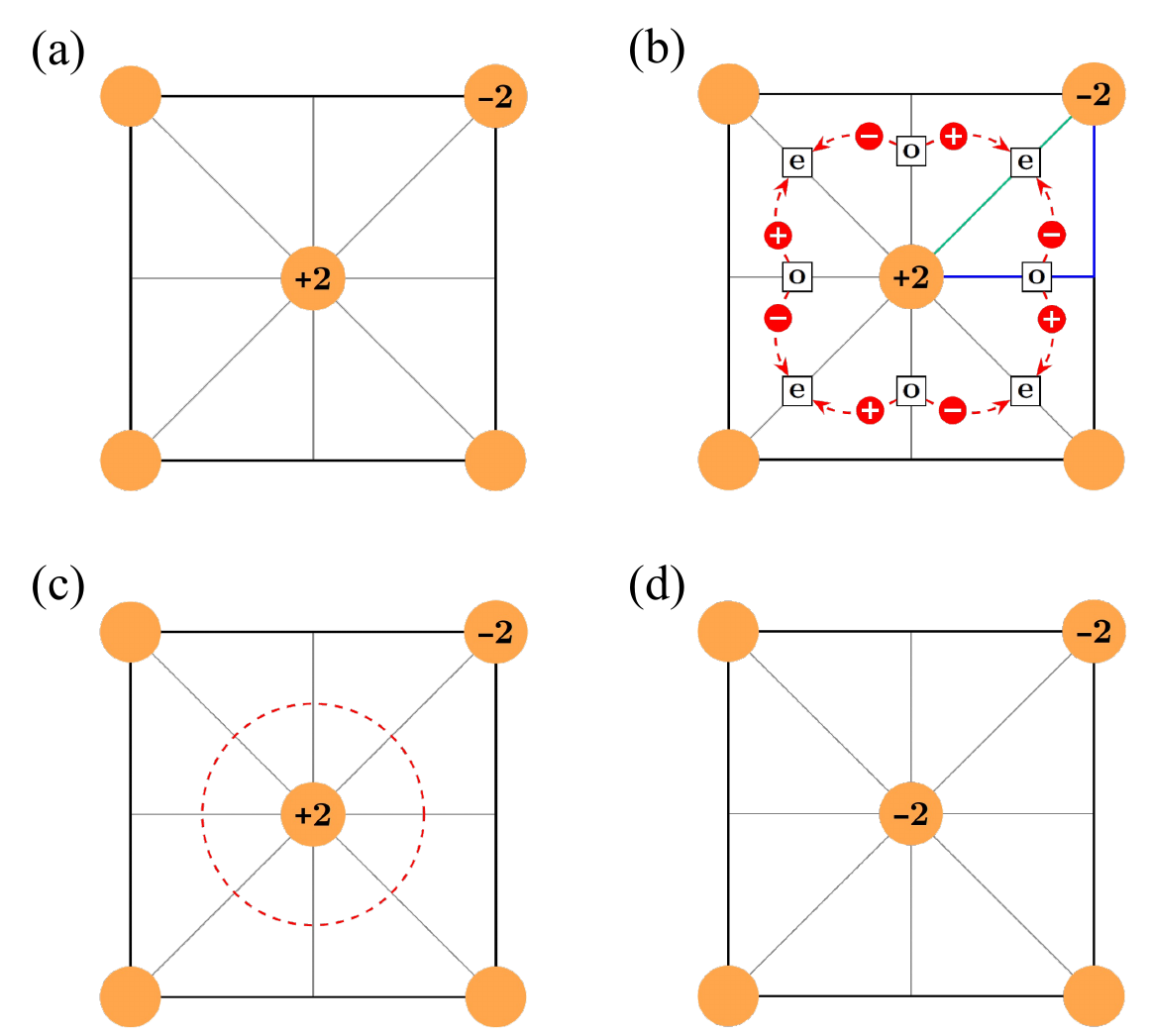}
\caption{
Distribution of nodal points in the $k_z = 0$ plane during the phase transition from $(e_2(0), e_2(\pi), \chi_{4,1}, \chi_{4,2}) = (0, 0, 0, 0)$ to $(e_2(0), e_2(\pi), \chi_{4,1}, \chi_{4,2}) = (2, 0, 1, 1)$.  
(a) Nodal points between occupied bands and their winding numbers in the insulating phase with $e_2(0) = 0$. The occupied bands are degenerate at the $C_{4z}T$-invariant points indicated in orange.
Among the nodal points related by the periodicity of the BZ, the winding number is indicated for only one of them.
(b) Nodal points in the Weyl semimetal phase mediating two insulating phases. The red circles indicate Weyl points (nodal points between occupied and unoccupied bands). The green (blue) line represents the $\mathrm{HMP}_{4,1}$ ($\mathrm{HMP}_{4,2}$). The sign $+$ $(-)$ represents Weyl points with positive (negative) chirality, and the symbol e (o) indicates that gap closure at the HMP takes place in the mirror-even (mirror-odd) subspace.
(c) Insulating phase after the pair annihilation of Weyl points. The red dashed lines represent branch cuts (Dirac strings).  
(d) Nodal points of occupied bands and their winding numbers in the insulating phase with $e_2(0) = 2$.  
}
\label{fig:C4_phase_transition}
\end{figure}

Equation~\eqref{eq:relation_e2bar_nu4bar_chi} provides a basis for a consistent and unified description of topological phase transitions associated with $e_2$ and $\chi_{4,i}$.
To explain this, we consider the following example: a topological phase transition from $(e_2(0), e_2(\pi), \chi_{4,1}, \chi_{4,2}) = (0, 0, 0, 0)$ to $(e_2(0), e_2(\pi), \chi_{4,1}, \chi_{4,2}) = (2, 0, 1, 1)$ through a Weyl semimetal phase.
Note that both initial and final insulating phases satisfy Eq.~\eqref{eq:relation_e2bar_nu4bar_chi}.
Since $e_2(\pi)$ remains unchanged between the two insulating phases, we only need to focus on the $k_z = 0$ plane, where all the action happens. 
In systems where $\chi_{4,i}$ is defined, the occupied bands form a 2D representation and have nodal points at the $C_{4z}T$-invariant points, each having a winding number of magnitude 2.
According to Eq.~\eqref{eq:e2_Nt}, in order to have $e_2(0)=0$, the winding numbers must have opposite signs, as illustrated in Fig.~\ref{fig:C4_phase_transition}(a).
For the halved mirror chirality to change, Weyl points must be pair-created at either $\mathrm{HMP}_{4,1}$ or $\mathrm{HMP}_{4,2}$ and subsequently pair-annihilated at the other, as depicted in Fig.~\ref{fig:C4_phase_transition}(b), in such a way that the trajectory of the Weyl points during this process forms a closed loop enclosing a $C_{4z}T$-invariant point, as shown in Fig.~\ref{fig:C4_phase_transition}(c).
In this case, the sign of the winding number of the enclosed nodal point is considered to have been reversed, which gives rise to the change of $e_2(0)$ to 2. 
This sign reversal occurs because, during the process of continuous contraction followed by disappearance, the loop of branch cuts must cross the enclosed nodal point, resulting in the state illustrated in Fig.~\ref{fig:C4_phase_transition}(d).

In this way, we integrate the discussion of topological phase transitions associated with $e_2$ and $\chi_{n,i}$.
Furthermore, this analysis allows us to infer the surface states on the (001) plane in the topological insulator phase.
The trajectories of the Weyl points in the $k_z = 0$ plane correspond to the Fermi arcs on the (001) surface~\cite{PhysRevB.83.205101,PhysRevB.89.235315}.
Thus, when the system undergoes the phase transition following the process shown in Fig.~\ref{fig:C4_phase_transition}, gapless surface states emerge on the (001) surface, with a Fermi surface that resembles the trajectory of the Weyl points in Fig.~\ref{fig:C4_phase_transition}(c).
In the case shown in Fig.~\ref{fig:C4_phase_transition}, symmetry arguments guarantee that the nodal points of the occupied bands coincide in position with the degeneracy points of the gapless surface states.
Additionally, the nodal point that is not enclosed by the Weyl-point trajectories in the transition process of Fig.~\ref{fig:C4_phase_transition} can become enclosed under a different choice of the unit cell, leading to the emergence of corresponding gapless surface states.
While such a strict correspondence generally does not hold, nodal points in the occupied bulk bands still indicate the approximate locations of nodal points in the surface bands for various unit cell definitions.

\section{Euler class in $\mathcal{C}_{6z}$-symmetric systems}
\label{sec:Euler_C6z}

In this section, we consider the sixfold rotation $\mathcal{C}_{6z}$ in spinless systems, whose transformation matrix in the 2D $k$-space is given by $R =(\sigma_0 - i \sqrt{3} \sigma_2)/2$.
The $\mathbb{Z}_2$ invariant protected by $\mathcal{C}_{4z}$ and $\mathcal{T}$ symmetry can be generalized to $\mathcal{C}_{6z}$-symmetric systems by modifying its definition~\cite{PhysRevLett.106.106802,PhysRevB.93.205104,PhysRevB.104.195114}.
Additionally, $\mathcal{C}_{6v}$-symmetric systems also possess one HMP, allowing the definition of the corresponding halved mirror chirality.
Following the calculation procedure in the previous section, we investigate the relationship among the Euler class, the $\mathbb{Z}_2$ invariant, and the halved mirror chirality, revealing that the sign of the Euler class plays an essential role.
To support our analytical results, we construct a tight-binding model and perform numerical calculations of the Wilson loop spectra and the surface states.

\subsection{Euler class in 2D systems}
\label{subsec:Euler_C6z_2D}

By applying the same approach as in Sec.~\ref{subsec:Euler_C4z_2D}, we obtain the following results.
When $\tilde{\mathcal{B}}_{C_{6z}}(\bm{k}) = \sigma_3 \exp[-i \phi_{C_{6z}}(\bm{k}) \sigma_2]$, we find that $\tilde{F}_z(\bm{k}) = - \tilde{F}_z(C_{6z}\bm{k})$, which results in $e_2 = 0$.
In contrast, when $\tilde{\mathcal{B}}_{C_{6z}}(\bm{k}) = \exp[-i \phi_{C_{6z}}(\bm{k}) \sigma_2]$, it follows that $\tilde{F}_z(\bm{k}) = \tilde{F}_z(C_{6z}\bm{k})$.
We can then restrict the integration domain to one-sixth of the BZ (sBZ) shown in Fig.~\ref{fig:BZ}(c), yielding
\begin{align}
    \label{eq:e2_C6z_det+1}
    e^{i \pi e_2/3} = e^{+i\phi_{C_{6z}}(\Gamma)} e^{-i\phi_{C_{2z}}(\mathrm{M})} e^{+i\phi_{C_{3z}}(\mathrm{K})},
\end{align}
where $\Gamma = (0, 0)$, $\mathrm{M} = (\pi, -\pi/\sqrt{3})$, and $\mathrm{K} = (4\pi/3, 0)$.
Equation~\eqref{eq:e2_C6z_det+1} relates $e_2$ to the $\mathcal{C}_{6z}$ eigenvalue at the $\Gamma$ point, $\mathcal{C}_{2z}$ eigenvalue at the M point, and the $\mathcal{C}_{3z}$ eigenvalue at the K point.
This formula is similar to the one for the Chern number~\cite{PhysRevB.86.115112}, but in Eq.~\eqref{eq:e2_C6z_det+1} only one rotational eigenvalue of the two occupied bands is considered at each high-symmetry point.
Thus, Eq.~\eqref{eq:e2_C6z_det+1} cannot determine $e_2$ when the $\mathcal{C}_{6z}$ and $\mathcal{C}_{3z}$ eigenvalues take complex values.
In this case, the bands transform in the same representation as an orbital doublet.
While $\mathcal{C}_{4z}$ symmetry permits only an orbital doublet with angular momentum $l = 1$, $\mathcal{C}_{6z}$ symmetry admits those with $l = 1$ and $l = 2$;
the doublet with $l = 2$ corresponds to the $\{d_{x^2-y^2}, d_{xy}\}$ orbitals.
By restricting our analysis to such orbital doublets, we derive an alternative expression for $e_2$ as follows.

Taking the real part of Eq.~\eqref{eq:e2_C6z_det+1} yields 
\begin{align}
    \label{eq:e2_C6z_det+1_re}
    \cos \qty(\dfrac{\pi}{3} e_2) &= e^{-i\phi_{C_{2z}}(\mathrm{M})} \Bigl\{\cos [\phi_{C_{6z}}(\Gamma)] \cos [\phi_{C_{3z}}(\mathrm{K})] \nonumber\\
    &\phantom{{}={}} - \sin [\phi_{C_{6z}}(\Gamma)] \sin [\phi_{C_{3z}}(\mathrm{K})] \Bigr\},
\end{align}
where the $\mathcal{C}_{2z}$ eigenvalue $e^{-i\phi_{C_{2z}}}$ is either $+1$ or $-1$.
The $\mathcal{C}_{3z}$ eigenvalues for the 2D representations are given by $\exp(\pm i 2\pi/3)$, which implies that $\cos[\phi_{C_{3z}}(\mathrm{K})] = -1/2$.
In addition, since the off-diagonal elements of the antisymmetric matrix $\tilde{\mathcal{B}}_{C_{3z}}^{\mathrm{A}}(\mathrm{K}) \coloneqq [\tilde{\mathcal{B}}_{C_{3z}}(\mathrm{K}) - \tilde{\mathcal{B}}^\mathrm{T}_{C_{3z}}(\mathrm{K})]/\sqrt{3}$ are given by $\pm (2/\sqrt{3}) \sin[\phi_{C_{3z}}(\mathrm{K})]$, we find
\begin{equation}
    \sin [\phi_{C_{3z}}(\mathrm{K})] = -\dfrac{\sqrt{3}}{2} \Pf \qty[\tilde{\mathcal{B}}_{C_{3z}}^{\mathrm{A}}(\mathrm{K})].
\end{equation}
For the $\Gamma$ point, the $\mathcal{C}_{6z}$ eigenvalue depends on the orbital angular momentum, denoted by $l_\Gamma$, of the orbital doublet forming the occupied bands. 
Then, we have
\begin{equation}
    \cos [\phi_{C_{6z}}(\Gamma)] = - (-1)^{l_{\Gamma}}/2,
    \quad
    \sin [\phi_{C_{6z}}(\Gamma)] =  -\dfrac{\sqrt{3}}{2} \Pf \qty[\tilde{\mathcal{B}}_{C_{6z}}^{\mathrm{A}}(\Gamma)],
\end{equation}
where $\tilde{\mathcal{B}}_{C_{6z}}^{\mathrm{A}}(\Gamma) \coloneqq [ \tilde{\mathcal{B}}_{C_{6z}}(\Gamma) - \tilde{\mathcal{B}}^\mathrm{T}_{C_{6z}}(\Gamma)]/\sqrt{3}$.
Thus, we obtain
\begin{align}
    \label{eq:e2_Pf_GK}
     \cos \qty(\dfrac{\pi}{3} e_2)
     &=
     \frac{e^{-i\phi_{C_{2z}}(\mathrm{M})}}{4} 
     \biggl(
     (-1)^{l_{\Gamma}}
     -3
     \Pf \qty[\tilde{\mathcal{B}}_{C_{6z}}^{\mathrm{A}}(\Gamma)]
     \Pf \qty[\tilde{\mathcal{B}}_{C_{3z}}^{\mathrm{A}}(\mathrm{K})] \biggr).
\end{align}

\subsection{Relationship between $e_2$ and $\nu_6$}

As stated in Sec.~\ref{subsec:relation_e2_nu4}, orbital doublets form a basis for a 2D representation of the $\mathcal{C}_{6z}$ operator in the presence of $\mathcal{T}$ symmetry. 
For bands composed of such orbital doublets, the $\mathbb{Z}_2$ invariant $\nu_6$ can be defined as~\cite{PhysRevLett.106.106802,PhysRevB.93.205104}
\begin{equation}
    \label{eq:nu6_def}
    (-1)^{\nu_{6}} = \exp\qty(i \int_{\Gamma}^{\mathrm{K}} d\bm{k} \cdot\bm{\mathcal{A}}(\bm{k})) \dfrac{\Pf \qty[w_6^{\mathrm{A}}(\mathrm{K})]}{\Pf \qty[w_6^{\mathrm{A}}(\Gamma)]},
\end{equation}
where $w_6^{\mathrm{A}}(\kinv)$ is an antisymmetric matrix defined by $w_6^{\mathrm{A}}(\kinv) = [w_6(\kinv) - w_6^{\mathrm{T}}(\kinv)]/\sqrt{3}$ with $\qty[w_6(\kinv)]_{mn} = \mel{u_m(\kinv)}{\mathcal{C}_{6z}\mathcal{T}}{u_n(\kinv)}$ evaluated at the $C_{6z}T$-invariant momenta $\kinv \in \{\Gamma, \mathrm{K}\}$.
Here, the denominator is chosen to be $\sqrt{3}$ in order to set the value of $\Pf \qty[w_6^{\mathrm{A}}(\kinv)]$ to $\pm 1$.
When we adopt the real-gauge in Eq.~\eqref{eq:real-gauge}, the U(1) Berry connection $\mathcal{A}(\bm{k})$ vanishes, and $w_6(\kinv)$ can be expressed as 
\begin{align}
    \qty[\tilde{w}_6(\kinv)]_{mn} &= \mel{\tilde{u}_m(\kinv)}{\mathcal{C}_{6z}\mathcal{C}_{2z}\mathcal{C}_{2z}\mathcal{T}}{\tilde{u}_n(\kinv)} \nonumber\\
    &= \mel{\tilde{u}_m(\kinv)}{\mathcal{C}_{6z}\mathcal{C}_{2z}}{\tilde{u}_n(\kinv)} \nonumber\\
    &= \qty[\tilde{\mathcal{B}}_{C_{3z}}^\dagger(\kinv)]_{mn} = \qty[\tilde{\mathcal{B}}_{C_{3z}}^{\mathrm{T}}(\kinv)]_{mn}.
\end{align}
In particular, for the $C_{6z}$-invariant $\Gamma$ point, we obtain 
\begin{equation}
    \qty[\tilde{w}_6(\Gamma)]_{mn} = \mel{\tilde{u}_m(\Gamma)}{\mathcal{C}_{6z}\mathcal{C}_{2z}}{\tilde{u}_n(\Gamma)} = (-1)^{l_{\Gamma}} \qty[\tilde{\mathcal{B}}_{C_{6z}}(\Gamma)]_{mn}.
\end{equation}
Therefore, we find
\begin{equation}
    \label{eq:nu6_Pf_GK}
    (-1)^{\nu_{6}} = \dfrac{-\Pf \qty[\tilde{\mathcal{B}}_{C_{3z}}^{\mathrm{A}}(\mathrm{K})]}{(-1)^{l_{\Gamma}} \Pf \qty[\tilde{\mathcal{B}}_{C_{6z}}^{\mathrm{A}}(\Gamma)]}.
\end{equation}

Given that $\Pf \qty[\tilde{\mathcal{B}}_{C_{6z}}^{\mathrm{A}}(\kinv)] = \pm 1$ and $\Pf \qty[\tilde{\mathcal{B}}_{C_{3z}}^{\mathrm{A}}(\kinv)] = \pm 1$, it follows from Eqs.~\eqref{eq:nu6_Pf_GK} and \eqref{eq:e2_Pf_GK} that
\begin{equation}
    \label{eq:relation_e2_nu6}
    \cos \qty(\dfrac{\pi}{3} e_2) = e^{-i\phi_{C_{2z}}(\mathrm{M})} (-1)^{l_{\Gamma}} \dfrac{1 + 3(-1)^{\nu_{6}}}{4}.
\end{equation}
This provides an explicit formula connecting the two topological invariants, $e_2$ and $\nu_6$.
It should be noted that $(-1)^{l_{\Gamma}}$ can be considered as the $\mathcal{C}_{2z}$ eigenvalue at the $\Gamma$ point.
This formula allows one to readily determine $e_2$ from $\nu_6$ and $\mathcal{C}_{2z}$ eigenvalues at the $\Gamma$ and M points, circumventing the direct computation of the Wilson loop spectra.
Table~\ref{tab:e2_nu6_values} presents the possible combinations of $e_2$ and $\nu_6$ values allowed by Eq.~\eqref{eq:relation_e2_nu6}.

\begin{table}[htp]
    \centering
    \caption{Possible combinations of $e_2$ and $\nu_{6}$ in 2D systems where both invariants can be defined. These combinations satisfy the relation in Eq.~\eqref{eq:relation_e2_nu6}, with the open (filled) circles indicating cases where the product of the $\mathcal{C}_{2z}$ eigenvalues at the $\Gamma$ and M points is $+1$ ($-1$).}
    \label{tab:e2_nu6_values}
    \begin{tabular}{|l||l|l|l|l|l|l|l|l|l|}
        \hline
        \diagbox[width=4em,height=2em]{$\nu_{6}$}{$e_2$} & 0 & 1 & 2 & 3 & 4 & 5 & 6 & 7 & $\cdots$ \\ \hline\hline
        0 & $\circ$ &   &   & $\bullet$ &   &   & $\circ$ &  & \\ \hline
        1 &   & $\bullet$ & $\circ$ &   & $\circ$ & $\bullet$ &   & $\bullet$ & \\ \hline
    \end{tabular}
\end{table}

\subsection{Euler class in 3D systems}
\label{subsec:Euler_C6z_3D}
We now shift our focus to 3D systems.
As in the case of the $\mathcal{C}_{4z}$-symmetric system, we denote the $\mathbb{Z}_2$ invariant $\nu_6$ defined on the $k_z = \bar{k}_z$ plane by $\nu_6(\bar{k}_z)$, where $\bar{k}_z \in \{0, \pi\}$, and define $\bar{\nu}_6 = \nu_6(\pi) - \nu_6(0) \mod 2$.
The 3D topological phases are characterized by $\bar{e}_2$ and $\bar{\nu}_6$ ~\cite{PhysRevLett.106.106802,PhysRevB.90.165114,PhysRevB.99.235125,PhysRevB.104.195114}.

Although $\bar{e}_2$ and $\bar{\nu}_6$ are related, the expression for their relation is more intricate than that given by Eq.~\eqref{eq:relation_e2bar_nu4bar} for $\mathcal{C}_{4z}$-symmetric systems.
Similarly to the derivation of Eq.~\eqref{eq:e2_Pf_GK} from the real part of Eq.~\eqref{eq:e2_C6z_det+1}, we consider the imaginary part of Eq.~\eqref{eq:e2_C6z_det+1} to obtain 
\begin{align}
    \sin \qty(\dfrac{\pi}{3} e_2) &= \dfrac{\sqrt{3}}{4} e^{-i\phi_{C_{2z}}(\mathrm{M})} \Pf \qty[\tilde{\mathcal{B}}_{C_{6z}}^{\mathrm{A}}(\Gamma)] \nonumber\\
    &\phantom{{}={}} \times\biggl(1 + (-1)^{l_{\Gamma}} \Pf \qty[\tilde{\mathcal{B}}_{C_{6z}}^{\mathrm{A}}(\Gamma)] \Pf \qty[\tilde{\mathcal{B}}_{C_{3z}}^{\mathrm{A}}(\mathrm{K})] \biggr).
\end{align}
Combining this equation with Eq.~\eqref{eq:nu6_Pf_GK} leads to
\begin{equation}
    \label{eq:relation_e2_nu6_sin}
    \sin \qty(\dfrac{\pi}{3} e_2) = \dfrac{\sqrt{3}}{4} e^{-i\phi_{C_{2z}}(\mathrm{M})} \Pf \qty[\tilde{\mathcal{B}}_{C_{6z}}^{\mathrm{A}}(\Gamma)] [1 - (-1)^{\nu_6}].
\end{equation}
It then follows from Eqs.~\eqref{eq:relation_e2_nu6} and \eqref{eq:relation_e2_nu6_sin} that
\begin{align}
    \label{eq:relation_e2bar_nu6bar}
    & \cos \qty(\dfrac{\pi}{3} \qty(\Pf \qty[\tilde{\mathcal{B}}_{C_{6z}}^{\mathrm{A}}(\mathrm{A})] e_2(\pi) - \Pf \qty[\tilde{\mathcal{B}}_{C_{6z}}^{\mathrm{A}}(\Gamma)] e_2(0))) \nonumber\\
    =& \cos \qty(\dfrac{\pi}{3} e_2(\pi)) \cos \qty(\dfrac{\pi}{3} e_2(0)) \nonumber\\
    \phantom{{}={}}& + \Pf \qty[\tilde{\mathcal{B}}_{C_{6z}}^{\mathrm{A}}(\Gamma)] \Pf \qty[\tilde{\mathcal{B}}_{C_{6z}}^{\mathrm{A}}(\mathrm{A})] \sin \qty(\dfrac{\pi}{3} e_2(\pi)) \sin \qty(\dfrac{\pi}{3} e_2(0)) \nonumber\\
    =& \dfrac{1 + 3(-1)^{\bar{\nu}_{6}}}{4},
\end{align}
where we have used the fact that $\Pf \qty[\tilde{\mathcal{B}}_{C_{6z}}^{\mathrm{A}}(\kinv)] = \pm 1$ and that the product of the $\mathcal{C}_{2z}$ eigenvalues at the $\Gamma$ and A points and that at the M and L points are both equal to unity.
Remarkably, in Eq.~(\ref{eq:relation_e2bar_nu6bar}), $\Pf \qty[\tilde{\mathcal{B}}_{C_{6z}}^{\mathrm{A}}]$ appears as a multiplicative factor of $e_2(\bar{k}_z)$.
This indicates that the sign of the Euler class, which is a $\mathbb{Z}$-valued invariant, plays a crucial role when considering it in $\mathcal{C}_{6z}$-symmetric 3D systems.
The second line of Eq.~\eqref{eq:relation_e2bar_nu6bar} reveals that the relative sign between $e_2(0)$ and $e_2(\pi)$ is encoded in the product $\Pf \qty[\tilde{\mathcal{B}}_{C_{6z}}^{\mathrm{A}}(\Gamma)] \Pf \qty[\tilde{\mathcal{B}}_{C_{6z}}^{\mathrm{A}}(\mathrm{A})]$, which equals $+1$ ($-1$) when $e_2(0)$ and $e_2(\pi)$ have the same (different) signs.
Here we have assumed that Eq.~\eqref{eq:relation_e2bar_nu6bar} relates $\bar{e}_2$ and $\bar{\nu}_6$.
We note that the quantities in Eq.~\eqref{eq:relation_e2bar_nu6bar} remain invariant under the simultaneous sign change $(e_2(0), e_2(\pi)) \to (-e_2(0), -e_2(\pi))$.
This reflects the fact that the sign of the Euler class is gauge-dependent, and only the relative sign between $e_2(0)$ and $e_2(\pi)$ matters.

\subsection{Tight-binding model}
\begin{figure}[htp]
\includegraphics[width=8.5cm]{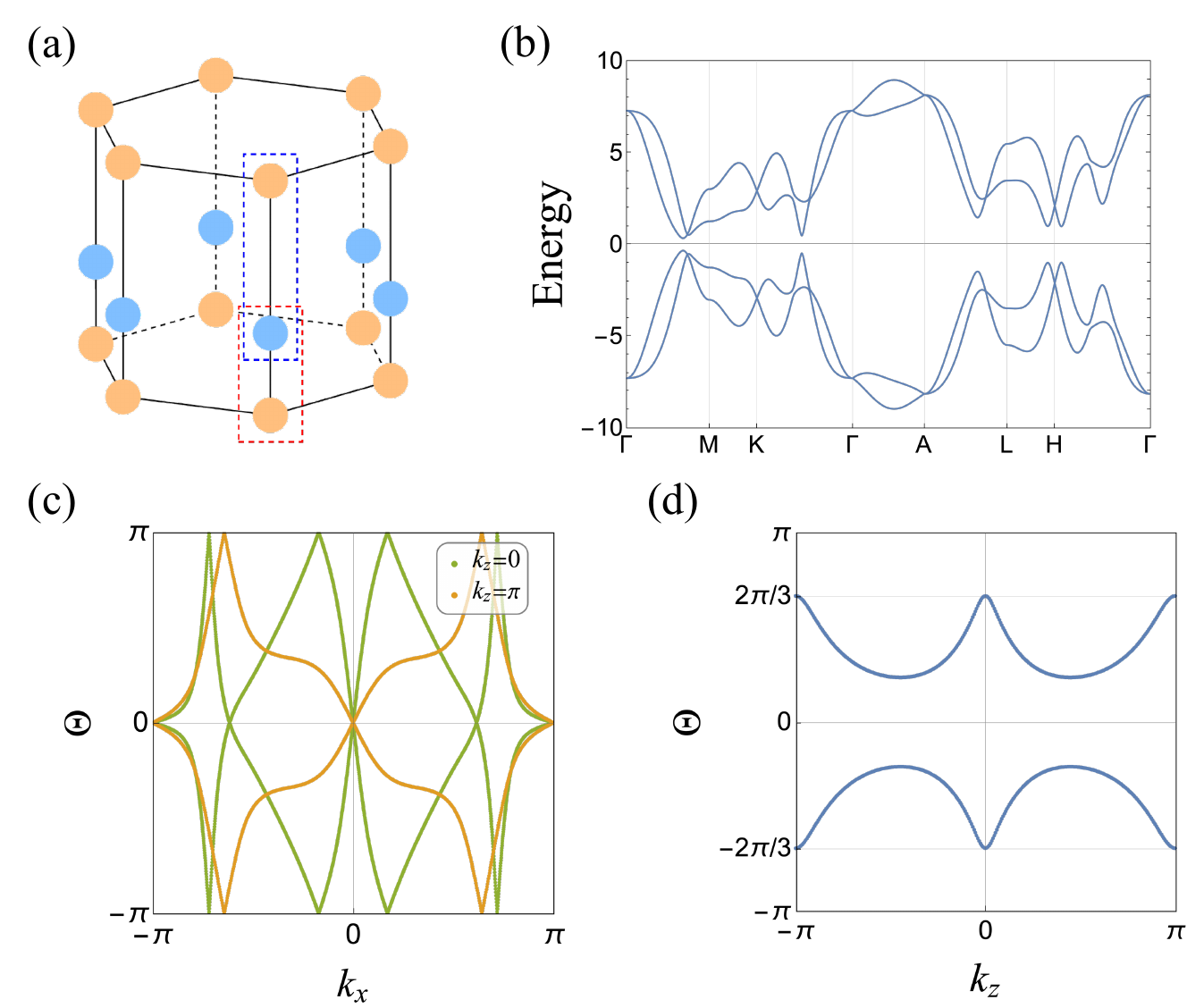}
\caption{
Lattice and topological properties of the Hamiltonian in Eq.~\eqref{eq:Hamil_3D_e2bar=6_nu6bar=0}, where the parameters are chosen as $(M, t_1, t_2, t_3, t_4, t_5, t_6, t_7, t_{1}', t_{2}', t_{3}', t_4') = (-0.7, -0.8, -0.7, 0.6, -0.2, 0.3, -0.6, -0.8, 0.3, 0.2, -0.6, -1.2)$.
(a) Hexagonal lattice structure with two atoms per unit cell.
The red and blue dashed lines indicate different possible choices for defining the unit cell.
(b) Bulk band structure along high-symmetry lines.
(c) Wilson loop spectra obtained by integrating along the $k_y$ direction while keeping $k_z$ fixed within each $C_{2z} T$-invariant plane.
(d) Wilson loop spectrum obtained by integrating along a polygonal path connecting $(k_x, k_y) = (-2\pi/3, 2\pi/\sqrt{3})$, $(0,0)$, and $(4\pi/3, 0)$ with fixed $k_z$.
}
\label{fig:3DTB_C6_bulk}
\end{figure}

Table~\ref{tab:e2_nu6_values} indicates that topological phases with $(\abs{e_2}, \nu_6) = (2, 1)$ and $(4, 1)$ can exist.
For a 3D system exhibiting these phases in the $k_z = 0$ and $k_z = \pi$ planes, respectively, Eq.~\eqref{eq:relation_e2bar_nu6bar} implies that $\Pf \qty[\tilde{\mathcal{B}}_{C_{6z}}^{\mathrm{A}}(\Gamma)] \Pf \qty[\tilde{\mathcal{B}}_{C_{6z}}^{\mathrm{A}}(\mathrm{A})] = -1$, i.e., $e_2(0) e_2(\pi) < 0$.
We confirm this sign difference through numerical calculations of surface states for a tight-binding model, based on the expectation that the difference between $e_2(0)$ and $e_2(\pi)$ determines the number of surface Dirac cones.
The Hamiltonian is given by
\begin{align}
    \label{eq:Hamil_3D_e2bar=6_nu6bar=0}
    &\phantom{{}={}} H(k_x, k_y, k_z) \nonumber\\
    &= \qty[M + 2t_1 \qty(\cos k_x + 2 \cos\dfrac{k_x}{2} \cos \dfrac{\sqrt{3}k_y}{2}) + 2t_1' \cos k_z] \Gamma_{3,0} \nonumber\\
    &\phantom{{}={}} + \qty[2(t_2 + 2t_2' \cos k_z) e_{21}(k_x, k_y) + 2t_3 e_{21}(2k_x, 2k_y)] \Gamma_{1,3} \nonumber\\
    &\phantom{{}={}} + \qty[2 (t_2 + 2t_2' \cos k_z) e_{22}(k_x, k_y) + 2 t_3 e_{22}(2k_x, 2k_y)] \Gamma_{1,1} \nonumber\\
    &\phantom{{}={}} + \qty[2t_4 e_{21}(k_x, k_y) + 2t_5 e_{21}(2k_x, 2k_y)] \Gamma_{3,3} \nonumber\\
    &\phantom{{}={}} + \qty[2 t_4 e_{22}(k_x, k_y) + 2 t_5 e_{22}(2k_x, 2k_y)] \Gamma_{3,1} \nonumber\\
    &\phantom{{}={}} + \qty[t_6 + 2t_7 \qty(\cos k_x + 2 \cos\dfrac{k_x}{2} \cos \dfrac{\sqrt{3}k_y}{2})] \Gamma_{2,2} \nonumber\\
    &\phantom{{}={}} + 2t_3' \sin k_z \Gamma_{2,0} + 2t_4' \sin k_z \Gamma_{1,2},
\end{align}
where
\begin{align}
    e_{21}(k_x, k_y) &= -\qty(\cos k_x - \cos\dfrac{k_x}{2} \cos \dfrac{\sqrt{3}k_y}{2}) \\
    e_{22}(k_x, k_y) &= \sqrt{3} \sin \dfrac{k_x}{2} \sin \dfrac{\sqrt{3}k_y}{2}.
\end{align}
This Hamiltonian has sixfold rotation symmetry
\begin{equation}
    \begin{aligned}
        &\mathcal{C}_{6z} H(k_x, k_y, k_z) \mathcal{C}_{6z}^\dagger = H\qty(\frac{k_x - \sqrt{3}k_y}{2}, \frac{\sqrt{3}k_x + k_y}{2}, k_z), \\
        &\mathcal{C}_{6z} = (\Gamma_{0,0} - i\sqrt{3}\Gamma_{0,2})/2
    \end{aligned}
\end{equation}
and time-reversal symmetry
\begin{equation}
    \mathcal{T} H(k_x, k_y, k_z) \mathcal{T}^\dagger = H(-k_x, -k_y, -k_z), \quad \mathcal{T} = \Gamma_{0,0} K.
\end{equation}
This model describes the $\{p_x, p_y\}$ orbitals on two sites located along the $z$-axis in a hexagonal lattice; see Fig.~\ref{fig:3DTB_C6_bulk}(a).
Thus, the bands form the basis for the real 2D representation, allowing $\nu_6$ to be defined for this model.

Figure~\ref{fig:3DTB_C6_bulk}(b) shows the bulk band structure of this model, where all $C_{6z}T$-invariant points exhibit twofold degeneracy.  
We now compute the bulk topological invariants using the Wilson loop method.
Figure~\ref{fig:3DTB_C6_bulk}(c) presents the Wilson loop spectra obtained by integrating along the $k_y$ direction while keeping $k_x$ fixed within each $C_{2z} T$-invariant plane, from which we obtain $e_2(0) = 4$ and $e_2(\pi) = 2$.
Figure~\ref{fig:3DTB_C6_bulk}(d) illustrates the Wilson loop spectrum obtained by integrating along a polygonal path connecting $(k_x, k_y) = (-2\pi/3, 2\pi/\sqrt{3})$, $(0,0)$, and $(4\pi/3, 0)$ with fixed $k_z$.
The $\mathbb{Z}_2$ invariant $\nu_6(\bar{k}_z)$ is given by the parity of half the number of eigenvalues $\Theta(\bar{k}_z) = \pm2\pi/3$~\cite{PhysRevB.93.205104}.  
Thus, we obtain $\nu_6(0) = \nu_6(\pi) = 1$.
The topological indices of the $k_z = 0, \pi$ planes computed by the Wilson loop method are both listed in Table~\ref{tab:e2_nu6_values}, which supports our symmetry-based analysis.

\begin{figure}[htp]
\includegraphics[width=7.5cm]{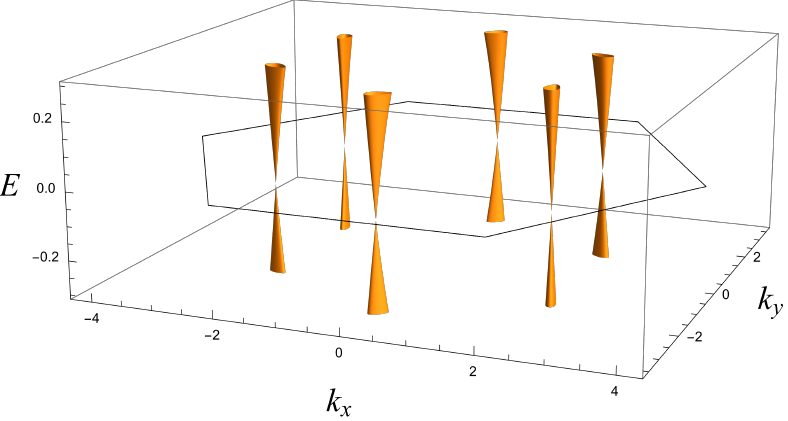}
\caption{
Surface states on the (001) surface of the Hamiltonian in Eq.~\eqref{eq:Hamil_3D_e2bar=6_nu6bar=0}.
The parameters are the same as those in Fig.~\ref{fig:3DTB_C6_bulk}.
The black hexagon indicates the surface BZ.
The calculation is performed for a slab consisting of 100 unit cells, and only the surface states on one side of the slab are shown.
}
\label{fig:3DTB_C6_surf}
\end{figure}

We now move on to the discussion of surface states on the (001) surface.
Figure~\ref{fig:3DTB_C6_surf} presents a 3D plot of midgap surface bands appearing on one side of the slab, where six Dirac cones are observed.
Recalling that the Euler class takes the values $\abs{e_2(0)} = 4$ and $\abs{e_2(\pi)} = 2$ and that the number of surface Dirac cones is expected to be determined by $\abs{\bar{e}_2}$, the observed number of Dirac cones implies that $e_2(0)$ and $e_2(\pi)$ have opposite signs.
The fact that $e_2(0)$ and $e_2(\pi)$ have opposite signs can also be seen from the bulk band structure at parameters yielding inversion symmetry, the details of which are provided in the Appendix~\ref{sec:C6TB_surf_PT}.
We note that these surface states are consistent with $\bar{\nu}_6 = 0$.
By analogy with the discussion of $\bar{\nu}_4$, it is inferred that $\bar{\nu}_6$ dictates the presence or absence of band-touching points in surface states at the $C_{6z}T$-invariant $\bar{\Gamma}$ or $\bar{\mathrm{K}}$ points.
Thus, the absence of degeneracies at both $\bar{\Gamma}$ and $\bar{\mathrm{K}}$ points is in agreement with $\bar{\nu}_6 = 0$.
Surface states for a different choice of unit cell are discussed in Appendix~\ref{sec:C6TB_surf_different_cell}.

\subsection{Relationship among $\bar{e}_2$, $\bar{\nu}_6$, and the halved mirror chirality}

We now turn to the discussion of halved mirror chirality.
In systems with $\mathcal{C}_{6v}$ symmetry, a single half-mirror plane can be defined, denoted as
\begin{align}
    \mathrm{HMP}_{6,1} = \left\{
        \qty(4\pi t_{6,1}/3, 0, k_z) 
        \;\middle|\;
        \begin{array}{l}
            0 \leq t_{6,1} \leq 1, \\
            -\pi \leq k_z \leq \pi
        \end{array}
    \right\},
\end{align}
and shown in Fig.~\ref{fig:HMP}(b).
This plane is referred to as $\mathrm{HMP}_1$ in Ref.~\cite{PhysRevLett.113.116403}.
With the gauge choice of Eq.~\eqref{eq:real-gauge_3D}, Eq.~\eqref{eq:B_eta_Stokes_real-gauge_2bands} follows from the same calculation as in Sec.~\ref{subsec:Euler_C4z_3D}.
As the analog of Eq.~\eqref{eq:C4v_u_-eta}, we consider the state defined by
\begin{equation}
    \label{eq:C6v_u_-eta}
    \ket{\tilde{u}^{-\eta}(\bar{t}, k_z)} = \dfrac{1}{\sqrt{3}} e^{i\varphi_\eta(\bar{t}, k_z)} \qty(\mathcal{C}_{3z} - \mathcal{C}_{3z}^{-1}) \ket{\tilde{u}^\eta(\bar{t}, k_z)},
\end{equation}
where $\bar{t} \in \{0, 1\}$ and $e^{i\varphi_\eta(\bar{t}, k_z)}$ accounts for the U(1) phase ambiguity.
This is a normalized state with the mirror eigenvalue opposite to that of $\ket{\tilde{u}^\eta(\bar{t}, k_z)}$, and it forms a 2D representation of the point group $\mathcal{C}_{6v}$ together with $\ket{\tilde{u}^\eta(\bar{t}, k_z)}$.
Using Eq.~\eqref{eq:C6v_u_-eta}, we see that Eq.~\eqref{eq:A_eta_C4v} is satisfied in this case as well.
We thus obtain
\begin{equation}
    \label{eq:chi_C6v}
    (-1)^{\chi_{6,1}} = e^{-i\varphi_{\mathrm{e}}(0, 0)} e^{i\varphi_{\mathrm{e}}(1, 0)} e^{i\varphi_{\mathrm{e}}(0, \pi)} e^{-i\varphi_{\mathrm{e}}(1, \pi)}
\end{equation}
which has the same form as Eq.~\eqref{eq:chi_C4v}.
The right-hand side can be expressed in terms of $\tilde{\mathcal{B}}_{C_{3z}}^{\mathrm{A}}$ as
\begin{align}
    e^{-i\varphi_{\mathrm{e}}(\bar{t}, \bar{k}_z)} &= \dfrac{1}{\sqrt{3}} \mel{\tilde{u}^{\mathrm{o}}(\bar{t}, \bar{k}_z)}{\qty(\mathcal{C}_{3z} - \mathcal{C}_{3z}^{-1})}{\tilde{u}^{\mathrm{e}}(\bar{t}, \bar{k}_z)} \nonumber\\
    &= \pm \Pf \qty[\tilde{\mathcal{B}}_{C_{3z}}^{\mathrm{A}}(\bar{t}, \bar{k}_z)].
\end{align}
The $\Gamma$ and A points, which correspond to $\bar{t} = 0$, possess $\mathcal{C}_{6z}$ symmetry.
At the $\Gamma$ point, we have
\begin{align}
    \mathcal{C}_{6z}^{-1} = \mathcal{C}_{2z} \mathcal{C}_{3z} = (-1)^{l_{\Gamma}} \mathcal{C}_{3z},
\end{align}
from which it follows that
\begin{equation}
    \Pf \qty[\tilde{\mathcal{B}}_{C_{3z}}^{\mathrm{A}}(0, 0)] = - (-1)^{l_{\Gamma}} \Pf \qty[\tilde{\mathcal{B}}_{C_{6z}}^{\mathrm{A}}(0, 0)].
\end{equation}
A similar relation holds at the A point.
If the system is insulating, the $\mathcal{C}_{2z}$ eigenvalues of the occupied bands are identical at the $\Gamma$ and A points. 
This implies that $(-1)^{l_{\Gamma}} = (-1)^{l_{\mathrm{A}}}$; otherwise the system becomes a spinless Dirac semimetal~\cite{Sato2024}.
From the foregoing discussion, we finally arrive at
\begin{align}
    \label{eq:chi61_Pf_GKAH}
    (-1)^{\chi_{6,1}} &= \dfrac{-(-1)^{l_{\Gamma}} \Pf \qty[\tilde{\mathcal{B}}_{C_{6z}}^{\mathrm{A}}(0, 0)]}{\Pf \qty[\tilde{\mathcal{B}}_{C_{3z}}^{\mathrm{A}}(1, 0)]} \dfrac{\Pf \qty[\tilde{\mathcal{B}}_{C_{3z}}^{\mathrm{A}}(1, \pi)]}{-(-1)^{l_{\mathrm{A}}} \Pf \qty[\tilde{\mathcal{B}}_{C_{6z}}^{\mathrm{A}}(0, \pi)]} \nonumber\\
    &= \dfrac{\Pf \qty[\tilde{\mathcal{B}}_{C_{6z}}^{\mathrm{A}}(\Gamma)]}{\Pf \qty[\tilde{\mathcal{B}}_{C_{3z}}^{\mathrm{A}}(\mathrm{K})]} \dfrac{\Pf \qty[\tilde{\mathcal{B}}_{C_{3z}}^{\mathrm{A}}(\mathrm{H})]}{\Pf \qty[\tilde{\mathcal{B}}_{C_{6z}}^{\mathrm{A}}(\mathrm{A})]}.
\end{align}

We are now in a position to derive an explicit relation connecting $\bar{e}_2$, $\bar{\nu}_6$, and $\chi_{6,1}$.
Using Eq.~\eqref{eq:nu6_Pf_GK} together with $(-1)^{l_{\Gamma}} = (-1)^{l_{\mathrm{A}}}$, we find that
\begin{align}
    \label{eq:nu6bar_Pf_GKAH}
    (-1)^{\bar{\nu}_6} &= \dfrac{\Pf \qty[\tilde{\mathcal{B}}_{C_{6z}}^{\mathrm{A}}(\Gamma)]}{\Pf \qty[\tilde{\mathcal{B}}_{C_{3z}}^{\mathrm{A}}(\mathrm{K})]} \dfrac{\Pf \qty[\tilde{\mathcal{B}}_{C_{3z}}^{\mathrm{A}}(\mathrm{H})]}{\Pf \qty[\tilde{\mathcal{B}}_{C_{6z}}^{\mathrm{A}}(\mathrm{A})]}
\end{align}
holds for $\bar{\nu}_6$.
We obtain from Eqs.~\eqref{eq:chi61_Pf_GKAH} and \eqref{eq:nu6bar_Pf_GKAH} the simple relation 
\begin{equation}
\label{eq:relation_nu6bar_chi}
(-1)^{\bar{\nu}_{6}} = (-1)^{\chi_{6,1}}, 
\end{equation}
which takes the same form as the relation between $\bar{\nu}_4$ and $\chi_{4,i}$ in $\mathcal{C}_{4z}$-symmetric systems.
This relationship between $\bar{\nu}_6$ and $\chi_{6,1}$ remains valid for an arbitrary number of bands in the doublet representations, which will be proven in Appendix~\ref{sec:nu6_chi6_Nbands}.

Consequently, combining Eqs.~\eqref{eq:relation_e2bar_nu6bar} and \eqref{eq:relation_nu6bar_chi}, we obtain
\begin{align}
    \label{eq:relation_e2bar_nu6bar_chi}
    & \cos \qty(\dfrac{\pi}{3} \qty(\Pf \qty[\tilde{\mathcal{B}}_{C_{6z}}^{\mathrm{A}}(\mathrm{A})] e_2(\pi) - \Pf \qty[\tilde{\mathcal{B}}_{C_{6z}}^{\mathrm{A}}(\Gamma)] e_2(0))) \nonumber\\
    =& \dfrac{1 + 3(-1)^{\bar{\nu}_{6}}}{4} = \dfrac{1 + 3(-1)^{\chi_{6,1}}}{4}
\end{align}
as the general relationship among the three types of topological invariants.

\subsection{Phase transitions}
\begin{figure}[htp]
\includegraphics[width=8.5cm]{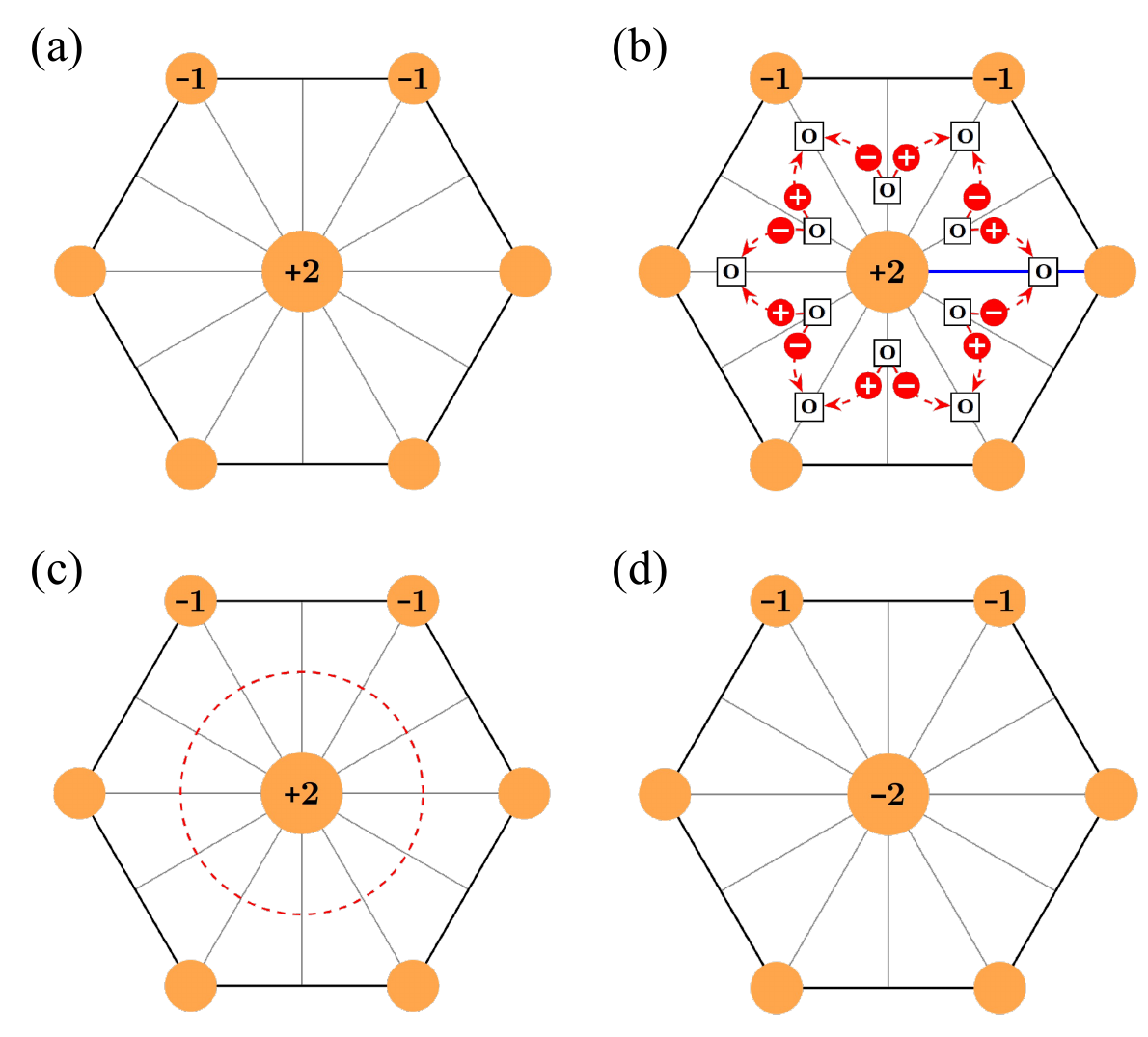}
\caption{
Distribution of nodal points in the $k_z = 0$ plane during the phase transition from $(e_2(0), e_2(\pi), \chi_{6,1}) = (0, 0, 0)$ to $(e_2(0), e_2(\pi), \chi_{6,1}) = (2, 0, 1)$. 
(a) Nodal points between occupied bands and their winding numbers in the insulating phase with $e_2(0) = 0$. The occupied bands are degenerate at the $C_{6z}T$-invariant points indicated in orange.
Among the nodal points related by the periodicity of the BZ, the winding number is indicated for only one of them.
(b) Nodal points in the Weyl semimetal phase mediating two insulating phases. The red circles indicate Weyl points (nodal points between occupied and unoccupied bands). The blue line represents the $\mathrm{HMP}_{6,1}$. 
The sign $+$ $(-)$ represents Weyl points with positive (negative) chirality, and the symbol o indicates that gap closure at the HMP takes place in the mirror-odd subspace.
(c) Insulating phase after the pair annihilation of Weyl points. The red dashed lines represent branch cuts (Dirac strings).  
(d) Nodal points of occupied bands and their winding numbers in the insulating phase with $e_2(0) = 2$.  
}
\label{fig:C6_phase_transition}
\end{figure}

As discussed for $\mathcal{C}_{4z}$-symmetric systems in Sec.~\ref{subsec:C4z_phase_transitions}, the relation between $\bar{e}_2$ and $\chi_{6,1}$ in Eq.~\eqref{eq:relation_e2bar_nu6bar_chi}, together with the phase transition processes considered separately in previous studies~\cite{PhysRevLett.113.116403,PhysRevX.9.021013}, allows for a unified description of the topological phase transitions associated with $\bar{e}_2$ and $\chi_{6,1}$.
To demonstrate this, we consider a phase transition from $(e_2(0), e_2(\pi), \chi_{6,1}) = (0, 0, 0)$ to $(e_2(0), e_2(\pi), \chi_{6,1}) = (2, 0, 1)$.
Since $e_2(\pi)$ remains unchanged during this transition, we only need to focus on the $k_z = 0$ plane. 

In systems where $\chi_{6,1}$ is defined, the occupied bands form 2D irreducible representations and exhibit band-touching points (nodal points) at the $C_{6z}T$-invariant momenta.  
The winding number associated with a nodal point depends on the symmetry of the high-symmetry point.
For example, band degeneracies at the $C_{6z}$-invariant $\Gamma$ point have a winding number of magnitude 2, whereas those at the $C_{3z}$-invariant $\mathrm{K}$ and $\mathrm{K}'$ points have a winding number of magnitude 1. 
Note that the $\mathrm{K}$ and $\mathrm{K}'$ points, which are related by $C_{6z}$ symmetry, always share the same winding number when the basis consists of an orbital pair labeled by an orbital angular momentum~\cite{PhysRevB.104.195114}.
As implied by Eq.~\eqref{eq:e2_Nt}, the configuration shown in Fig.~\ref{fig:C6_phase_transition}(a) has $e_2(0) = 0$, as the band degeneracies (nodal points) at $\Gamma$ and $(\mathrm{K}, \mathrm{K}')$ possess winding numbers of opposite sign.

In $\mathcal{C}_{6v}$-symmetric systems in which the halved mirror chirality can take nontrivial values, phase transitions in $\chi_{6,1}$ are known to occur through the transfer of Weyl points between distinct mirror-invariant planes, as illustrated in Fig.~\ref{fig:C6_phase_transition}(b).
During this process, as shown in Fig.~\ref{fig:C6_phase_transition}(c), the trajectory of the Weyl points, which produce branch cuts, forms a closed loop enclosing the $C_{6z}$-invariant $\Gamma$ point, thereby completing the phase transition to $e_2(0) = 2$.
This loop may shrink and eventually disappear, leading to the state illustrated in Fig.~\ref{fig:C6_phase_transition}(d).
The same phase transition can also occur when Weyl points move in a manner that encloses the $C_{3z}$-invariant $\mathrm{K}$ and $\mathrm{K}'$ points.
We end this section by noting that the correspondence between the nodal points of the occupied bulk bands and those of the midgap surface bands also holds in $\mathcal{C}_{6z}$-symmetric systems.

\section{Conclusion}
We have investigated the relationship between the Euler class $e_2$ and rotational symmetries in spinless insulators with two occupied bands.  
We have demonstrated that the transformation properties of the real Berry connection and curvature under point group operations depend on the form of the corresponding sewing matrices.
By applying these transformation rules to $\mathcal{C}_{4z}$ or $\mathcal{C}_{6z}$ in spinless systems, we have established explicit relations between the value of $e_2$ and the rotation eigenvalues at high-symmetry points in the BZ.  
While these formulas are not always applicable to computing $e_2$, we have shown that the condition for their applicability is equivalent to the condition for defining the representation-protected $\mathbb{Z}_2$ invariant $\nu_n$, and we have derived the closed-form formulas relating $e_2$ to $\nu_n$.  
As a consequence, our results enable the determination of $e_2$ or $\nu_n$ without resorting to the computation of Wilson loop spectra.

We have further extended our analysis to three-dimensional systems, focusing on the differences of topological indices calculated at the two $C_{2z}T$-invariant planes, $\bar{e}_2$ and $\bar{\nu}_n$.
We have derived the explicit relations between $\bar{e}_2$ and $\bar{\nu}_n$, and have confirmed that these relations differ significantly between $\mathcal{C}_{4z}$- and $\mathcal{C}_{6z}$-symmetric systems.
Our results are consistent with numerical findings reported in previous studies and further suggest the existence of previously unexplored topological phases.
Remarkably, we have identified cases in $\mathcal{C}_{6z}$-symmetric systems where the Euler classes on the $k_z = 0$ and $k_z = \pi$ planes have opposite signs.
To substantiate the existence of these phases, we have constructed tight-binding models that realize them and, by calculating the surface states, have gained key insight into the bulk-boundary correspondence.
Although $\bar{\nu}_n = 0$ in our models, gapless surface states appear as a consequence of the nontrivial values of $\bar{e}_2$.
We have further studied systems with mirror symmetry and have obtained relations between $\chi_{n,i}$ and the other invariants in both $\mathcal{C}_{4v}$- and $\mathcal{C}_{6v}$-symmetric systems.

Furthermore, we have pointed out that phase transition processes, which have been discussed separately for each topological invariant in previous studies, can be understood within a unified framework.  
By tracking the surface states during the phase transitions, we have also shown that the nodal points between the occupied bulk bands and those between the gapless surface bands nearly coincide in position in momentum space.

Finally, we comment on the effect of the spin-orbit coupling (SOC). 
In the presence of the SOC, the spin (up and down) sectors are no longer decoupled, and the full Hamiltonian must be considered. 
As a result, the number of bands doubles due to the spin degrees of freedom, rendering the Euler class, applicable only to two-band subspaces, undefined for the occupied bands.
Instead, the system is characterized by the $\mathbb{Z}_2$-valued second Stiefel–Whitney class. 
If the Euler class is even, the second Stiefel–Whitney class vanishes, indicating the trivial topology.
Accordingly, the SOC opens a gap in the surface band structure.

\begin{acknowledgments}
This work was supported by JST SPRING, Grant No.\ JPMJSP2108 and JST CREST (Grant No.\ JPMJCR19T2).
S.K. acknowledges financial support from JSPS KAKENHI Grants No.\ JP22K03478, No.\ JP24K00557, and No.\ 25K07161.
\end{acknowledgments}

\appendix{}

\section{2D tight-binding models with $\mathcal{C}_{4z}$ symmetry}
\label{sec:2DTBs_e2_nu4}
According to Table~\ref{tab:e2_nu4_values}, in cases where the $\mathcal{C}_{2z}$ eigenvalue at the X point is $+1$, there exist topological phases where either $e_2$ or $\nu_4$ is trivial while the other is nontrivial.
In this Appendix, we present 2D tight-binding models that realize such phases.

\begin{figure}[htp]
\includegraphics[width=8.5cm]{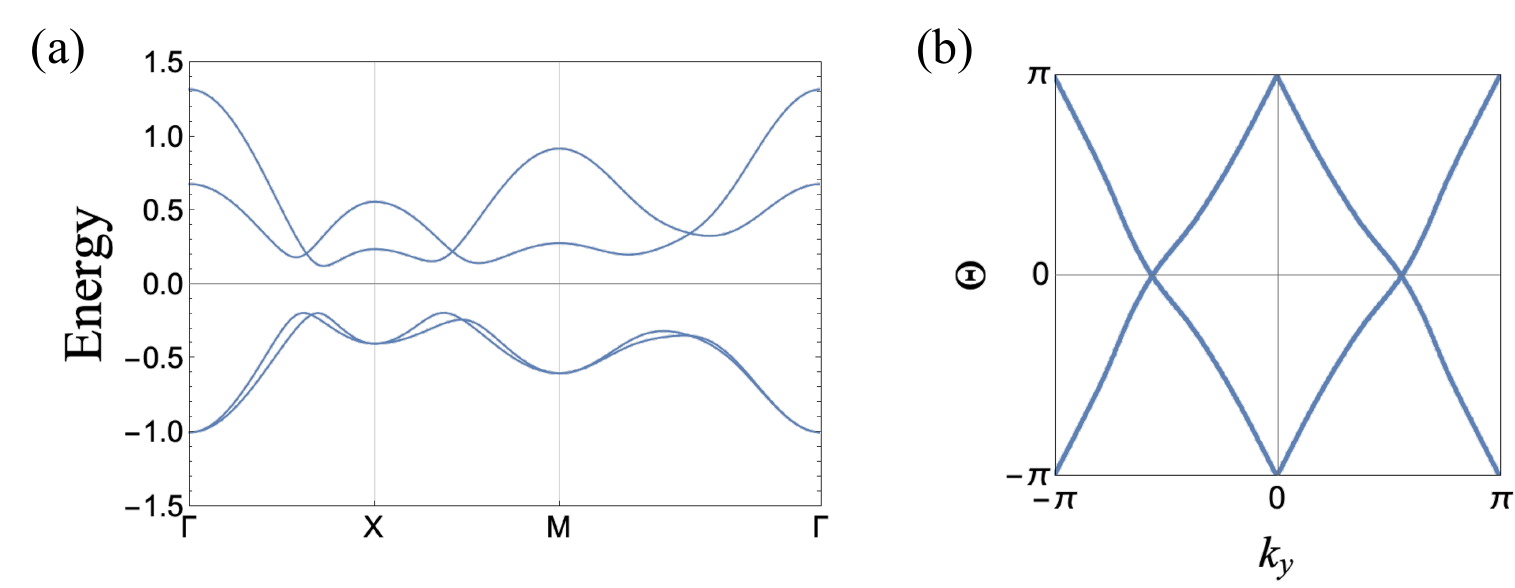}
\caption{
Bulk band structure and topology of the Hamiltonian in Eq.~\eqref{eq:Hamil_2D_e2=2_nu=0}, where the parameters are chosen as $(M, t_1, t_2, t_3, t_4, t_5) = (0.2, 0.1, 0.6, 0.08, 0.04, 0.2)$.
(a) Bulk band structure along high-symmetry lines.
(b) Wilson loop spectra obtained by integrating along the $k_x$ direction at fixed $k_y$.
}
\label{fig:2DTB1_C4_bulk}
\end{figure}

First, we introduce a tight-binding model that exhibits $e_2 = 2$ and $\nu_4 = 0$.
The Hamiltonian is given by
\begin{align}
    \label{eq:Hamil_2D_e2=2_nu=0}
    H(k_x, k_y) &= \qty[M + t_1(\cos k_x + \cos k_y) + t_2 \cos k_x \cos k_y] \Gamma_{3,0} \nonumber\\
    &\phantom{{}={}} + t_3 (\cos k_x + \cos k_y) \qty(\Gamma_{0,1} + \Gamma_{3,1}) \nonumber\\
    &\phantom{{}={}} + t_4 (\cos k_x - \cos k_y) \qty(\Gamma_{0,1} - \Gamma_{3,1}) \nonumber\\
    &\phantom{{}={}} + t_5 \qty(\sin k_x \Gamma_{2,1} - \sin k_y \Gamma_{2,3}).
\end{align}
The basis of this model consists of two orbitals with $\mathcal{C}_{2z}$ eigenvalue $+1$ and two with $-1$, all located on the same rotation axis.
The Hamiltonian in Eq.~\eqref{eq:Hamil_2D_e2=2_nu=0} has fourfold rotation symmetry
\begin{equation}
    \begin{aligned}
        &\mathcal{C}_{4z} H(k_x, k_y) \mathcal{C}_{4z}^\dagger = H(-k_y, k_x), \\
        &\mathcal{C}_{4z} = \qty[\Gamma_{0,0} + \Gamma_{3,0} - i(\Gamma_{0,2} - \Gamma_{3,2})]/2
    \end{aligned}
\end{equation}
and time-reversal symmetry
\begin{equation}
    \mathcal{T} H(k_x, k_y) \mathcal{T}^\dagger = H(-k_x, -k_y), \quad \mathcal{T} = \Gamma_{0,0} K.
\end{equation}

Figure~\ref{fig:2DTB1_C4_bulk}(a) shows the bulk band structure of this model, where we can observe the band inversion at the X point.
The $\mathcal{C}_{2z}$ eigenvalues at the $\Gamma$ and M points are $-1$, which allows $\nu_4$ to be defined.
However, due to the band inversion, the $\mathcal{C}_{2z}$ eigenvalue at the X point becomes $+1$.
As shown in Fig.~\ref{fig:2DTB1_C4_bulk}(b), the Wilson loop spectra calculated along the $k_x$ direction consist of the curves with the winding number $+2$ and $-2$.
Thus, the Euler class of this model is determined to be $e_2 = 2$.
However, the eigenvalues of the Wilson loop operator calculated along the bent loop connecting $C_{4z}$-invariant points are $(0, 0)$, which means $\nu_4 = 0$.

\begin{figure}[htp]
\includegraphics[width=8.5cm]{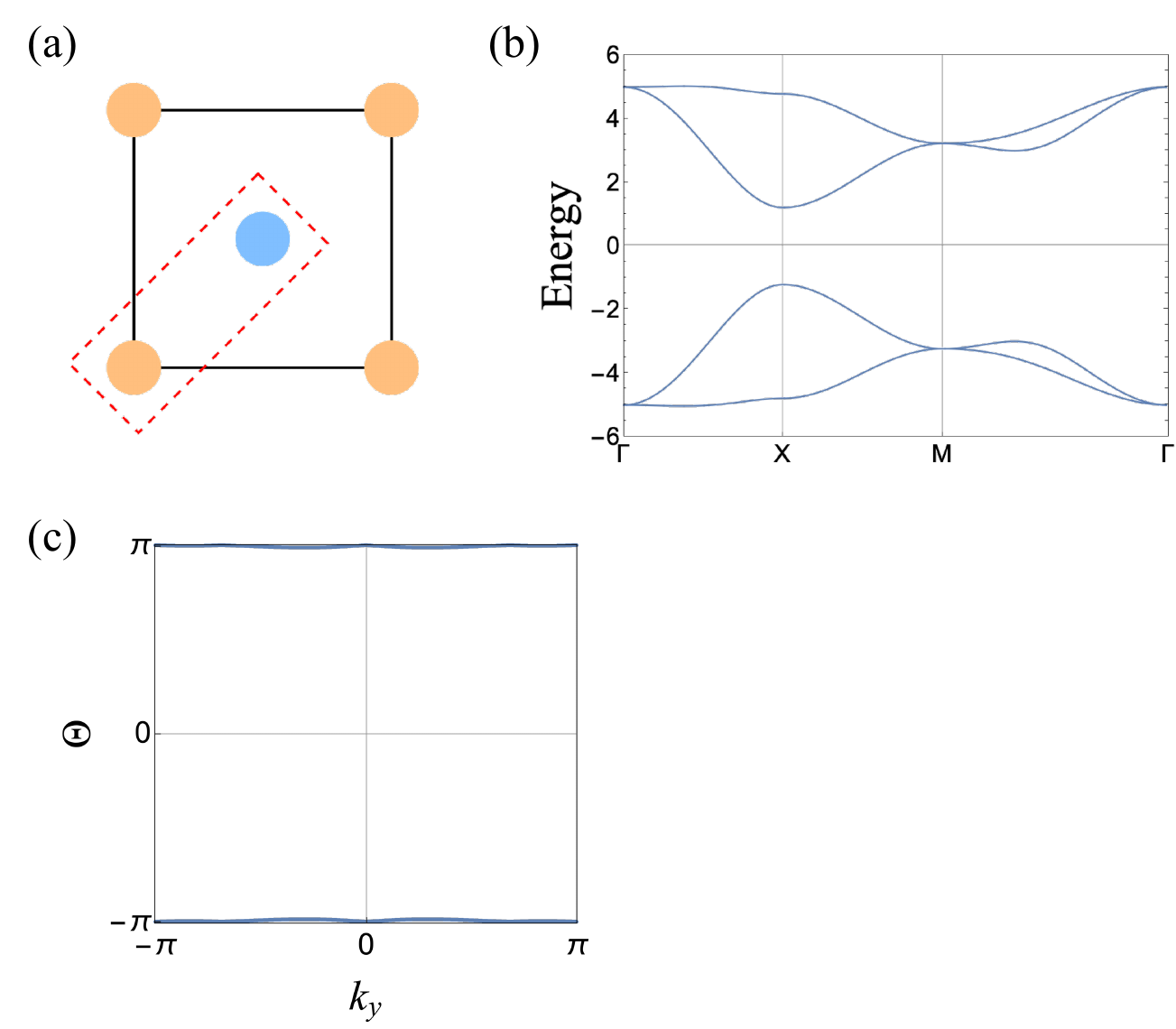}
\caption{
Bulk band structure and topology of the Hamiltonian in Eq.~\eqref{eq:Hamil_2D_e2=0_nu=1}, where the parameters are chosen as $(M, t_1, t_2, t_3, t_4, t_5) = (3.0, 1.0, 0.3, 0.4, 0.2)$.
(a) Tetragonal lattice structure with two atoms per unit cell.
The red dashed line indicates the unit cell.
(b) Bulk band structure along high-symmetry lines.
(c) Wilson loop spectra obtained by integrating along the $k_x$ direction at fixed $k_y$.
}
\label{fig:2DTB2_C4_bulk}
\end{figure}

Next, we introduce a tight-binding model that exhibits $e_2 = 0$ and $\nu_4 = 1$.
Unlike the other models which we discussed, this model utilizes sublattice degrees of freedom in the $xy$-plane.
The basis of this model consists of $\{p_x, p_y\}$ orbitals at the vertex and center of the tetragonal crystal (see Fig.~\ref{fig:2DTB2_C4_bulk}(a)).
The Hamiltonian is given by
\begin{align}
    \label{eq:Hamil_2D_e2=0_nu=1}
    H(k_x, k_y) &= M \Gamma_{3,0} + 4t_1 \cos\dfrac{k_x}{2} \cos\dfrac{k_y}{2} \Gamma_{1,0} \nonumber\\
    &\phantom{{}={}} + 4t_2 \sin\dfrac{k_x}{2} \sin\dfrac{k_y}{2} \Gamma_{1,1} \nonumber\\
    &\phantom{{}={}} + 2 t_3 (\cos k_x - \cos k_y) \Gamma_{0,3} \nonumber\\
    &\phantom{{}={}} + 2 t_4 (\cos k_x - \cos k_y) \Gamma_{0,1},
\end{align}
where the Fourier transformation is performed taking into account the atomic positions within the unit cell.
The Hamiltonian in Eq.~\eqref{eq:Hamil_2D_e2=0_nu=1} has fourfold rotation symmetry
\begin{equation}
    \mathcal{C}_{4z} H(k_x, k_y) \mathcal{C}_{4z}^\dagger = H(-k_y, k_x), \quad \mathcal{C}_{4z} = -i \Gamma_{0,2}
\end{equation}
and time-reversal symmetry
\begin{equation}
    \mathcal{T} H(k_x, k_y) \mathcal{T}^\dagger = H(-k_x, -k_y), \quad \mathcal{T} = \Gamma_{0,0} K.
\end{equation}
The matrix representations of the symmetry operations are identical to those in the Fu model~\cite{PhysRevLett.106.106802,PhysRevB.104.195114}, which enables the definition of $\nu_4$.
Under a translation by a reciprocal lattice vector $\bm{G}$, the Hamiltonian transforms as
\begin{equation}
    \begin{aligned}
        & H(\bm{k}+\bm{G}) = V^{-1}(\bm{G}) H(\bm{k}) V(\bm{G}), \\
        &V(\bm{G}) = \operatorname{diag} \qty(1, 1, \exp\qty[i \bm{G} \cdot \qty(\dfrac{1}{2}, \dfrac{1}{2})], \exp\qty[i \bm{G} \cdot \qty(\dfrac{1}{2}, \dfrac{1}{2})]).
    \end{aligned}
\end{equation}
Since $\det V(\bm{G}) = \exp\qty[i \bm{G} \cdot \qty(1, 1)] = +1$, the Hamiltonian is glued with an orientation-preserving transformation at the BZ boundary.
Thus, the Euler class is well-defined in this model~\cite{PhysRevX.9.021013}.

As shown in Fig.~\ref{fig:2DTB2_C4_bulk}(b), band inversion does not occur in this model.
The occupied bands originate from the site with lower energy in the entire BZ.
Correspondingly, as shown in Fig.~\ref{fig:2DTB2_C4_bulk}(c), the Wilson loop spectra remain nearly constant for any $k_y$, with values corresponding to the position of the lower-energy site.
Since the spectra show no winding, we obtain $e_2 = 0$.
However, the eigenvalues of the Wilson loop operator calculated along the bent loop connecting $C_{4z}$-invariant points are $(\pi, \pi)$, which means $\nu_4 = 1$.

\section{Generalization of the relationship between $\bar{\nu}_4$ and $\chi_{4,i}$ for an arbitrary number of bands}
\label{sec:nu4_chi4_Nbands}
In Sec.~\ref{subsec:Euler_C4z_3D}, we considered systems with two occupied bands, which allows the definition of the Euler class.
In this Appendix, we shift our focus away from the Euler class and instead extend the relationship between $\bar{\nu}_4$ and $\chi_{4,i}$ in Eq.~\eqref{eq:relation_e2bar_nu4bar_chi} to insulating systems in which the occupied bands consist of $N$ pairs, each belonging to the doublet representation.
For the following calculations, we adopt the gauge choice in Eq.~\eqref{eq:real-gauge_3D}.

First, we consider $\nu_4$, which is defined in Eq.~\eqref{eq:nu4_def}.
Since it is defined in the $C_{2z}T$-invariant plane, we adopt the real-gauge satisfying $\mathcal{C}_{2z}\mathcal{T} \ket{\tilde{u}_n(\bm{k})} = \ket{\tilde{u}_n(\bm{k})}$.
Even when the number of bands increases, the properties $\bm{\mathcal{A}}(\bm{k}) = \bm{0}$ and $\tilde{w}_4(\kinv) = - \tilde{\mathcal{B}}_{C_{4z}}(\kinv)$ remain valid, where $\kinv$ denotes the $C_{4z}T$-invariant momenta.
Noting that the relation
\begin{equation} 
\Pf \qty[-\tilde{\mathcal{B}}_{C_{4z}}(\kinv)] = (-1)^N \Pf \qty[\tilde{\mathcal{B}}_{C_{4z}}(\kinv)] 
\end{equation}
holds for the $2N \times 2N$ antisymmetric matrix $\tilde{\mathcal{B}}_{C_{4z}}(\kinv)$, we find that the expression
\begin{equation}
    (-1)^{\nu_{4}} = \dfrac{\Pf \qty[\tilde{\mathcal{B}}_{C_{4z}}(\mathrm{M})]}{\Pf \qty[\tilde{\mathcal{B}}_{C_{4z}}(\Gamma)]}
\end{equation}
is applicable to an arbitrary number of bands.
Thus, we obtain
\begin{equation}
    \label{eq:nu4bar_Pf_GMZA_Nbands}
    (-1)^{\bar{\nu}_4} = \dfrac{\Pf \qty[\tilde{\mathcal{B}}_{C_{4z}}(\mathrm{A})]}{\Pf \qty[\tilde{\mathcal{B}}_{C_{4z}}(\mathrm{Z})]} \dfrac{\Pf \qty[\tilde{\mathcal{B}}_{C_{4z}}(\Gamma)]}{\Pf \qty[\tilde{\mathcal{B}}_{C_{4z}}(\mathrm{M})]},
\end{equation}
which coincides with Eq.~\eqref{eq:nu4bar_Pf_GMZA}.

Next, we consider $\chi_{4,i}$, which is defined in Eq.~\eqref{eq:chi_def}.
Based on our assumption, each HMP contains $N$ bands with mirror eigenvalue $+1$ and $N$ bands with mirror eigenvalue $-1$.
Following the derivation of Eq.~\eqref{eq:B_eta_Stokes_real-gauge_2bands}, we obtain
\begin{align}
    \dfrac{1}{2} B_\eta^{(n,i)} &= \int_{
        \substack{
            0 \leq t_{n,i} \leq 1 \\
            0 \leq k_z \leq \pi
        }
        } dt_{n,i} dk_z \; \mathcal{F}_{\eta} \nonumber\\
    &= \int_{0}^\pi dk_z \; \tilde{\mathcal{A}}^\eta_z(1, k_z) - \int_{0}^\pi dk_z \; \tilde{\mathcal{A}}^\eta_z(0, k_z),
\end{align}
where $\tilde{\mathcal{A}}^\eta_\mu$ is the U(1) Berry connection in the mirror subspace specified by $\eta$, given by
\begin{equation}
    \tilde{\mathcal{A}}^\eta_\mu(t, k_z) = -i \sum_{n}^{\mathrm{occ.}} \braket{\tilde{u}_n^\eta(t, k_z)}{\partial_\mu \tilde{u}_n^\eta(t, k_z)}.
\end{equation}
As in the main text, we set $\bar{t} \in \qty{0,1}$, which represents the rotation-invariant line in the BZ.
For a state $\ket{\tilde{u}^\eta_m(\bar{t}, k_z)}$ in the occupied subspace with mirror eigenvalue $\eta$, it follows from the relation $\mathcal{M}_{4,i} \mathcal{C}_{4z} \mathcal{M}_{4,i}^{-1} = \mathcal{C}_{4z}^{-1}$ that
\begin{align}
    \label{eq:u_-eta_C4v_Nbands}
    \ket{\tilde{u}^{-\eta}_n(\bar{t}, k_z)} &= \sum_m U_{mn}^\eta(\bar{t}, k_z) \dfrac{1}{2} \qty(\mathcal{C}_{4z} - \mathcal{C}_{4z}^{-1}) \ket{\tilde{u}^\eta_m(\bar{t}, k_z)} \nonumber\\
    &= \sum_m U_{mn}^\eta(\bar{t}, k_z) \mathcal{C}_{4z} \ket{\tilde{u}^\eta_m(\bar{t}, k_z)}
\end{align}
belongs to the occupied subspace with mirror eigenvalue $-\eta$.  
Here, $U^\eta(\bar{t}, k_z)$ is an $N \times N$ unitary matrix that must satisfy $[U^\eta(\bar{t}, k_z)]^* = U^\eta(\bar{t}, -k_z)$ to ensure the gauge choice in Eq.~\eqref{eq:real-gauge_3D} is preserved.  
It should be noted that, unlike in the case of two occupied bands, neither $\ket{\tilde{u}^\eta_m(\bar{t}, k_z)}$ nor $\ket{\tilde{u}^{-\eta}_m(\bar{t}, k_z)}$ is necessarily an energy eigenstate.

It can be derived from Eq.~\eqref{eq:u_-eta_C4v_Nbands} that
\begin{align}
    \label{eq:Berry_connection_mirror_Nbands}
    \tilde{\mathcal{A}}^{-\eta}_z(\bar{t}, k_z) &= \tilde{\mathcal{A}}^{\eta}_z(\bar{t}, k_z) - i \Tr \qty[(U^\eta(\bar{t}, k_z))^\dagger \partial_{k_z} U^\eta(\bar{t}, k_z)] \nonumber\\
    &= \tilde{\mathcal{A}}^{\eta}_z(\bar{t}, k_z) - i \partial_{k_z} \log \det \qty[U^\eta(\bar{t}, k_z)].
\end{align}
Thus, we obtain
\begin{align}
    \dfrac{1}{2} \chi_{4,i} &= \dfrac{1}{2\pi} \dfrac{1}{2} \qty(B_{\mathrm{e}}^{(4,i)} - B_{\mathrm{o}}^{(4,i)}) \nonumber\\
    &= \dfrac{i}{2\pi} \Bigl( \log \det \qty[U^{\mathrm{e}}(1, \pi)] - \log \det \qty[U^{\mathrm{e}}(1, 0)] \nonumber\\
    &\phantom{{}={}} + \log \det \qty[U^{\mathrm{e}}(0, 0)] - \log \det \qty[U^{\mathrm{e}}(0, \pi)] \Bigr) \mod 1,
\end{align}
which leads to
\begin{align}
    \label{eq:chi_C4v_Nbands}
    (-1)^{\chi_{4,i}} = \dfrac{\det \qty[U^{\mathrm{e}}(1,0)]}{\det \qty[U^{\mathrm{e}}(0,0)]} \dfrac{\det \qty[U^{\mathrm{e}}(0,\pi)]}{\det \qty[U^{\mathrm{e}}(1,\pi)]}.
\end{align}

From Eq.~\eqref{eq:u_-eta_C4v_Nbands}, we derive  
\begin{align}
    \delta_{l,n} &= \sum_m U^{\mathrm{e}}_{mn}(\bar{t}, k_z) \mel{\tilde{u}^{\mathrm{o}}_l(\bar{t}, k_z)}{\mathcal{C}_{4z}}{\tilde{u}^{\mathrm{e}}_m(\bar{t}, k_z)} \nonumber\\
    &= \sum_m U^{\mathrm{e}}_{mn}(\bar{t}, k_z) \qty[\tilde{\mathcal{B}}_{C_{4z}}(\bar{t}, k_z)]_{(\mathrm{o},l)(\mathrm{e},m)},
\end{align}  
where $(\mathrm{o},l)$ (resp.\ $(\mathrm{e},m)$) denotes the $l$-th (resp.\ $m$-th) occupied band in the mirror-odd (resp.\ mirror-even) subspace.
Defining the $N \times N$ matrix $\tilde{b}_{C_{4z}}(\bar{t}, k_z)$ as  
\begin{equation}
    \label{eq:b_C4v_def}
    \qty[\tilde{b}_{C_{4z}}(\bar{t}, k_z)]_{mn} = \qty[\tilde{\mathcal{B}}_{C_{4z}}(\bar{t}, k_z)]_{(\mathrm{o},m)(\mathrm{e},n)},
\end{equation}  
we obtain  
\begin{equation}
    I_N = \tilde{b}_{C_{4z}}(\bar{t}, k_z) U^{\mathrm{e}}(\bar{t}, k_z),
\end{equation}  
where $I_N$ is the $N \times N$ identity matrix.  
Taking the determinant of both sides, we find 
\begin{equation}
    \det\qty[U^{\mathrm{e}}(\bar{t}, k_z)] = \dfrac{1}{\det\qty[\tilde{b}_{C_{4z}}(\bar{t}, k_z)]},
\end{equation}  
which allows us to compute the right-hand side of Eq.~\eqref{eq:chi_C4v_Nbands}.
For the 2D representation of $\mathcal{C}_{4v}$ considered here, we have $\mathcal{C}_{2z} = -I$.  
This implies that the sewing matrix of the $\mathcal{C}_{4z}$ operation at the $C_{4z}T$-invariant momenta is antisymmetric.  
Furthermore, when representing $\mathcal{C}_{4z}$ in the basis of mirror eigenstates, the matrix elements between states with the same mirror eigenvalue vanish.  
Thus, when setting $\bar{k}_z \in \qty{0, \pi}$, the sewing matrix $\tilde{\mathcal{B}}_{C_{4z}}(\bar{t}, \bar{k}_z)$ can be expressed in block form with respect to the mirror sectors as
\begin{equation}
    \tilde{\mathcal{B}}_{C_{4z}}(\bar{t}, \bar{k}_z) = 
    \begin{pNiceMatrix}[first-row,first-col]
            & {\mathrm{e}}    & {\mathrm{o}}    \\[4pt]
    {\mathrm{e}}    & \text{\Large{0}}  & -\tilde{b}_{C_{4z}}^{\mathrm{T}}(\bar{t}, \bar{k}_z)   \\[6pt]
    {\mathrm{o}}   & \,\tilde{b}_{C_{4z}}(\bar{t}, \bar{k}_z)  & \text{\Large{0}}
    \end{pNiceMatrix}.
\end{equation}  
Then, it follows that  
\begin{equation}
    \Pf \qty[\tilde{\mathcal{B}}_{C_{4z}}(\bar{t}, \bar{k}_z)] = (-1)^{\frac{N(N+1)}{2}} \det \qty[\tilde{b}_{C_{4z}}(\bar{t}, \bar{k}_z)].
\end{equation}  
Consequently, we obtain
\begin{equation}
    (-1)^{\chi_{4,i}} = \dfrac{\Pf \qty[\tilde{\mathcal{B}}_{C_{4z}}(0, 0)]}{\Pf \qty[\tilde{\mathcal{B}}_{C_{4z}}(1, 0)]} \dfrac{\Pf \qty[\tilde{\mathcal{B}}_{C_{4z}}(1, \pi)]}{\Pf \qty[\tilde{\mathcal{B}}_{C_{4z}}(0, \pi)]}.
\end{equation}  
Specifically, we have
\begin{equation}
    \label{eq:chi41_Pf_GMZA_Nbands}
    (-1)^{\chi_{4,1}} = \dfrac{\Pf \qty[\tilde{\mathcal{B}}_{C_{4z}}(\mathrm{M})]}{\Pf \qty[\tilde{\mathcal{B}}_{C_{4z}}(\Gamma)]} \dfrac{\Pf \qty[\tilde{\mathcal{B}}_{C_{4z}}(\mathrm{Z})]}{\Pf \qty[\tilde{\mathcal{B}}_{C_{4z}}(\mathrm{A})]}
\end{equation}
and
\begin{equation}
    \label{eq:chi42_Pf_GMZA_Nbands}
    (-1)^{\chi_{4,2}} = \dfrac{\Pf \qty[\tilde{\mathcal{B}}_{C_{4z}}(\Gamma)]}{\Pf \qty[\tilde{\mathcal{B}}_{C_{4z}}(\mathrm{M})]} \dfrac{\Pf \qty[\tilde{\mathcal{B}}_{C_{4z}}(\mathrm{A})]}{\Pf \qty[\tilde{\mathcal{B}}_{C_{4z}}(\mathrm{Z})]},
\end{equation}
which coincides with Eqs.~\eqref{eq:chi41_Pf_GMZA} and \eqref{eq:chi42_Pf_GMZA}, respectively.

From Eqs.~\eqref{eq:nu4bar_Pf_GMZA_Nbands}, \eqref{eq:chi41_Pf_GMZA_Nbands}, and \eqref{eq:chi42_Pf_GMZA_Nbands}, we demonstrate that the relation
\begin{equation}
    (-1)^{\bar{\nu}_4} = (-1)^{\chi_{4,1}} = (-1)^{\chi_{4,2}}
\end{equation}
remains valid for systems with an arbitrary number of occupied bands.

\section{Surface states of the $\mathcal{C}_{6z}$-symmetric tight-binding model under $\mathcal{PT}$ symmetry}
\label{sec:C6TB_surf_PT}
In this Appendix, as supporting evidence that $e_2(0)$ and $e_2(\pi)$ have opposite signs in the tight-binding model in Eq.~\eqref{eq:Hamil_3D_e2bar=6_nu6bar=0}, we calculate the bulk band structure at parameters that give rise to inversion symmetry.
When we set $t_3' = t_4' = 0$, inversion symmetry, represented as $\mathcal{P} = -\Gamma_{0,0}$, emerges. 
Note that since $t_3'$ and $t_4'$ appear in the Hamiltonian multiplied by $\sin k_z$, they do not affect the $k_z = 0$ and $k_z = \pi$ planes, and thus $\mathcal{P}$ symmetry can be realized while preserving $e_2(0)$ and $e_2(\pi)$.

\begin{figure}[htp]
\includegraphics[width=8.5cm]{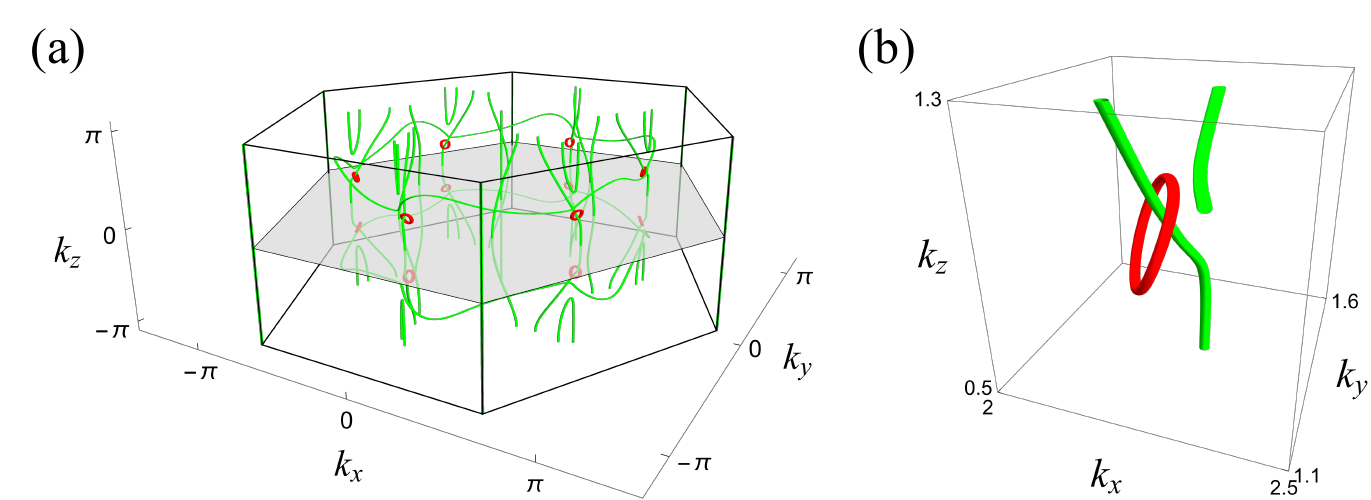}
\caption{
Linked nodal lines of the Hamiltonian in Eq.~\eqref{eq:Hamil_3D_e2bar=6_nu6bar=0} under $\mathcal{PT}$ symmetry. 
The parameters are chosen as $(M, t_1, t_2, t_3, t_4, t_5, t_6, t_7, t_{1}', t_{2}', t_{3}', t_4') = (-0.7, -0.8, -0.7, 0.6, -0.2, 0.3, -0.6, -0.8, 0.3, 0.2, 0.0, 0.0)$. 
(a) Linked nodal lines in the entire BZ. The red lines represent nodal lines between the occupied and unoccupied bands, while the green lines represent nodal lines between the occupied bands. 
The light gray plane indicates the $k_z = 0$ plane.
(b) Enlarged view of one of the nodal lines in (a).
}
\label{fig:C6_PT_linking}
\end{figure}

In the presence of $\mathcal{PT}$ symmetry, the model exhibits six groups of linking nodal lines in the region $0 < k_z < \pi$, as shown in Fig.~\ref{fig:C6_PT_linking}(a).
Moreover, each of these closed nodal lines is pierced by a nodal line between the occupied bands, as can be clearly seen in Fig.~\ref{fig:C6_PT_linking}(b).
This linking structure is known to arise when the $k_z = 0$ and $k_z = \pi$ planes have different Euler classes, and its linking number is considered to equal the difference $\bar{e}_2$~\cite{PhysRevLett.121.106403,bouhon2022multigaptopologicalconversioneuler}.
From the presence of six nodal lines together with $\abs{e_2(0)} = 4$ and $\abs{e_2(\pi)} = 2$, it follows that $e_2(0)$ and $e_2(\pi)$ must have opposite signs.

\section{Surface states of the $\mathcal{C}_{6z}$-symmetric tight-binding model with a different unit cell}
\label{sec:C6TB_surf_different_cell}
In Sec.~\ref{subsec:Euler_C6z_3D}, we show that the surface states of the tight-binding model in Eq.~\eqref{eq:Hamil_3D_e2bar=6_nu6bar=0} have six Dirac cones, as illustrated in Fig.~\ref{fig:3DTB_C6_surf}.
In this Appendix, we examine surface states for a different unit cell; see Fig.~\ref{fig:3DTB_C6_bulk}(a).

\begin{figure}[htp]
\includegraphics[width=7.5cm]{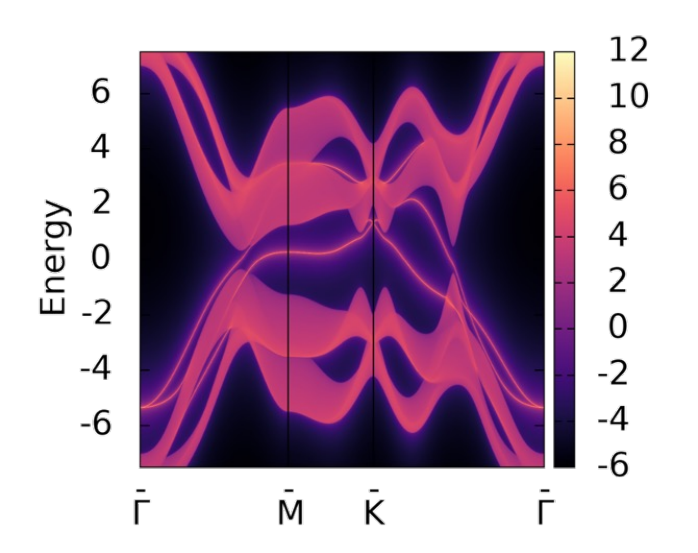}
\caption{
Surface states on the (001) surface of the Hamiltonian in Eq.~\eqref{eq:Hamil_3D_e2bar=6_nu6bar=0}, obtained using the Green's function method~\cite{PhysRevB.28.4397, MPLopezSancho_1984, MPLopezSancho_1985} implemented in WannierTools~\cite{WU2017}.
The parameters are set to be the same as those in Figs.~\ref{fig:3DTB_C6_bulk} and \ref{fig:3DTB_C6_surf}, but the unit cell is chosen differently from that in Fig.~\ref{fig:3DTB_C6_surf}.
}
\label{fig:3DTB_C6_surf_different_cell}
\end{figure}

When a different unit cell is employed, degeneracies of surface bands appear at the $\bar{\Gamma}$ and $\bar{\mathrm{K}}$ points, as shown in Fig.~\ref{fig:3DTB_C6_surf_different_cell}.  
At the $C_{6z}$-invariant $\bar{\Gamma}$ point, the quadratic band touching appears, while at the $C_{3z}$-invariant $\bar{\mathrm{K}}$ point, the band crossing with linear dispersion emerges.
These features are consistent with the symmetries of each high-symmetry point.
The occurrence of degeneracies at both the $\bar{\Gamma}$ and $\bar{\mathrm{K}}$ points is in agreement with the bulk topology characterized by $\bar{\nu}_6 = 0$.
Note that, due to $\mathcal{C}_{6z}$ symmetry, a Dirac cone also appears at $\bar{\mathrm{K}}' = (-2\pi/3, 2\pi/\sqrt{3})$, which has the same dispersion as that at $\bar{\mathrm{K}}$ and is therefore counted as a single contribution when considering the correspondence with $\bar{\nu}_6$.
In addition to the degeneracies at the $\bar{\Gamma}$ and $\bar{\mathrm{K}}$ points, Fig.~\ref{fig:3DTB_C6_surf_different_cell} shows the other band crossing along the $\bar{\Gamma}$–$\bar{\mathrm{K}}$ line.
Due to $\mathcal{C}_{6z}$ symmetry, there are six equivalent surface Dirac cones within the surface BZ, contributing $\pm 6$ to the total winding number.
Therefore, assuming that the degeneracies at the $\bar{\Gamma}$ and $(\bar{\mathrm{K}}, \bar{\mathrm{K}}')$ points carry winding numbers of opposite sign, the surface states shown in Fig.~\ref{fig:3DTB_C6_surf_different_cell} are also consistent with $\bar{e}_2 = 6$.

\section{Generalization of the relationship between $\bar{\nu}_6$ and $\chi_{6,1}$ for an arbitrary number of bands}
\label{sec:nu6_chi6_Nbands}
Following the argument in Appendix~\ref{sec:nu4_chi4_Nbands}, we generalize the relationship between $\bar{\nu}_6$ and $\chi_{6,1}$ in Eq.~\eqref{eq:relation_nu6bar_chi}, derived in Sec.~\ref{subsec:Euler_C6z_3D}, to insulating systems where the occupied bands are composed of $N$ doublets transforming under the 2D irreducible representations.
For the following calculations, we adopt the gauge choice in Eq.~\eqref{eq:real-gauge_3D}.

We first examine $\nu_6$, which is defined in Eq.~\eqref{eq:nu6_def}.
Since it is defined in the $C_{2z}T$-invariant plane, we employ the real-gauge that satisfies $\mathcal{C}_{2z}\mathcal{T} \ket{\tilde{u}_n(\bm{k})} = \ket{\tilde{u}_n(\bm{k})}$.
Even in the case with an arbitrary number of occupied doublet bands, the properties $\bm{\mathcal{A}}(\bm{k}) = \bm{0}$ and $\tilde{w}_6(\kinv) = \tilde{\mathcal{B}}_{C_{3z}}^{\mathrm{T}}(\kinv)$ remain valid, where $\kinv$ denotes the $C_{6z}T$-invariant momenta.
Noting that the relation
\begin{equation}
    \Pf \qty[-\tilde{\mathcal{B}}_{C_{3z}}^{\mathrm{A}}(\kinv)] = (-1)^N \Pf \qty[\tilde{\mathcal{B}}_{C_{3z}}^{\mathrm{A}}(\kinv)] 
\end{equation}
holds for the $2N \times 2N$ antisymmetric matrix $\tilde{\mathcal{B}}_{C_{3z}}^{\mathrm{A}}(\kinv)$, we find
\begin{equation}
    (-1)^{\nu_{6}} = \dfrac{\Pf \qty[\tilde{\mathcal{B}}_{C_{3z}}^{\mathrm{A}}(\mathrm{K})]}{\Pf \qty[\tilde{\mathcal{B}}_{C_{3z}}^{\mathrm{A}}(\Gamma)]}.
\end{equation}
Thus, we obtain
\begin{equation}
    \label{eq:nu6bar_Pf_GKAH_Nbands}
    (-1)^{\bar{\nu}_6} = \dfrac{\Pf \qty[\tilde{\mathcal{B}}_{C_{3z}}^{\mathrm{A}}(\mathrm{H})]}{\Pf \qty[\tilde{\mathcal{B}}_{C_{3z}}^{\mathrm{A}}(\mathrm{A})]} \dfrac{\Pf \qty[\tilde{\mathcal{B}}_{C_{3z}}^{\mathrm{A}}(\Gamma)]}{\Pf \qty[\tilde{\mathcal{B}}_{C_{3z}}^{\mathrm{A}}(\mathrm{K})]}.
\end{equation}
Although this expression differs from Eq.~\eqref{eq:nu6bar_Pf_GKAH}, it still allows us to obtain the relation between $\bar{\nu}_6$ and $\chi_{6,1}$.

Next, we consider $\chi_{6,1}$, which is defined in Eq.~\eqref{eq:chi_def}.
Based on our assumption, each HMP contains $N$ bands with mirror eigenvalue $+1$ and $N$ bands with mirror eigenvalue $-1$.
As the counterpart of Eq.~\eqref{eq:u_-eta_C4v_Nbands}, we here consider the state defined by
\begin{equation}
    \label{eq:u_-eta_C6v_Nbands}
    \ket{\tilde{u}^{-\eta}_n(\bar{t}, k_z)} = \dfrac{1}{\sqrt{3}} \sum_m U^\eta_{mn}(\bar{t}, k_z) \qty(\mathcal{C}_{3z} - \mathcal{C}_{3z}^{-1}) \ket{\tilde{u}^\eta_m(\bar{t}, k_z)},
\end{equation}
where $\bar{t} \in \qty{0,1}$ and $\ket{\tilde{u}^\eta_m(\bar{t}, k_z)}$ is a state in the occupied subspace with mirror eigenvalue $\eta$.
This state also belongs to the occupied subspace and has mirror eigenvalue $-\eta$.
Here, $U^\eta(\bar{t}, k_z)$ is an $N \times N$ unitary matrix that must satisfy $(U^\eta(\bar{t}, k_z))^* = U^\eta(\bar{t}, -k_z)$ to ensure that the gauge choice in Eq.~\eqref{eq:real-gauge_3D} is preserved.  
It follows from Eq.~\eqref{eq:u_-eta_C6v_Nbands} that Eq.~\eqref{eq:Berry_connection_mirror_Nbands} holds in this case as well, which leads to
\begin{align}
    \label{eq:chi_C6v_Nbands}
    (-1)^{\chi_{6,1}} = \dfrac{\det \qty[U^{\mathrm{e}}(1,0)]}{\det \qty[U^{\mathrm{e}}(0,0)]} \dfrac{\det \qty[U^{\mathrm{e}}(0,\pi)]}{\det \qty[U^{\mathrm{e}}(1,\pi)]}.
\end{align}

Referring to Eq.~\eqref{eq:b_C4v_def}, we define the $N \times N$ matrix $\tilde{b}_{C_{3z}}(\bar{t}, k_z)$ via
\begin{equation}
    \qty[\tilde{b}_{C_{3z}}(\bar{t}, k_z)]_{mn} = \qty[(\tilde{\mathcal{B}}_{C_{3z}}(\bar{t}, k_z) - \tilde{\mathcal{B}}_{C_{3z}}^\dagger(\bar{t}, k_z))/\sqrt{3}]_{(\mathrm{o},m)(\mathrm{e},n)}.
\end{equation}
It can be confirmed that $\tilde{b}_{C_{3z}}(\bar{t}, k_z)$ serves as the inverse of $U(\bar{t}, k_z)$.
At $k_z = \bar{k}_z \in \{0, \pi\}$, the matrix $(\tilde{\mathcal{B}}_{C_{3z}}(\bar{t}, \bar{k}_z) - \tilde{\mathcal{B}}_{C_{3z}}^\dagger(\bar{t}, \bar{k}_z))/\sqrt{3}$ becomes the antisymmetric matrix $\tilde{\mathcal{B}}_{C_{3z}}^{\mathrm{A}}(\bar{t}, \bar{k}_z)$ and it can be expressed as
\begin{equation}
    \Pf \qty[\tilde{\mathcal{B}}_{C_{3z}}^{\mathrm{A}}(\bar{t}, \bar{k}_z)] = \;
    \begin{pNiceMatrix}[first-row,first-col]
            & \mathrm{e}    & \mathrm{o}    \\[4pt]
    \mathrm{e}    & \text{\Large{0}}  & -\tilde{b}_{C_{3z}}^{\mathrm{T}}(\bar{t}, \bar{k}_z)   \\[6pt]
    \mathrm{o}   & \,\tilde{b}_{C_{3z}}(\bar{t}, \bar{k}_z)  & \text{\Large{0}}
    \end{pNiceMatrix}.
\end{equation}
Thus, we obtain
\begin{align}
    \Pf \qty[\tilde{\mathcal{B}}_{C_{3z}}^{\mathrm{A}}(\bar{t}, \bar{k}_z)] &= (-1)^{\frac{N(N+1)}{2}} \det \qty[\tilde{b}_{C_{3z}}(\bar{t}, \bar{k}_z)] \nonumber\\
    &= (-1)^{\frac{N(N+1)}{2}} / \det \qty[U^{\mathrm{e}}(\bar{t}, \bar{k}_z)],
\end{align}
which results in
\begin{align}
    \label{eq:chi61_Pf_GKAH_Nbands}
    (-1)^{\chi_{6,1}} &= \dfrac{(-1)^{\frac{N(N+1)}{2}} \Pf \qty[\tilde{\mathcal{B}}_{C_{3z}}^{\mathrm{A}}(0, 0)]}{(-1)^{\frac{N(N+1)}{2}} \Pf \qty[\tilde{\mathcal{B}}_{C_{3z}}^{\mathrm{A}}(1, 0)]} \dfrac{(-1)^{\frac{N(N+1)}{2}} \Pf \qty[\tilde{\mathcal{B}}_{C_{3z}}^{\mathrm{A}}(1, \pi)]}{(-1)^{\frac{N(N+1)}{2}} \Pf \qty[\tilde{\mathcal{B}}_{C_{3z}}^{\mathrm{A}}(0, \pi)]} \nonumber\\
    &= \dfrac{\Pf \qty[\tilde{\mathcal{B}}_{C_{3z}}^{\mathrm{A}}(\Gamma)]}{\Pf \qty[\tilde{\mathcal{B}}_{C_{3z}}^{\mathrm{A}}(\mathrm{K})]} \dfrac{\Pf \qty[\tilde{\mathcal{B}}_{C_{3z}}^{\mathrm{A}}(\mathrm{H})]}{\Pf \qty[\tilde{\mathcal{B}}_{C_{3z}}^{\mathrm{A}}(\mathrm{A})]}.
\end{align}

From Eqs.~\eqref{eq:nu6bar_Pf_GKAH_Nbands} and \eqref{eq:chi61_Pf_GKAH_Nbands}, we demonstrate that the relation
\begin{equation}
    (-1)^{\bar{\nu}_6} = (-1)^{\chi_{6,1}}
\end{equation}
remains valid for systems with an arbitrary number of occupied bands.


\begin{thebibliography}{79}%
\makeatletter
\providecommand \@ifxundefined [1]{%
 \@ifx{#1\undefined}
}%
\providecommand \@ifnum [1]{%
 \ifnum #1\expandafter \@firstoftwo
 \else \expandafter \@secondoftwo
 \fi
}%
\providecommand \@ifx [1]{%
 \ifx #1\expandafter \@firstoftwo
 \else \expandafter \@secondoftwo
 \fi
}%
\providecommand \natexlab [1]{#1}%
\providecommand \enquote  [1]{``#1''}%
\providecommand \bibnamefont  [1]{#1}%
\providecommand \bibfnamefont [1]{#1}%
\providecommand \citenamefont [1]{#1}%
\providecommand \href@noop [0]{\@secondoftwo}%
\providecommand \href [0]{\begingroup \@sanitize@url \@href}%
\providecommand \@href[1]{\@@startlink{#1}\@@href}%
\providecommand \@@href[1]{\endgroup#1\@@endlink}%
\providecommand \@sanitize@url [0]{\catcode `\\12\catcode `\$12\catcode `\&12\catcode `\#12\catcode `\^12\catcode `\_12\catcode `\%12\relax}%
\providecommand \@@startlink[1]{}%
\providecommand \@@endlink[0]{}%
\providecommand \url  [0]{\begingroup\@sanitize@url \@url }%
\providecommand \@url [1]{\endgroup\@href {#1}{\urlprefix }}%
\providecommand \urlprefix  [0]{URL }%
\providecommand \Eprint [0]{\href }%
\providecommand \doibase [0]{https://doi.org/}%
\providecommand \selectlanguage [0]{\@gobble}%
\providecommand \bibinfo  [0]{\@secondoftwo}%
\providecommand \bibfield  [0]{\@secondoftwo}%
\providecommand \translation [1]{[#1]}%
\providecommand \BibitemOpen [0]{}%
\providecommand \bibitemStop [0]{}%
\providecommand \bibitemNoStop [0]{.\EOS\space}%
\providecommand \EOS [0]{\spacefactor3000\relax}%
\providecommand \BibitemShut  [1]{\csname bibitem#1\endcsname}%
\let\auto@bib@innerbib\@empty
\bibitem [{\citenamefont {Qi}\ and\ \citenamefont {Zhang}(2011)}]{RevModPhys.83.1057}%
  \BibitemOpen
  \bibfield  {author} {\bibinfo {author} {\bibfnamefont {X.-L.}\ \bibnamefont {Qi}}\ and\ \bibinfo {author} {\bibfnamefont {S.-C.}\ \bibnamefont {Zhang}},\ }\bibfield  {title} {\bibinfo {title} {Topological insulators and superconductors},\ }\href {https://doi.org/10.1103/RevModPhys.83.1057} {\bibfield  {journal} {\bibinfo  {journal} {Rev. Mod. Phys.}\ }\textbf {\bibinfo {volume} {83}},\ \bibinfo {pages} {1057} (\bibinfo {year} {2011})}\BibitemShut {NoStop}%
\bibitem [{\citenamefont {Hasan}\ and\ \citenamefont {Kane}(2010)}]{RevModPhys.82.3045}%
  \BibitemOpen
  \bibfield  {author} {\bibinfo {author} {\bibfnamefont {M.~Z.}\ \bibnamefont {Hasan}}\ and\ \bibinfo {author} {\bibfnamefont {C.~L.}\ \bibnamefont {Kane}},\ }\bibfield  {title} {\bibinfo {title} {{Colloquium: Topological insulators}},\ }\href {https://doi.org/10.1103/RevModPhys.82.3045} {\bibfield  {journal} {\bibinfo  {journal} {Rev. Mod. Phys.}\ }\textbf {\bibinfo {volume} {82}},\ \bibinfo {pages} {3045} (\bibinfo {year} {2010})}\BibitemShut {NoStop}%
\bibitem [{\citenamefont {Schnyder}\ \emph {et~al.}(2008)\citenamefont {Schnyder}, \citenamefont {Ryu}, \citenamefont {Furusaki},\ and\ \citenamefont {Ludwig}}]{PhysRevB.78.195125}%
  \BibitemOpen
  \bibfield  {author} {\bibinfo {author} {\bibfnamefont {A.~P.}\ \bibnamefont {Schnyder}}, \bibinfo {author} {\bibfnamefont {S.}~\bibnamefont {Ryu}}, \bibinfo {author} {\bibfnamefont {A.}~\bibnamefont {Furusaki}},\ and\ \bibinfo {author} {\bibfnamefont {A.~W.~W.}\ \bibnamefont {Ludwig}},\ }\bibfield  {title} {\bibinfo {title} {Classification of topological insulators and superconductors in three spatial dimensions},\ }\href {https://doi.org/10.1103/PhysRevB.78.195125} {\bibfield  {journal} {\bibinfo  {journal} {Phys. Rev. B}\ }\textbf {\bibinfo {volume} {78}},\ \bibinfo {pages} {195125} (\bibinfo {year} {2008})}\BibitemShut {NoStop}%
\bibitem [{\citenamefont {Kitaev}(2009)}]{10.1063/1.3149495}%
  \BibitemOpen
  \bibfield  {author} {\bibinfo {author} {\bibfnamefont {A.}~\bibnamefont {Kitaev}},\ }\bibfield  {title} {\bibinfo {title} {Periodic table for topological insulators and superconductors},\ }\href {https://doi.org/10.1063/1.3149495} {\bibfield  {journal} {\bibinfo  {journal} {AIP Conference Proceedings}\ }\textbf {\bibinfo {volume} {1134}},\ \bibinfo {pages} {22} (\bibinfo {year} {2009})},\ \Eprint {https://arxiv.org/abs/https://pubs.aip.org/aip/acp/article-pdf/1134/1/22/11584243/22\_1\_online.pdf} {https://pubs.aip.org/aip/acp/article-pdf/1134/1/22/11584243/22\_1\_online.pdf} \BibitemShut {NoStop}%
\bibitem [{\citenamefont {Ryu}\ \emph {et~al.}(2010)\citenamefont {Ryu}, \citenamefont {Schnyder}, \citenamefont {Furusaki},\ and\ \citenamefont {Ludwig}}]{Ryu_2010}%
  \BibitemOpen
  \bibfield  {author} {\bibinfo {author} {\bibfnamefont {S.}~\bibnamefont {Ryu}}, \bibinfo {author} {\bibfnamefont {A.~P.}\ \bibnamefont {Schnyder}}, \bibinfo {author} {\bibfnamefont {A.}~\bibnamefont {Furusaki}},\ and\ \bibinfo {author} {\bibfnamefont {A.~W.~W.}\ \bibnamefont {Ludwig}},\ }\bibfield  {title} {\bibinfo {title} {Topological insulators and superconductors: tenfold way and dimensional hierarchy},\ }\href {https://doi.org/10.1088/1367-2630/12/6/065010} {\bibfield  {journal} {\bibinfo  {journal} {New Journal of Physics}\ }\textbf {\bibinfo {volume} {12}},\ \bibinfo {pages} {065010} (\bibinfo {year} {2010})}\BibitemShut {NoStop}%
\bibitem [{\citenamefont {Bradlyn}\ \emph {et~al.}(2017)\citenamefont {Bradlyn}, \citenamefont {Elcoro}, \citenamefont {Cano}, \citenamefont {Vergniory}, \citenamefont {Wang}, \citenamefont {Felser}, \citenamefont {Aroyo},\ and\ \citenamefont {Bernevig}}]{Bradlyn2017}%
  \BibitemOpen
  \bibfield  {author} {\bibinfo {author} {\bibfnamefont {B.}~\bibnamefont {Bradlyn}}, \bibinfo {author} {\bibfnamefont {L.}~\bibnamefont {Elcoro}}, \bibinfo {author} {\bibfnamefont {J.}~\bibnamefont {Cano}}, \bibinfo {author} {\bibfnamefont {M.~G.}\ \bibnamefont {Vergniory}}, \bibinfo {author} {\bibfnamefont {Z.}~\bibnamefont {Wang}}, \bibinfo {author} {\bibfnamefont {C.}~\bibnamefont {Felser}}, \bibinfo {author} {\bibfnamefont {M.~I.}\ \bibnamefont {Aroyo}},\ and\ \bibinfo {author} {\bibfnamefont {B.~A.}\ \bibnamefont {Bernevig}},\ }\bibfield  {title} {\bibinfo {title} {Topological quantum chemistry},\ }\href {https://doi.org/10.1038/nature23268} {\bibfield  {journal} {\bibinfo  {journal} {Nature}\ }\textbf {\bibinfo {volume} {547}},\ \bibinfo {pages} {298} (\bibinfo {year} {2017})}\BibitemShut {NoStop}%
\bibitem [{\citenamefont {Bradlyn}\ \emph {et~al.}(2018)\citenamefont {Bradlyn}, \citenamefont {Elcoro}, \citenamefont {Vergniory}, \citenamefont {Cano}, \citenamefont {Wang}, \citenamefont {Felser}, \citenamefont {Aroyo},\ and\ \citenamefont {Bernevig}}]{PhysRevB.97.035138}%
  \BibitemOpen
  \bibfield  {author} {\bibinfo {author} {\bibfnamefont {B.}~\bibnamefont {Bradlyn}}, \bibinfo {author} {\bibfnamefont {L.}~\bibnamefont {Elcoro}}, \bibinfo {author} {\bibfnamefont {M.~G.}\ \bibnamefont {Vergniory}}, \bibinfo {author} {\bibfnamefont {J.}~\bibnamefont {Cano}}, \bibinfo {author} {\bibfnamefont {Z.}~\bibnamefont {Wang}}, \bibinfo {author} {\bibfnamefont {C.}~\bibnamefont {Felser}}, \bibinfo {author} {\bibfnamefont {M.~I.}\ \bibnamefont {Aroyo}},\ and\ \bibinfo {author} {\bibfnamefont {B.~A.}\ \bibnamefont {Bernevig}},\ }\bibfield  {title} {\bibinfo {title} {{Band connectivity for topological quantum chemistry: Band structures as a graph theory problem}},\ }\href {https://doi.org/10.1103/PhysRevB.97.035138} {\bibfield  {journal} {\bibinfo  {journal} {Phys. Rev. B}\ }\textbf {\bibinfo {volume} {97}},\ \bibinfo {pages} {035138} (\bibinfo {year} {2018})}\BibitemShut {NoStop}%
\bibitem [{\citenamefont {Cano}\ \emph {et~al.}(2018{\natexlab{a}})\citenamefont {Cano}, \citenamefont {Bradlyn}, \citenamefont {Wang}, \citenamefont {Elcoro}, \citenamefont {Vergniory}, \citenamefont {Felser}, \citenamefont {Aroyo},\ and\ \citenamefont {Bernevig}}]{PhysRevB.97.035139}%
  \BibitemOpen
  \bibfield  {author} {\bibinfo {author} {\bibfnamefont {J.}~\bibnamefont {Cano}}, \bibinfo {author} {\bibfnamefont {B.}~\bibnamefont {Bradlyn}}, \bibinfo {author} {\bibfnamefont {Z.}~\bibnamefont {Wang}}, \bibinfo {author} {\bibfnamefont {L.}~\bibnamefont {Elcoro}}, \bibinfo {author} {\bibfnamefont {M.~G.}\ \bibnamefont {Vergniory}}, \bibinfo {author} {\bibfnamefont {C.}~\bibnamefont {Felser}}, \bibinfo {author} {\bibfnamefont {M.~I.}\ \bibnamefont {Aroyo}},\ and\ \bibinfo {author} {\bibfnamefont {B.~A.}\ \bibnamefont {Bernevig}},\ }\bibfield  {title} {\bibinfo {title} {{Building blocks of topological quantum chemistry: Elementary band representations}},\ }\href {https://doi.org/10.1103/PhysRevB.97.035139} {\bibfield  {journal} {\bibinfo  {journal} {Phys. Rev. B}\ }\textbf {\bibinfo {volume} {97}},\ \bibinfo {pages} {035139} (\bibinfo {year} {2018}{\natexlab{a}})}\BibitemShut {NoStop}%
\bibitem [{\citenamefont {Cano}\ and\ \citenamefont {Bradlyn}(2021)}]{annurev:/content/journals/10.1146/annurev-conmatphys-041720-124134}%
  \BibitemOpen
  \bibfield  {author} {\bibinfo {author} {\bibfnamefont {J.}~\bibnamefont {Cano}}\ and\ \bibinfo {author} {\bibfnamefont {B.}~\bibnamefont {Bradlyn}},\ }\bibfield  {title} {\bibinfo {title} {{Band Representations and Topological Quantum Chemistry}},\ }\href {https://doi.org/https://doi.org/10.1146/annurev-conmatphys-041720-124134} {\bibfield  {journal} {\bibinfo  {journal} {Annual Review of Condensed Matter Physics}\ }\textbf {\bibinfo {volume} {12}},\ \bibinfo {pages} {225} (\bibinfo {year} {2021})}\BibitemShut {NoStop}%
\bibitem [{\citenamefont {Kruthoff}\ \emph {et~al.}(2017)\citenamefont {Kruthoff}, \citenamefont {de~Boer}, \citenamefont {van Wezel}, \citenamefont {Kane},\ and\ \citenamefont {Slager}}]{PhysRevX.7.041069}%
  \BibitemOpen
  \bibfield  {author} {\bibinfo {author} {\bibfnamefont {J.}~\bibnamefont {Kruthoff}}, \bibinfo {author} {\bibfnamefont {J.}~\bibnamefont {de~Boer}}, \bibinfo {author} {\bibfnamefont {J.}~\bibnamefont {van Wezel}}, \bibinfo {author} {\bibfnamefont {C.~L.}\ \bibnamefont {Kane}},\ and\ \bibinfo {author} {\bibfnamefont {R.-J.}\ \bibnamefont {Slager}},\ }\bibfield  {title} {\bibinfo {title} {{Topological Classification of Crystalline Insulators through Band Structure Combinatorics}},\ }\href {https://doi.org/10.1103/PhysRevX.7.041069} {\bibfield  {journal} {\bibinfo  {journal} {Phys. Rev. X}\ }\textbf {\bibinfo {volume} {7}},\ \bibinfo {pages} {041069} (\bibinfo {year} {2017})}\BibitemShut {NoStop}%
\bibitem [{\citenamefont {Po}\ \emph {et~al.}(2017)\citenamefont {Po}, \citenamefont {Vishwanath},\ and\ \citenamefont {Watanabe}}]{Po2017}%
  \BibitemOpen
  \bibfield  {author} {\bibinfo {author} {\bibfnamefont {H.~C.}\ \bibnamefont {Po}}, \bibinfo {author} {\bibfnamefont {A.}~\bibnamefont {Vishwanath}},\ and\ \bibinfo {author} {\bibfnamefont {H.}~\bibnamefont {Watanabe}},\ }\bibfield  {title} {\bibinfo {title} {Symmetry-based indicators of band topology in the 230 space groups},\ }\href {https://doi.org/10.1038/s41467-017-00133-2} {\bibfield  {journal} {\bibinfo  {journal} {Nature Communications}\ }\textbf {\bibinfo {volume} {8}},\ \bibinfo {pages} {50} (\bibinfo {year} {2017})}\BibitemShut {NoStop}%
\bibitem [{\citenamefont {Khalaf}\ \emph {et~al.}(2018)\citenamefont {Khalaf}, \citenamefont {Po}, \citenamefont {Vishwanath},\ and\ \citenamefont {Watanabe}}]{PhysRevX.8.031070}%
  \BibitemOpen
  \bibfield  {author} {\bibinfo {author} {\bibfnamefont {E.}~\bibnamefont {Khalaf}}, \bibinfo {author} {\bibfnamefont {H.~C.}\ \bibnamefont {Po}}, \bibinfo {author} {\bibfnamefont {A.}~\bibnamefont {Vishwanath}},\ and\ \bibinfo {author} {\bibfnamefont {H.}~\bibnamefont {Watanabe}},\ }\bibfield  {title} {\bibinfo {title} {{Symmetry Indicators and Anomalous Surface States of Topological Crystalline Insulators}},\ }\href {https://doi.org/10.1103/PhysRevX.8.031070} {\bibfield  {journal} {\bibinfo  {journal} {Phys. Rev. X}\ }\textbf {\bibinfo {volume} {8}},\ \bibinfo {pages} {031070} (\bibinfo {year} {2018})}\BibitemShut {NoStop}%
\bibitem [{\citenamefont {Po}(2020)}]{Po_2020}%
  \BibitemOpen
  \bibfield  {author} {\bibinfo {author} {\bibfnamefont {H.~C.}\ \bibnamefont {Po}},\ }\bibfield  {title} {\bibinfo {title} {Symmetry indicators of band topology},\ }\href {https://doi.org/10.1088/1361-648X/ab7adb} {\bibfield  {journal} {\bibinfo  {journal} {Journal of Physics: Condensed Matter}\ }\textbf {\bibinfo {volume} {32}},\ \bibinfo {pages} {263001} (\bibinfo {year} {2020})}\BibitemShut {NoStop}%
\bibitem [{\citenamefont {Tang}\ \emph {et~al.}(2019{\natexlab{a}})\citenamefont {Tang}, \citenamefont {Po}, \citenamefont {Vishwanath},\ and\ \citenamefont {Wan}}]{Tang2019}%
  \BibitemOpen
  \bibfield  {author} {\bibinfo {author} {\bibfnamefont {F.}~\bibnamefont {Tang}}, \bibinfo {author} {\bibfnamefont {H.~C.}\ \bibnamefont {Po}}, \bibinfo {author} {\bibfnamefont {A.}~\bibnamefont {Vishwanath}},\ and\ \bibinfo {author} {\bibfnamefont {X.}~\bibnamefont {Wan}},\ }\bibfield  {title} {\bibinfo {title} {Efficient topological materials discovery using symmetry indicators},\ }\href {https://doi.org/10.1038/s41567-019-0418-7} {\bibfield  {journal} {\bibinfo  {journal} {Nature Physics}\ }\textbf {\bibinfo {volume} {15}},\ \bibinfo {pages} {470} (\bibinfo {year} {2019}{\natexlab{a}})}\BibitemShut {NoStop}%
\bibitem [{\citenamefont {Zhang}\ \emph {et~al.}(2019)\citenamefont {Zhang}, \citenamefont {Jiang}, \citenamefont {Song}, \citenamefont {Huang}, \citenamefont {He}, \citenamefont {Fang}, \citenamefont {Weng},\ and\ \citenamefont {Fang}}]{Zhang2019}%
  \BibitemOpen
  \bibfield  {author} {\bibinfo {author} {\bibfnamefont {T.}~\bibnamefont {Zhang}}, \bibinfo {author} {\bibfnamefont {Y.}~\bibnamefont {Jiang}}, \bibinfo {author} {\bibfnamefont {Z.}~\bibnamefont {Song}}, \bibinfo {author} {\bibfnamefont {H.}~\bibnamefont {Huang}}, \bibinfo {author} {\bibfnamefont {Y.}~\bibnamefont {He}}, \bibinfo {author} {\bibfnamefont {Z.}~\bibnamefont {Fang}}, \bibinfo {author} {\bibfnamefont {H.}~\bibnamefont {Weng}},\ and\ \bibinfo {author} {\bibfnamefont {C.}~\bibnamefont {Fang}},\ }\bibfield  {title} {\bibinfo {title} {Catalogue of topological electronic materials},\ }\href {https://doi.org/10.1038/s41586-019-0944-6} {\bibfield  {journal} {\bibinfo  {journal} {Nature}\ }\textbf {\bibinfo {volume} {566}},\ \bibinfo {pages} {475} (\bibinfo {year} {2019})}\BibitemShut {NoStop}%
\bibitem [{\citenamefont {Vergniory}\ \emph {et~al.}(2019)\citenamefont {Vergniory}, \citenamefont {Elcoro}, \citenamefont {Felser}, \citenamefont {Regnault}, \citenamefont {Bernevig},\ and\ \citenamefont {Wang}}]{Vergniory2019}%
  \BibitemOpen
  \bibfield  {author} {\bibinfo {author} {\bibfnamefont {M.~G.}\ \bibnamefont {Vergniory}}, \bibinfo {author} {\bibfnamefont {L.}~\bibnamefont {Elcoro}}, \bibinfo {author} {\bibfnamefont {C.}~\bibnamefont {Felser}}, \bibinfo {author} {\bibfnamefont {N.}~\bibnamefont {Regnault}}, \bibinfo {author} {\bibfnamefont {B.~A.}\ \bibnamefont {Bernevig}},\ and\ \bibinfo {author} {\bibfnamefont {Z.}~\bibnamefont {Wang}},\ }\bibfield  {title} {\bibinfo {title} {A complete catalogue of high-quality topological materials},\ }\href {https://doi.org/10.1038/s41586-019-0954-4} {\bibfield  {journal} {\bibinfo  {journal} {Nature}\ }\textbf {\bibinfo {volume} {566}},\ \bibinfo {pages} {480} (\bibinfo {year} {2019})}\BibitemShut {NoStop}%
\bibitem [{\citenamefont {Tang}\ \emph {et~al.}(2019{\natexlab{b}})\citenamefont {Tang}, \citenamefont {Po}, \citenamefont {Vishwanath},\ and\ \citenamefont {Wan}}]{doi:10.1126/sciadv.aau8725}%
  \BibitemOpen
  \bibfield  {author} {\bibinfo {author} {\bibfnamefont {F.}~\bibnamefont {Tang}}, \bibinfo {author} {\bibfnamefont {H.~C.}\ \bibnamefont {Po}}, \bibinfo {author} {\bibfnamefont {A.}~\bibnamefont {Vishwanath}},\ and\ \bibinfo {author} {\bibfnamefont {X.}~\bibnamefont {Wan}},\ }\bibfield  {title} {\bibinfo {title} {Topological materials discovery by large-order symmetry indicators},\ }\href {https://doi.org/10.1126/sciadv.aau8725} {\bibfield  {journal} {\bibinfo  {journal} {Science Advances}\ }\textbf {\bibinfo {volume} {5}},\ \bibinfo {pages} {eaau8725} (\bibinfo {year} {2019}{\natexlab{b}})},\ \Eprint {https://arxiv.org/abs/https://www.science.org/doi/pdf/10.1126/sciadv.aau8725} {https://www.science.org/doi/pdf/10.1126/sciadv.aau8725} \BibitemShut {NoStop}%
\bibitem [{\citenamefont {Wang}\ \emph {et~al.}(2019)\citenamefont {Wang}, \citenamefont {Tang}, \citenamefont {Ji}, \citenamefont {Zhang}, \citenamefont {Vishwanath}, \citenamefont {Po},\ and\ \citenamefont {Wan}}]{PhysRevB.100.195108}%
  \BibitemOpen
  \bibfield  {author} {\bibinfo {author} {\bibfnamefont {D.}~\bibnamefont {Wang}}, \bibinfo {author} {\bibfnamefont {F.}~\bibnamefont {Tang}}, \bibinfo {author} {\bibfnamefont {J.}~\bibnamefont {Ji}}, \bibinfo {author} {\bibfnamefont {W.}~\bibnamefont {Zhang}}, \bibinfo {author} {\bibfnamefont {A.}~\bibnamefont {Vishwanath}}, \bibinfo {author} {\bibfnamefont {H.~C.}\ \bibnamefont {Po}},\ and\ \bibinfo {author} {\bibfnamefont {X.}~\bibnamefont {Wan}},\ }\bibfield  {title} {\bibinfo {title} {Two-dimensional topological materials discovery by symmetry-indicator method},\ }\href {https://doi.org/10.1103/PhysRevB.100.195108} {\bibfield  {journal} {\bibinfo  {journal} {Phys. Rev. B}\ }\textbf {\bibinfo {volume} {100}},\ \bibinfo {pages} {195108} (\bibinfo {year} {2019})}\BibitemShut {NoStop}%
\bibitem [{\citenamefont {Vergniory}\ \emph {et~al.}(2022)\citenamefont {Vergniory}, \citenamefont {Wieder}, \citenamefont {Elcoro}, \citenamefont {Parkin}, \citenamefont {Felser}, \citenamefont {Bernevig},\ and\ \citenamefont {Regnault}}]{doi:10.1126/science.abg9094}%
  \BibitemOpen
  \bibfield  {author} {\bibinfo {author} {\bibfnamefont {M.~G.}\ \bibnamefont {Vergniory}}, \bibinfo {author} {\bibfnamefont {B.~J.}\ \bibnamefont {Wieder}}, \bibinfo {author} {\bibfnamefont {L.}~\bibnamefont {Elcoro}}, \bibinfo {author} {\bibfnamefont {S.~S.~P.}\ \bibnamefont {Parkin}}, \bibinfo {author} {\bibfnamefont {C.}~\bibnamefont {Felser}}, \bibinfo {author} {\bibfnamefont {B.~A.}\ \bibnamefont {Bernevig}},\ and\ \bibinfo {author} {\bibfnamefont {N.}~\bibnamefont {Regnault}},\ }\bibfield  {title} {\bibinfo {title} {All topological bands of all nonmagnetic stoichiometric materials},\ }\href {https://doi.org/10.1126/science.abg9094} {\bibfield  {journal} {\bibinfo  {journal} {Science}\ }\textbf {\bibinfo {volume} {376}},\ \bibinfo {pages} {eabg9094} (\bibinfo {year} {2022})},\ \Eprint {https://arxiv.org/abs/https://www.science.org/doi/pdf/10.1126/science.abg9094} {https://www.science.org/doi/pdf/10.1126/science.abg9094} \BibitemShut {NoStop}%
\bibitem [{\citenamefont {Fu}(2011)}]{PhysRevLett.106.106802}%
  \BibitemOpen
  \bibfield  {author} {\bibinfo {author} {\bibfnamefont {L.}~\bibnamefont {Fu}},\ }\bibfield  {title} {\bibinfo {title} {{Topological Crystalline Insulators}},\ }\href {https://doi.org/10.1103/PhysRevLett.106.106802} {\bibfield  {journal} {\bibinfo  {journal} {Phys. Rev. Lett.}\ }\textbf {\bibinfo {volume} {106}},\ \bibinfo {pages} {106802} (\bibinfo {year} {2011})}\BibitemShut {NoStop}%
\bibitem [{\citenamefont {Kim}\ \emph {et~al.}(2022)\citenamefont {Kim}, \citenamefont {Wang}, \citenamefont {Yang}, \citenamefont {Teo}, \citenamefont {Rho},\ and\ \citenamefont {Zhang}}]{Kim2022}%
  \BibitemOpen
  \bibfield  {author} {\bibinfo {author} {\bibfnamefont {M.}~\bibnamefont {Kim}}, \bibinfo {author} {\bibfnamefont {Z.}~\bibnamefont {Wang}}, \bibinfo {author} {\bibfnamefont {Y.}~\bibnamefont {Yang}}, \bibinfo {author} {\bibfnamefont {H.~T.}\ \bibnamefont {Teo}}, \bibinfo {author} {\bibfnamefont {J.}~\bibnamefont {Rho}},\ and\ \bibinfo {author} {\bibfnamefont {B.}~\bibnamefont {Zhang}},\ }\bibfield  {title} {\bibinfo {title} {Three-dimensional photonic topological insulator without spin-orbit coupling},\ }\href {https://doi.org/10.1038/s41467-022-30909-0} {\bibfield  {journal} {\bibinfo  {journal} {Nature Communications}\ }\textbf {\bibinfo {volume} {13}},\ \bibinfo {pages} {3499} (\bibinfo {year} {2022})}\BibitemShut {NoStop}%
\bibitem [{\citenamefont {Alexandradinata}\ \emph {et~al.}(2020)\citenamefont {Alexandradinata}, \citenamefont {H\"oller}, \citenamefont {Wang}, \citenamefont {Cheng},\ and\ \citenamefont {Lu}}]{PhysRevB.102.115117}%
  \BibitemOpen
  \bibfield  {author} {\bibinfo {author} {\bibfnamefont {A.}~\bibnamefont {Alexandradinata}}, \bibinfo {author} {\bibfnamefont {J.}~\bibnamefont {H\"oller}}, \bibinfo {author} {\bibfnamefont {C.}~\bibnamefont {Wang}}, \bibinfo {author} {\bibfnamefont {H.}~\bibnamefont {Cheng}},\ and\ \bibinfo {author} {\bibfnamefont {L.}~\bibnamefont {Lu}},\ }\bibfield  {title} {\bibinfo {title} {Crystallographic splitting theorem for band representations and fragile topological photonic crystals},\ }\href {https://doi.org/10.1103/PhysRevB.102.115117} {\bibfield  {journal} {\bibinfo  {journal} {Phys. Rev. B}\ }\textbf {\bibinfo {volume} {102}},\ \bibinfo {pages} {115117} (\bibinfo {year} {2020})}\BibitemShut {NoStop}%
\bibitem [{\citenamefont {Song}\ \emph {et~al.}(2020)\citenamefont {Song}, \citenamefont {Elcoro}, \citenamefont {Xu}, \citenamefont {Regnault},\ and\ \citenamefont {Bernevig}}]{PhysRevX.10.031001}%
  \BibitemOpen
  \bibfield  {author} {\bibinfo {author} {\bibfnamefont {Z.-D.}\ \bibnamefont {Song}}, \bibinfo {author} {\bibfnamefont {L.}~\bibnamefont {Elcoro}}, \bibinfo {author} {\bibfnamefont {Y.-F.}\ \bibnamefont {Xu}}, \bibinfo {author} {\bibfnamefont {N.}~\bibnamefont {Regnault}},\ and\ \bibinfo {author} {\bibfnamefont {B.~A.}\ \bibnamefont {Bernevig}},\ }\bibfield  {title} {\bibinfo {title} {{Fragile Phases as Affine Monoids: Classification and Material Examples}},\ }\href {https://doi.org/10.1103/PhysRevX.10.031001} {\bibfield  {journal} {\bibinfo  {journal} {Phys. Rev. X}\ }\textbf {\bibinfo {volume} {10}},\ \bibinfo {pages} {031001} (\bibinfo {year} {2020})}\BibitemShut {NoStop}%
\bibitem [{\citenamefont {Brouwer}\ and\ \citenamefont {Dwivedi}(2023)}]{PhysRevB.108.155137}%
  \BibitemOpen
  \bibfield  {author} {\bibinfo {author} {\bibfnamefont {P.~W.}\ \bibnamefont {Brouwer}}\ and\ \bibinfo {author} {\bibfnamefont {V.}~\bibnamefont {Dwivedi}},\ }\bibfield  {title} {\bibinfo {title} {{Homotopic classification of band structures: Stable, fragile, delicate, and stable representation-protected topology}},\ }\href {https://doi.org/10.1103/PhysRevB.108.155137} {\bibfield  {journal} {\bibinfo  {journal} {Phys. Rev. B}\ }\textbf {\bibinfo {volume} {108}},\ \bibinfo {pages} {155137} (\bibinfo {year} {2023})}\BibitemShut {NoStop}%
\bibitem [{\citenamefont {Kobayashi}\ and\ \citenamefont {Furusaki}(2024)}]{PhysRevB.110.L100508}%
  \BibitemOpen
  \bibfield  {author} {\bibinfo {author} {\bibfnamefont {S.}~\bibnamefont {Kobayashi}}\ and\ \bibinfo {author} {\bibfnamefont {A.}~\bibnamefont {Furusaki}},\ }\bibfield  {title} {\bibinfo {title} {{Representation-protected topology of spin-singlet $s$-wave superconductors}},\ }\href {https://doi.org/10.1103/PhysRevB.110.L100508} {\bibfield  {journal} {\bibinfo  {journal} {Phys. Rev. B}\ }\textbf {\bibinfo {volume} {110}},\ \bibinfo {pages} {L100508} (\bibinfo {year} {2024})}\BibitemShut {NoStop}%
\bibitem [{\citenamefont {Alexandradinata}\ \emph {et~al.}(2014{\natexlab{a}})\citenamefont {Alexandradinata}, \citenamefont {Fang}, \citenamefont {Gilbert},\ and\ \citenamefont {Bernevig}}]{PhysRevLett.113.116403}%
  \BibitemOpen
  \bibfield  {author} {\bibinfo {author} {\bibfnamefont {A.}~\bibnamefont {Alexandradinata}}, \bibinfo {author} {\bibfnamefont {C.}~\bibnamefont {Fang}}, \bibinfo {author} {\bibfnamefont {M.~J.}\ \bibnamefont {Gilbert}},\ and\ \bibinfo {author} {\bibfnamefont {B.~A.}\ \bibnamefont {Bernevig}},\ }\bibfield  {title} {\bibinfo {title} {{Spin-Orbit-Free Topological Insulators without Time-Reversal Symmetry}},\ }\href {https://doi.org/10.1103/PhysRevLett.113.116403} {\bibfield  {journal} {\bibinfo  {journal} {Phys. Rev. Lett.}\ }\textbf {\bibinfo {volume} {113}},\ \bibinfo {pages} {116403} (\bibinfo {year} {2014}{\natexlab{a}})}\BibitemShut {NoStop}%
\bibitem [{\citenamefont {Kobayashi}\ and\ \citenamefont {Furusaki}(2021)}]{PhysRevB.104.195114}%
  \BibitemOpen
  \bibfield  {author} {\bibinfo {author} {\bibfnamefont {S.}~\bibnamefont {Kobayashi}}\ and\ \bibinfo {author} {\bibfnamefont {A.}~\bibnamefont {Furusaki}},\ }\bibfield  {title} {\bibinfo {title} {Fragile topological insulators protected by rotation symmetry without spin-orbit coupling},\ }\href {https://doi.org/10.1103/PhysRevB.104.195114} {\bibfield  {journal} {\bibinfo  {journal} {Phys. Rev. B}\ }\textbf {\bibinfo {volume} {104}},\ \bibinfo {pages} {195114} (\bibinfo {year} {2021})}\BibitemShut {NoStop}%
\bibitem [{\citenamefont {Ahn}\ \emph {et~al.}(2019{\natexlab{a}})\citenamefont {Ahn}, \citenamefont {Park},\ and\ \citenamefont {Yang}}]{PhysRevX.9.021013}%
  \BibitemOpen
  \bibfield  {author} {\bibinfo {author} {\bibfnamefont {J.}~\bibnamefont {Ahn}}, \bibinfo {author} {\bibfnamefont {S.}~\bibnamefont {Park}},\ and\ \bibinfo {author} {\bibfnamefont {B.-J.}\ \bibnamefont {Yang}},\ }\bibfield  {title} {\bibinfo {title} {{Failure of Nielsen-Ninomiya Theorem and Fragile Topology in Two-Dimensional Systems with Space-Time Inversion Symmetry: Application to Twisted Bilayer Graphene at Magic Angle}},\ }\href {https://doi.org/10.1103/PhysRevX.9.021013} {\bibfield  {journal} {\bibinfo  {journal} {Phys. Rev. X}\ }\textbf {\bibinfo {volume} {9}},\ \bibinfo {pages} {021013} (\bibinfo {year} {2019}{\natexlab{a}})}\BibitemShut {NoStop}%
\bibitem [{\citenamefont {Ahn}\ \emph {et~al.}(2019{\natexlab{b}})\citenamefont {Ahn}, \citenamefont {Park}, \citenamefont {Kim}, \citenamefont {Kim},\ and\ \citenamefont {Yang}}]{Ahn_2019}%
  \BibitemOpen
  \bibfield  {author} {\bibinfo {author} {\bibfnamefont {J.}~\bibnamefont {Ahn}}, \bibinfo {author} {\bibfnamefont {S.}~\bibnamefont {Park}}, \bibinfo {author} {\bibfnamefont {D.}~\bibnamefont {Kim}}, \bibinfo {author} {\bibfnamefont {Y.}~\bibnamefont {Kim}},\ and\ \bibinfo {author} {\bibfnamefont {B.-J.}\ \bibnamefont {Yang}},\ }\bibfield  {title} {\bibinfo {title} {{Stiefel–Whitney classes and topological phases in band theory}},\ }\href {https://doi.org/10.1088/1674-1056/ab4d3b} {\bibfield  {journal} {\bibinfo  {journal} {Chinese Physics B}\ }\textbf {\bibinfo {volume} {28}},\ \bibinfo {pages} {117101} (\bibinfo {year} {2019}{\natexlab{b}})}\BibitemShut {NoStop}%
\bibitem [{\citenamefont {Bouhon}\ \emph {et~al.}(2020{\natexlab{a}})\citenamefont {Bouhon}, \citenamefont {Bzdu{\ifmmode \check{s}\else \v{s}\fi{}}ek},\ and\ \citenamefont {Slager}}]{PhysRevB.102.115135}%
  \BibitemOpen
  \bibfield  {author} {\bibinfo {author} {\bibfnamefont {A.}~\bibnamefont {Bouhon}}, \bibinfo {author} {\bibfnamefont {T.}~\bibnamefont {Bzdu{\ifmmode \check{s}\else \v{s}\fi{}}ek}},\ and\ \bibinfo {author} {\bibfnamefont {R.-J.}\ \bibnamefont {Slager}},\ }\bibfield  {title} {\bibinfo {title} {Geometric approach to fragile topology beyond symmetry indicators},\ }\href {https://doi.org/10.1103/PhysRevB.102.115135} {\bibfield  {journal} {\bibinfo  {journal} {Phys. Rev. B}\ }\textbf {\bibinfo {volume} {102}},\ \bibinfo {pages} {115135} (\bibinfo {year} {2020}{\natexlab{a}})}\BibitemShut {NoStop}%
\bibitem [{\citenamefont {Hatcher}()}]{hatcherVBKT}%
  \BibitemOpen
  \bibfield  {author} {\bibinfo {author} {\bibfnamefont {A.}~\bibnamefont {Hatcher}},\ }\href@noop {} {\bibinfo {title} {Vector bundles and {$K$}-theory}},\ \bibinfo {howpublished} {\url{http://pi.math.cornell.edu/~hatcher/VBKT/VB.pdf}}\BibitemShut {NoStop}%
\bibitem [{\citenamefont {Fang}\ and\ \citenamefont {Fu}(2015)}]{PhysRevB.91.161105}%
  \BibitemOpen
  \bibfield  {author} {\bibinfo {author} {\bibfnamefont {C.}~\bibnamefont {Fang}}\ and\ \bibinfo {author} {\bibfnamefont {L.}~\bibnamefont {Fu}},\ }\bibfield  {title} {\bibinfo {title} {{New classes of three-dimensional topological crystalline insulators: Nonsymmorphic and magnetic}},\ }\href {https://doi.org/10.1103/PhysRevB.91.161105} {\bibfield  {journal} {\bibinfo  {journal} {Phys. Rev. B}\ }\textbf {\bibinfo {volume} {91}},\ \bibinfo {pages} {161105} (\bibinfo {year} {2015})}\BibitemShut {NoStop}%
\bibitem [{\citenamefont {Zhao}\ and\ \citenamefont {Lu}(2017)}]{PhysRevLett.118.056401}%
  \BibitemOpen
  \bibfield  {author} {\bibinfo {author} {\bibfnamefont {Y.~X.}\ \bibnamefont {Zhao}}\ and\ \bibinfo {author} {\bibfnamefont {Y.}~\bibnamefont {Lu}},\ }\bibfield  {title} {\bibinfo {title} {{$PT$-Symmetric Real Dirac Fermions and Semimetals}},\ }\href {https://doi.org/10.1103/PhysRevLett.118.056401} {\bibfield  {journal} {\bibinfo  {journal} {Phys. Rev. Lett.}\ }\textbf {\bibinfo {volume} {118}},\ \bibinfo {pages} {056401} (\bibinfo {year} {2017})}\BibitemShut {NoStop}%
\bibitem [{\citenamefont {Ahn}\ and\ \citenamefont {Yang}(2017)}]{PhysRevLett.118.156401}%
  \BibitemOpen
  \bibfield  {author} {\bibinfo {author} {\bibfnamefont {J.}~\bibnamefont {Ahn}}\ and\ \bibinfo {author} {\bibfnamefont {B.-J.}\ \bibnamefont {Yang}},\ }\bibfield  {title} {\bibinfo {title} {{Unconventional Topological Phase Transition in Two-Dimensional Systems with Space-Time Inversion Symmetry}},\ }\href {https://doi.org/10.1103/PhysRevLett.118.156401} {\bibfield  {journal} {\bibinfo  {journal} {Phys. Rev. Lett.}\ }\textbf {\bibinfo {volume} {118}},\ \bibinfo {pages} {156401} (\bibinfo {year} {2017})}\BibitemShut {NoStop}%
\bibitem [{\citenamefont {Wu}\ \emph {et~al.}(2019)\citenamefont {Wu}, \citenamefont {Soluyanov},\ and\ \citenamefont {Bzdušek}}]{doi:10.1126/science.aau8740}%
  \BibitemOpen
  \bibfield  {author} {\bibinfo {author} {\bibfnamefont {Q.}~\bibnamefont {Wu}}, \bibinfo {author} {\bibfnamefont {A.~A.}\ \bibnamefont {Soluyanov}},\ and\ \bibinfo {author} {\bibfnamefont {T.}~\bibnamefont {Bzdušek}},\ }\bibfield  {title} {\bibinfo {title} {{Non-Abelian band topology in noninteracting metals}},\ }\href {https://doi.org/10.1126/science.aau8740} {\bibfield  {journal} {\bibinfo  {journal} {Science}\ }\textbf {\bibinfo {volume} {365}},\ \bibinfo {pages} {1273} (\bibinfo {year} {2019})},\ \Eprint {https://arxiv.org/abs/https://www.science.org/doi/pdf/10.1126/science.aau8740} {https://www.science.org/doi/pdf/10.1126/science.aau8740} \BibitemShut {NoStop}%
\bibitem [{\citenamefont {Bouhon}\ \emph {et~al.}(2020{\natexlab{b}})\citenamefont {Bouhon}, \citenamefont {Wu}, \citenamefont {Slager}, \citenamefont {Weng}, \citenamefont {Yazyev},\ and\ \citenamefont {Bzdu{\v{s}}ek}}]{Bouhon2020}%
  \BibitemOpen
  \bibfield  {author} {\bibinfo {author} {\bibfnamefont {A.}~\bibnamefont {Bouhon}}, \bibinfo {author} {\bibfnamefont {Q.}~\bibnamefont {Wu}}, \bibinfo {author} {\bibfnamefont {R.-J.}\ \bibnamefont {Slager}}, \bibinfo {author} {\bibfnamefont {H.}~\bibnamefont {Weng}}, \bibinfo {author} {\bibfnamefont {O.~V.}\ \bibnamefont {Yazyev}},\ and\ \bibinfo {author} {\bibfnamefont {T.}~\bibnamefont {Bzdu{\v{s}}ek}},\ }\bibfield  {title} {\bibinfo {title} {{Non-Abelian reciprocal braiding of Weyl points and its manifestation in ZrTe}},\ }\href {https://doi.org/10.1038/s41567-020-0967-9} {\bibfield  {journal} {\bibinfo  {journal} {Nature Physics}\ }\textbf {\bibinfo {volume} {16}},\ \bibinfo {pages} {1137} (\bibinfo {year} {2020}{\natexlab{b}})}\BibitemShut {NoStop}%
\bibitem [{\citenamefont {K\"onye}\ \emph {et~al.}(2021)\citenamefont {K\"onye}, \citenamefont {Bouhon}, \citenamefont {Fulga}, \citenamefont {Slager}, \citenamefont {van~den Brink},\ and\ \citenamefont {Facio}}]{PhysRevResearch.3.L042017}%
  \BibitemOpen
  \bibfield  {author} {\bibinfo {author} {\bibfnamefont {V.}~\bibnamefont {K\"onye}}, \bibinfo {author} {\bibfnamefont {A.}~\bibnamefont {Bouhon}}, \bibinfo {author} {\bibfnamefont {I.~C.}\ \bibnamefont {Fulga}}, \bibinfo {author} {\bibfnamefont {R.-J.}\ \bibnamefont {Slager}}, \bibinfo {author} {\bibfnamefont {J.}~\bibnamefont {van~den Brink}},\ and\ \bibinfo {author} {\bibfnamefont {J.~I.}\ \bibnamefont {Facio}},\ }\bibfield  {title} {\bibinfo {title} {{Chirality flip of Weyl nodes and its manifestation in strained ${\mathrm{MoTe}}_{2}$}},\ }\href {https://doi.org/10.1103/PhysRevResearch.3.L042017} {\bibfield  {journal} {\bibinfo  {journal} {Phys. Rev. Res.}\ }\textbf {\bibinfo {volume} {3}},\ \bibinfo {pages} {L042017} (\bibinfo {year} {2021})}\BibitemShut {NoStop}%
\bibitem [{\citenamefont {Chen}\ \emph {et~al.}(2022)\citenamefont {Chen}, \citenamefont {Bouhon}, \citenamefont {Slager},\ and\ \citenamefont {Monserrat}}]{PhysRevB.105.L081117}%
  \BibitemOpen
  \bibfield  {author} {\bibinfo {author} {\bibfnamefont {S.}~\bibnamefont {Chen}}, \bibinfo {author} {\bibfnamefont {A.}~\bibnamefont {Bouhon}}, \bibinfo {author} {\bibfnamefont {R.-J.}\ \bibnamefont {Slager}},\ and\ \bibinfo {author} {\bibfnamefont {B.}~\bibnamefont {Monserrat}},\ }\bibfield  {title} {\bibinfo {title} {{Non-Abelian braiding of Weyl nodes via symmetry-constrained phase transitions}},\ }\href {https://doi.org/10.1103/PhysRevB.105.L081117} {\bibfield  {journal} {\bibinfo  {journal} {Phys. Rev. B}\ }\textbf {\bibinfo {volume} {105}},\ \bibinfo {pages} {L081117} (\bibinfo {year} {2022})}\BibitemShut {NoStop}%
\bibitem [{\citenamefont {Breach}\ \emph {et~al.}(2024)\citenamefont {Breach}, \citenamefont {Slager},\ and\ \citenamefont {\"Unal}}]{PhysRevLett.133.093404}%
  \BibitemOpen
  \bibfield  {author} {\bibinfo {author} {\bibfnamefont {O.}~\bibnamefont {Breach}}, \bibinfo {author} {\bibfnamefont {R.-J.}\ \bibnamefont {Slager}},\ and\ \bibinfo {author} {\bibfnamefont {F.~N.}\ \bibnamefont {\"Unal}},\ }\bibfield  {title} {\bibinfo {title} {{Interferometry of Non-Abelian Band Singularities and Euler Class Topology}},\ }\href {https://doi.org/10.1103/PhysRevLett.133.093404} {\bibfield  {journal} {\bibinfo  {journal} {Phys. Rev. Lett.}\ }\textbf {\bibinfo {volume} {133}},\ \bibinfo {pages} {093404} (\bibinfo {year} {2024})}\BibitemShut {NoStop}%
\bibitem [{\citenamefont {Lee}\ \emph {et~al.}(2025)\citenamefont {Lee}, \citenamefont {Qian},\ and\ \citenamefont {Yang}}]{lee2024eulerbandtopologyspinorbit}%
  \BibitemOpen
  \bibfield  {author} {\bibinfo {author} {\bibfnamefont {S.~H.}\ \bibnamefont {Lee}}, \bibinfo {author} {\bibfnamefont {Y.}~\bibnamefont {Qian}},\ and\ \bibinfo {author} {\bibfnamefont {B.-J.}\ \bibnamefont {Yang}},\ }\bibfield  {title} {\bibinfo {title} {Euler band topology in spin-orbit coupled magnetic systems},\ }\href {https://doi.org/10.1103/xnqg-3bgh} {\bibfield  {journal} {\bibinfo  {journal} {Phys. Rev. B}\ }\textbf {\bibinfo {volume} {111}},\ \bibinfo {pages} {245127} (\bibinfo {year} {2025})}\BibitemShut {NoStop}%
\bibitem [{\citenamefont {Park}\ \emph {et~al.}(2021{\natexlab{a}})\citenamefont {Park}, \citenamefont {Hwang}, \citenamefont {Choi},\ and\ \citenamefont {Yang}}]{Park2021}%
  \BibitemOpen
  \bibfield  {author} {\bibinfo {author} {\bibfnamefont {S.}~\bibnamefont {Park}}, \bibinfo {author} {\bibfnamefont {Y.}~\bibnamefont {Hwang}}, \bibinfo {author} {\bibfnamefont {H.~C.}\ \bibnamefont {Choi}},\ and\ \bibinfo {author} {\bibfnamefont {B.-J.}\ \bibnamefont {Yang}},\ }\bibfield  {title} {\bibinfo {title} {Topological acoustic triple point},\ }\href {https://doi.org/10.1038/s41467-021-27158-y} {\bibfield  {journal} {\bibinfo  {journal} {Nature Communications}\ }\textbf {\bibinfo {volume} {12}},\ \bibinfo {pages} {6781} (\bibinfo {year} {2021}{\natexlab{a}})}\BibitemShut {NoStop}%
\bibitem [{\citenamefont {Peng}\ \emph {et~al.}(2022{\natexlab{a}})\citenamefont {Peng}, \citenamefont {Bouhon}, \citenamefont {Monserrat},\ and\ \citenamefont {Slager}}]{Peng2022}%
  \BibitemOpen
  \bibfield  {author} {\bibinfo {author} {\bibfnamefont {B.}~\bibnamefont {Peng}}, \bibinfo {author} {\bibfnamefont {A.}~\bibnamefont {Bouhon}}, \bibinfo {author} {\bibfnamefont {B.}~\bibnamefont {Monserrat}},\ and\ \bibinfo {author} {\bibfnamefont {R.-J.}\ \bibnamefont {Slager}},\ }\bibfield  {title} {\bibinfo {title} {{Phonons as a platform for non-Abelian braiding and its manifestation in layered silicates}},\ }\href {https://doi.org/10.1038/s41467-022-28046-9} {\bibfield  {journal} {\bibinfo  {journal} {Nature Communications}\ }\textbf {\bibinfo {volume} {13}},\ \bibinfo {pages} {423} (\bibinfo {year} {2022}{\natexlab{a}})}\BibitemShut {NoStop}%
\bibitem [{\citenamefont {Peng}\ \emph {et~al.}(2022{\natexlab{b}})\citenamefont {Peng}, \citenamefont {Bouhon}, \citenamefont {Slager},\ and\ \citenamefont {Monserrat}}]{PhysRevB.105.085115}%
  \BibitemOpen
  \bibfield  {author} {\bibinfo {author} {\bibfnamefont {B.}~\bibnamefont {Peng}}, \bibinfo {author} {\bibfnamefont {A.}~\bibnamefont {Bouhon}}, \bibinfo {author} {\bibfnamefont {R.-J.}\ \bibnamefont {Slager}},\ and\ \bibinfo {author} {\bibfnamefont {B.}~\bibnamefont {Monserrat}},\ }\bibfield  {title} {\bibinfo {title} {{Multigap topology and non-Abelian braiding of phonons from first principles}},\ }\href {https://doi.org/10.1103/PhysRevB.105.085115} {\bibfield  {journal} {\bibinfo  {journal} {Phys. Rev. B}\ }\textbf {\bibinfo {volume} {105}},\ \bibinfo {pages} {085115} (\bibinfo {year} {2022}{\natexlab{b}})}\BibitemShut {NoStop}%
\bibitem [{\citenamefont {\"Unal}\ \emph {et~al.}(2020)\citenamefont {\"Unal}, \citenamefont {Bouhon},\ and\ \citenamefont {Slager}}]{PhysRevLett.125.053601}%
  \BibitemOpen
  \bibfield  {author} {\bibinfo {author} {\bibfnamefont {F.~N.}\ \bibnamefont {\"Unal}}, \bibinfo {author} {\bibfnamefont {A.}~\bibnamefont {Bouhon}},\ and\ \bibinfo {author} {\bibfnamefont {R.-J.}\ \bibnamefont {Slager}},\ }\bibfield  {title} {\bibinfo {title} {{Topological Euler Class as a Dynamical Observable in Optical Lattices}},\ }\href {https://doi.org/10.1103/PhysRevLett.125.053601} {\bibfield  {journal} {\bibinfo  {journal} {Phys. Rev. Lett.}\ }\textbf {\bibinfo {volume} {125}},\ \bibinfo {pages} {053601} (\bibinfo {year} {2020})}\BibitemShut {NoStop}%
\bibitem [{\citenamefont {Ezawa}(2021)}]{PhysRevB.103.205303}%
  \BibitemOpen
  \bibfield  {author} {\bibinfo {author} {\bibfnamefont {M.}~\bibnamefont {Ezawa}},\ }\bibfield  {title} {\bibinfo {title} {{Topological Euler insulators and their electric circuit realization}},\ }\href {https://doi.org/10.1103/PhysRevB.103.205303} {\bibfield  {journal} {\bibinfo  {journal} {Phys. Rev. B}\ }\textbf {\bibinfo {volume} {103}},\ \bibinfo {pages} {205303} (\bibinfo {year} {2021})}\BibitemShut {NoStop}%
\bibitem [{\citenamefont {Guo}\ \emph {et~al.}(2021)\citenamefont {Guo}, \citenamefont {Jiang}, \citenamefont {Zhang}, \citenamefont {Zhang}, \citenamefont {Zhang}, \citenamefont {Yang}, \citenamefont {Zhang},\ and\ \citenamefont {Chan}}]{Guo2021}%
  \BibitemOpen
  \bibfield  {author} {\bibinfo {author} {\bibfnamefont {Q.}~\bibnamefont {Guo}}, \bibinfo {author} {\bibfnamefont {T.}~\bibnamefont {Jiang}}, \bibinfo {author} {\bibfnamefont {R.-Y.}\ \bibnamefont {Zhang}}, \bibinfo {author} {\bibfnamefont {L.}~\bibnamefont {Zhang}}, \bibinfo {author} {\bibfnamefont {Z.-Q.}\ \bibnamefont {Zhang}}, \bibinfo {author} {\bibfnamefont {B.}~\bibnamefont {Yang}}, \bibinfo {author} {\bibfnamefont {S.}~\bibnamefont {Zhang}},\ and\ \bibinfo {author} {\bibfnamefont {C.~T.}\ \bibnamefont {Chan}},\ }\bibfield  {title} {\bibinfo {title} {{Experimental observation of non-Abelian topological charges and edge states}},\ }\href {https://doi.org/10.1038/s41586-021-03521-3} {\bibfield  {journal} {\bibinfo  {journal} {Nature}\ }\textbf {\bibinfo {volume} {594}},\ \bibinfo {pages} {195} (\bibinfo {year} {2021})}\BibitemShut {NoStop}%
\bibitem [{\citenamefont {Jiang}\ \emph {et~al.}(2021{\natexlab{a}})\citenamefont {Jiang}, \citenamefont {Bouhon}, \citenamefont {Lin}, \citenamefont {Zhou}, \citenamefont {Hou}, \citenamefont {Li}, \citenamefont {Slager},\ and\ \citenamefont {Jiang}}]{Jiang2021}%
  \BibitemOpen
  \bibfield  {author} {\bibinfo {author} {\bibfnamefont {B.}~\bibnamefont {Jiang}}, \bibinfo {author} {\bibfnamefont {A.}~\bibnamefont {Bouhon}}, \bibinfo {author} {\bibfnamefont {Z.-K.}\ \bibnamefont {Lin}}, \bibinfo {author} {\bibfnamefont {X.}~\bibnamefont {Zhou}}, \bibinfo {author} {\bibfnamefont {B.}~\bibnamefont {Hou}}, \bibinfo {author} {\bibfnamefont {F.}~\bibnamefont {Li}}, \bibinfo {author} {\bibfnamefont {R.-J.}\ \bibnamefont {Slager}},\ and\ \bibinfo {author} {\bibfnamefont {J.-H.}\ \bibnamefont {Jiang}},\ }\bibfield  {title} {\bibinfo {title} {{Experimental observation of non-Abelian topological acoustic semimetals and their phase transitions}},\ }\href {https://doi.org/10.1038/s41567-021-01340-x} {\bibfield  {journal} {\bibinfo  {journal} {Nature Physics}\ }\textbf {\bibinfo {volume} {17}},\ \bibinfo {pages} {1239} (\bibinfo {year} {2021}{\natexlab{a}})}\BibitemShut {NoStop}%
\bibitem [{\citenamefont {Jiang}\ \emph {et~al.}(2021{\natexlab{b}})\citenamefont {Jiang}, \citenamefont {Guo}, \citenamefont {Zhang}, \citenamefont {Zhang}, \citenamefont {Yang},\ and\ \citenamefont {Chan}}]{Jiang2021.2}%
  \BibitemOpen
  \bibfield  {author} {\bibinfo {author} {\bibfnamefont {T.}~\bibnamefont {Jiang}}, \bibinfo {author} {\bibfnamefont {Q.}~\bibnamefont {Guo}}, \bibinfo {author} {\bibfnamefont {R.-Y.}\ \bibnamefont {Zhang}}, \bibinfo {author} {\bibfnamefont {Z.-Q.}\ \bibnamefont {Zhang}}, \bibinfo {author} {\bibfnamefont {B.}~\bibnamefont {Yang}},\ and\ \bibinfo {author} {\bibfnamefont {C.~T.}\ \bibnamefont {Chan}},\ }\bibfield  {title} {\bibinfo {title} {{Four-band non-Abelian topological insulator and its experimental realization}},\ }\href {https://doi.org/10.1038/s41467-021-26763-1} {\bibfield  {journal} {\bibinfo  {journal} {Nature Communications}\ }\textbf {\bibinfo {volume} {12}},\ \bibinfo {pages} {6471} (\bibinfo {year} {2021}{\natexlab{b}})}\BibitemShut {NoStop}%
\bibitem [{\citenamefont {Park}\ \emph {et~al.}(2021{\natexlab{b}})\citenamefont {Park}, \citenamefont {Wong}, \citenamefont {Zhang},\ and\ \citenamefont {Oh}}]{Park2021.ACS.Photo}%
  \BibitemOpen
  \bibfield  {author} {\bibinfo {author} {\bibfnamefont {H.}~\bibnamefont {Park}}, \bibinfo {author} {\bibfnamefont {S.}~\bibnamefont {Wong}}, \bibinfo {author} {\bibfnamefont {X.}~\bibnamefont {Zhang}},\ and\ \bibinfo {author} {\bibfnamefont {S.~S.}\ \bibnamefont {Oh}},\ }\bibfield  {title} {\bibinfo {title} {{Non-Abelian Charged Nodal Links in a Dielectric Photonic Crystal}},\ }\href {https://doi.org/10.1021/acsphotonics.1c00876} {\bibfield  {journal} {\bibinfo  {journal} {ACS Photonics}\ }\textbf {\bibinfo {volume} {8}},\ \bibinfo {pages} {2746} (\bibinfo {year} {2021}{\natexlab{b}})}\BibitemShut {NoStop}%
\bibitem [{\citenamefont {Park}\ \emph {et~al.}(2022)\citenamefont {Park}, \citenamefont {Wong}, \citenamefont {Bouhon}, \citenamefont {Slager},\ and\ \citenamefont {Oh}}]{PhysRevB.105.214108}%
  \BibitemOpen
  \bibfield  {author} {\bibinfo {author} {\bibfnamefont {H.}~\bibnamefont {Park}}, \bibinfo {author} {\bibfnamefont {S.}~\bibnamefont {Wong}}, \bibinfo {author} {\bibfnamefont {A.}~\bibnamefont {Bouhon}}, \bibinfo {author} {\bibfnamefont {R.-J.}\ \bibnamefont {Slager}},\ and\ \bibinfo {author} {\bibfnamefont {S.~S.}\ \bibnamefont {Oh}},\ }\bibfield  {title} {\bibinfo {title} {{Topological phase transitions of non-Abelian charged nodal lines in spring-mass systems}},\ }\href {https://doi.org/10.1103/PhysRevB.105.214108} {\bibfield  {journal} {\bibinfo  {journal} {Phys. Rev. B}\ }\textbf {\bibinfo {volume} {105}},\ \bibinfo {pages} {214108} (\bibinfo {year} {2022})}\BibitemShut {NoStop}%
\bibitem [{\citenamefont {Zhao}\ \emph {et~al.}(2022)\citenamefont {Zhao}, \citenamefont {Yang}, \citenamefont {Jiang}, \citenamefont {Mao}, \citenamefont {Guo}, \citenamefont {Qiu}, \citenamefont {Wang}, \citenamefont {Yao}, \citenamefont {He}, \citenamefont {Zhou}, \citenamefont {Xu},\ and\ \citenamefont {Duan}}]{Zhao2022}%
  \BibitemOpen
  \bibfield  {author} {\bibinfo {author} {\bibfnamefont {W.}~\bibnamefont {Zhao}}, \bibinfo {author} {\bibfnamefont {Y.-B.}\ \bibnamefont {Yang}}, \bibinfo {author} {\bibfnamefont {Y.}~\bibnamefont {Jiang}}, \bibinfo {author} {\bibfnamefont {Z.}~\bibnamefont {Mao}}, \bibinfo {author} {\bibfnamefont {W.}~\bibnamefont {Guo}}, \bibinfo {author} {\bibfnamefont {L.}~\bibnamefont {Qiu}}, \bibinfo {author} {\bibfnamefont {G.}~\bibnamefont {Wang}}, \bibinfo {author} {\bibfnamefont {L.}~\bibnamefont {Yao}}, \bibinfo {author} {\bibfnamefont {L.}~\bibnamefont {He}}, \bibinfo {author} {\bibfnamefont {Z.}~\bibnamefont {Zhou}}, \bibinfo {author} {\bibfnamefont {Y.}~\bibnamefont {Xu}},\ and\ \bibinfo {author} {\bibfnamefont {L.}~\bibnamefont {Duan}},\ }\bibfield  {title} {\bibinfo {title} {{Quantum simulation for topological Euler insulators}},\ }\href {https://doi.org/10.1038/s42005-022-01001-2} {\bibfield  {journal} {\bibinfo  {journal} {Communications Physics}\ }\textbf {\bibinfo {volume} {5}},\ \bibinfo {pages} {223}
  (\bibinfo {year} {2022})}\BibitemShut {NoStop}%
\bibitem [{\citenamefont {Qiu}\ \emph {et~al.}(2023)\citenamefont {Qiu}, \citenamefont {Zhang}, \citenamefont {Liu}, \citenamefont {Fan}, \citenamefont {Zhang},\ and\ \citenamefont {Qiu}}]{Qiu2023}%
  \BibitemOpen
  \bibfield  {author} {\bibinfo {author} {\bibfnamefont {H.}~\bibnamefont {Qiu}}, \bibinfo {author} {\bibfnamefont {Q.}~\bibnamefont {Zhang}}, \bibinfo {author} {\bibfnamefont {T.}~\bibnamefont {Liu}}, \bibinfo {author} {\bibfnamefont {X.}~\bibnamefont {Fan}}, \bibinfo {author} {\bibfnamefont {F.}~\bibnamefont {Zhang}},\ and\ \bibinfo {author} {\bibfnamefont {C.}~\bibnamefont {Qiu}},\ }\bibfield  {title} {\bibinfo {title} {Minimal non-abelian nodal braiding in ideal metamaterials},\ }\href {https://doi.org/10.1038/s41467-023-36952-9} {\bibfield  {journal} {\bibinfo  {journal} {Nature Communications}\ }\textbf {\bibinfo {volume} {14}},\ \bibinfo {pages} {1261} (\bibinfo {year} {2023})}\BibitemShut {NoStop}%
\bibitem [{\citenamefont {Jiang}\ \emph {et~al.}(2024)\citenamefont {Jiang}, \citenamefont {Bouhon}, \citenamefont {Wu}, \citenamefont {Kong}, \citenamefont {Lin}, \citenamefont {Slager},\ and\ \citenamefont {Jiang}}]{JIANG20241653}%
  \BibitemOpen
  \bibfield  {author} {\bibinfo {author} {\bibfnamefont {B.}~\bibnamefont {Jiang}}, \bibinfo {author} {\bibfnamefont {A.}~\bibnamefont {Bouhon}}, \bibinfo {author} {\bibfnamefont {S.-Q.}\ \bibnamefont {Wu}}, \bibinfo {author} {\bibfnamefont {Z.-L.}\ \bibnamefont {Kong}}, \bibinfo {author} {\bibfnamefont {Z.-K.}\ \bibnamefont {Lin}}, \bibinfo {author} {\bibfnamefont {R.-J.}\ \bibnamefont {Slager}},\ and\ \bibinfo {author} {\bibfnamefont {J.-H.}\ \bibnamefont {Jiang}},\ }\bibfield  {title} {\bibinfo {title} {{Observation of an acoustic topological Euler insulator with meronic waves}},\ }\href {https://doi.org/https://doi.org/10.1016/j.scib.2024.04.009} {\bibfield  {journal} {\bibinfo  {journal} {Science Bulletin}\ }\textbf {\bibinfo {volume} {69}},\ \bibinfo {pages} {1653} (\bibinfo {year} {2024})}\BibitemShut {NoStop}%
\bibitem [{\citenamefont {Davoyan}\ \emph {et~al.}(2024)\citenamefont {Davoyan}, \citenamefont {Jankowski}, \citenamefont {Bouhon},\ and\ \citenamefont {Slager}}]{PhysRevB.109.165125}%
  \BibitemOpen
  \bibfield  {author} {\bibinfo {author} {\bibfnamefont {Z.}~\bibnamefont {Davoyan}}, \bibinfo {author} {\bibfnamefont {W.~J.}\ \bibnamefont {Jankowski}}, \bibinfo {author} {\bibfnamefont {A.}~\bibnamefont {Bouhon}},\ and\ \bibinfo {author} {\bibfnamefont {R.-J.}\ \bibnamefont {Slager}},\ }\bibfield  {title} {\bibinfo {title} {{Three-dimensional $\mathcal{PT}$-symmetric topological phases with a Pontryagin index}},\ }\href {https://doi.org/10.1103/PhysRevB.109.165125} {\bibfield  {journal} {\bibinfo  {journal} {Phys. Rev. B}\ }\textbf {\bibinfo {volume} {109}},\ \bibinfo {pages} {165125} (\bibinfo {year} {2024})}\BibitemShut {NoStop}%
\bibitem [{\citenamefont {Yang}\ \emph {et~al.}(2024)\citenamefont {Yang}, \citenamefont {Yang}, \citenamefont {Ma}, \citenamefont {Li}, \citenamefont {Zhang},\ and\ \citenamefont {Chan}}]{doi:10.1126/science.adf9621}%
  \BibitemOpen
  \bibfield  {author} {\bibinfo {author} {\bibfnamefont {Y.}~\bibnamefont {Yang}}, \bibinfo {author} {\bibfnamefont {B.}~\bibnamefont {Yang}}, \bibinfo {author} {\bibfnamefont {G.}~\bibnamefont {Ma}}, \bibinfo {author} {\bibfnamefont {J.}~\bibnamefont {Li}}, \bibinfo {author} {\bibfnamefont {S.}~\bibnamefont {Zhang}},\ and\ \bibinfo {author} {\bibfnamefont {C.~T.}\ \bibnamefont {Chan}},\ }\bibfield  {title} {\bibinfo {title} {{Non-Abelian physics in light and sound}},\ }\href {https://doi.org/10.1126/science.adf9621} {\bibfield  {journal} {\bibinfo  {journal} {Science}\ }\textbf {\bibinfo {volume} {383}},\ \bibinfo {pages} {eadf9621} (\bibinfo {year} {2024})},\ \Eprint {https://arxiv.org/abs/https://www.science.org/doi/pdf/10.1126/science.adf9621} {https://www.science.org/doi/pdf/10.1126/science.adf9621} \BibitemShut {NoStop}%
\bibitem [{\citenamefont {Liu}\ \emph {et~al.}(2025)\citenamefont {Liu}, \citenamefont {Wang}, \citenamefont {Yang},\ and\ \citenamefont {Zhang}}]{doi:10.1126/sciadv.ads5081}%
  \BibitemOpen
  \bibfield  {author} {\bibinfo {author} {\bibfnamefont {W.}~\bibnamefont {Liu}}, \bibinfo {author} {\bibfnamefont {H.}~\bibnamefont {Wang}}, \bibinfo {author} {\bibfnamefont {B.}~\bibnamefont {Yang}},\ and\ \bibinfo {author} {\bibfnamefont {S.}~\bibnamefont {Zhang}},\ }\bibfield  {title} {\bibinfo {title} {{Correspondence between Euler charges and nodal-line topology in Euler semimetals}},\ }\href {https://doi.org/10.1126/sciadv.ads5081} {\bibfield  {journal} {\bibinfo  {journal} {Science Advances}\ }\textbf {\bibinfo {volume} {11}},\ \bibinfo {pages} {eads5081} (\bibinfo {year} {2025})},\ \Eprint {https://arxiv.org/abs/https://www.science.org/doi/pdf/10.1126/sciadv.ads5081} {https://www.science.org/doi/pdf/10.1126/sciadv.ads5081} \BibitemShut {NoStop}%
\bibitem [{\citenamefont {Karle}\ \emph {et~al.}(2024)\citenamefont {Karle}, \citenamefont {Lemeshko}, \citenamefont {Bouhon}, \citenamefont {Slager},\ and\ \citenamefont {Ünal}}]{karle2024anomalousmultigaptopologicalphases}%
  \BibitemOpen
  \bibfield  {author} {\bibinfo {author} {\bibfnamefont {V.}~\bibnamefont {Karle}}, \bibinfo {author} {\bibfnamefont {M.}~\bibnamefont {Lemeshko}}, \bibinfo {author} {\bibfnamefont {A.}~\bibnamefont {Bouhon}}, \bibinfo {author} {\bibfnamefont {R.-J.}\ \bibnamefont {Slager}},\ and\ \bibinfo {author} {\bibfnamefont {F.~N.}\ \bibnamefont {Ünal}},\ }\href {https://arxiv.org/abs/2408.16848} {\bibinfo {title} {Anomalous multi-gap topological phases in periodically driven quantum rotors}} (\bibinfo {year} {2024}),\ \Eprint {https://arxiv.org/abs/2408.16848} {arXiv:2408.16848 [quant-ph]} \BibitemShut {NoStop}%
\bibitem [{\citenamefont {Alexandradinata}\ and\ \citenamefont {Bernevig}(2016)}]{PhysRevB.93.205104}%
  \BibitemOpen
  \bibfield  {author} {\bibinfo {author} {\bibfnamefont {A.}~\bibnamefont {Alexandradinata}}\ and\ \bibinfo {author} {\bibfnamefont {B.~A.}\ \bibnamefont {Bernevig}},\ }\bibfield  {title} {\bibinfo {title} {Berry-phase description of topological crystalline insulators},\ }\href {https://doi.org/10.1103/PhysRevB.93.205104} {\bibfield  {journal} {\bibinfo  {journal} {Phys. Rev. B}\ }\textbf {\bibinfo {volume} {93}},\ \bibinfo {pages} {205104} (\bibinfo {year} {2016})}\BibitemShut {NoStop}%
\bibitem [{\citenamefont {Ahn}\ \emph {et~al.}(2018)\citenamefont {Ahn}, \citenamefont {Kim}, \citenamefont {Kim},\ and\ \citenamefont {Yang}}]{PhysRevLett.121.106403}%
  \BibitemOpen
  \bibfield  {author} {\bibinfo {author} {\bibfnamefont {J.}~\bibnamefont {Ahn}}, \bibinfo {author} {\bibfnamefont {D.}~\bibnamefont {Kim}}, \bibinfo {author} {\bibfnamefont {Y.}~\bibnamefont {Kim}},\ and\ \bibinfo {author} {\bibfnamefont {B.-J.}\ \bibnamefont {Yang}},\ }\bibfield  {title} {\bibinfo {title} {{Band Topology and Linking Structure of Nodal Line Semimetals with ${Z}_{2}$ Monopole Charges}},\ }\href {https://doi.org/10.1103/PhysRevLett.121.106403} {\bibfield  {journal} {\bibinfo  {journal} {Phys. Rev. Lett.}\ }\textbf {\bibinfo {volume} {121}},\ \bibinfo {pages} {106403} (\bibinfo {year} {2018})}\BibitemShut {NoStop}%
\bibitem [{\citenamefont {Bouhon}\ \emph {et~al.}(2019)\citenamefont {Bouhon}, \citenamefont {Black-Schaffer},\ and\ \citenamefont {Slager}}]{PhysRevB.100.195135}%
  \BibitemOpen
  \bibfield  {author} {\bibinfo {author} {\bibfnamefont {A.}~\bibnamefont {Bouhon}}, \bibinfo {author} {\bibfnamefont {A.~M.}\ \bibnamefont {Black-Schaffer}},\ and\ \bibinfo {author} {\bibfnamefont {R.-J.}\ \bibnamefont {Slager}},\ }\bibfield  {title} {\bibinfo {title} {Wilson loop approach to fragile topology of split elementary band representations and topological crystalline insulators with time-reversal symmetry},\ }\href {https://doi.org/10.1103/PhysRevB.100.195135} {\bibfield  {journal} {\bibinfo  {journal} {Phys. Rev. B}\ }\textbf {\bibinfo {volume} {100}},\ \bibinfo {pages} {195135} (\bibinfo {year} {2019})}\BibitemShut {NoStop}%
\bibitem [{\citenamefont {Yu}\ \emph {et~al.}(2011)\citenamefont {Yu}, \citenamefont {Qi}, \citenamefont {Bernevig}, \citenamefont {Fang},\ and\ \citenamefont {Dai}}]{PhysRevB.84.075119}%
  \BibitemOpen
  \bibfield  {author} {\bibinfo {author} {\bibfnamefont {R.}~\bibnamefont {Yu}}, \bibinfo {author} {\bibfnamefont {X.~L.}\ \bibnamefont {Qi}}, \bibinfo {author} {\bibfnamefont {A.}~\bibnamefont {Bernevig}}, \bibinfo {author} {\bibfnamefont {Z.}~\bibnamefont {Fang}},\ and\ \bibinfo {author} {\bibfnamefont {X.}~\bibnamefont {Dai}},\ }\bibfield  {title} {\bibinfo {title} {{Equivalent expression of ${\mathbb{Z}}_{2}$ topological invariant for band insulators using the non-Abelian Berry connection}},\ }\href {https://doi.org/10.1103/PhysRevB.84.075119} {\bibfield  {journal} {\bibinfo  {journal} {Phys. Rev. B}\ }\textbf {\bibinfo {volume} {84}},\ \bibinfo {pages} {075119} (\bibinfo {year} {2011})}\BibitemShut {NoStop}%
\bibitem [{\citenamefont {Alexandradinata}\ \emph {et~al.}(2014{\natexlab{b}})\citenamefont {Alexandradinata}, \citenamefont {Dai},\ and\ \citenamefont {Bernevig}}]{PhysRevB.89.155114}%
  \BibitemOpen
  \bibfield  {author} {\bibinfo {author} {\bibfnamefont {A.}~\bibnamefont {Alexandradinata}}, \bibinfo {author} {\bibfnamefont {X.}~\bibnamefont {Dai}},\ and\ \bibinfo {author} {\bibfnamefont {B.~A.}\ \bibnamefont {Bernevig}},\ }\bibfield  {title} {\bibinfo {title} {Wilson-loop characterization of inversion-symmetric topological insulators},\ }\href {https://doi.org/10.1103/PhysRevB.89.155114} {\bibfield  {journal} {\bibinfo  {journal} {Phys. Rev. B}\ }\textbf {\bibinfo {volume} {89}},\ \bibinfo {pages} {155114} (\bibinfo {year} {2014}{\natexlab{b}})}\BibitemShut {NoStop}%
\bibitem [{\citenamefont {Alexandradinata}\ \emph {et~al.}(2016)\citenamefont {Alexandradinata}, \citenamefont {Wang},\ and\ \citenamefont {Bernevig}}]{PhysRevX.6.021008}%
  \BibitemOpen
  \bibfield  {author} {\bibinfo {author} {\bibfnamefont {A.}~\bibnamefont {Alexandradinata}}, \bibinfo {author} {\bibfnamefont {Z.}~\bibnamefont {Wang}},\ and\ \bibinfo {author} {\bibfnamefont {B.~A.}\ \bibnamefont {Bernevig}},\ }\bibfield  {title} {\bibinfo {title} {{Topological Insulators from Group Cohomology}},\ }\href {https://doi.org/10.1103/PhysRevX.6.021008} {\bibfield  {journal} {\bibinfo  {journal} {Phys. Rev. X}\ }\textbf {\bibinfo {volume} {6}},\ \bibinfo {pages} {021008} (\bibinfo {year} {2016})}\BibitemShut {NoStop}%
\bibitem [{\citenamefont {Cano}\ \emph {et~al.}(2018{\natexlab{b}})\citenamefont {Cano}, \citenamefont {Bradlyn}, \citenamefont {Wang}, \citenamefont {Elcoro}, \citenamefont {Vergniory}, \citenamefont {Felser}, \citenamefont {Aroyo},\ and\ \citenamefont {Bernevig}}]{PhysRevLett.120.266401}%
  \BibitemOpen
  \bibfield  {author} {\bibinfo {author} {\bibfnamefont {J.}~\bibnamefont {Cano}}, \bibinfo {author} {\bibfnamefont {B.}~\bibnamefont {Bradlyn}}, \bibinfo {author} {\bibfnamefont {Z.}~\bibnamefont {Wang}}, \bibinfo {author} {\bibfnamefont {L.}~\bibnamefont {Elcoro}}, \bibinfo {author} {\bibfnamefont {M.~G.}\ \bibnamefont {Vergniory}}, \bibinfo {author} {\bibfnamefont {C.}~\bibnamefont {Felser}}, \bibinfo {author} {\bibfnamefont {M.~I.}\ \bibnamefont {Aroyo}},\ and\ \bibinfo {author} {\bibfnamefont {B.~A.}\ \bibnamefont {Bernevig}},\ }\bibfield  {title} {\bibinfo {title} {{Topology of Disconnected Elementary Band Representations}},\ }\href {https://doi.org/10.1103/PhysRevLett.120.266401} {\bibfield  {journal} {\bibinfo  {journal} {Phys. Rev. Lett.}\ }\textbf {\bibinfo {volume} {120}},\ \bibinfo {pages} {266401} (\bibinfo {year} {2018}{\natexlab{b}})}\BibitemShut {NoStop}%
\bibitem [{\citenamefont {Bradlyn}\ \emph {et~al.}(2019)\citenamefont {Bradlyn}, \citenamefont {Wang}, \citenamefont {Cano},\ and\ \citenamefont {Bernevig}}]{PhysRevB.99.045140}%
  \BibitemOpen
  \bibfield  {author} {\bibinfo {author} {\bibfnamefont {B.}~\bibnamefont {Bradlyn}}, \bibinfo {author} {\bibfnamefont {Z.}~\bibnamefont {Wang}}, \bibinfo {author} {\bibfnamefont {J.}~\bibnamefont {Cano}},\ and\ \bibinfo {author} {\bibfnamefont {B.~A.}\ \bibnamefont {Bernevig}},\ }\bibfield  {title} {\bibinfo {title} {{Disconnected elementary band representations, fragile topology, and Wilson loops as topological indices: An example on the triangular lattice}},\ }\href {https://doi.org/10.1103/PhysRevB.99.045140} {\bibfield  {journal} {\bibinfo  {journal} {Phys. Rev. B}\ }\textbf {\bibinfo {volume} {99}},\ \bibinfo {pages} {045140} (\bibinfo {year} {2019})}\BibitemShut {NoStop}%
\bibitem [{\citenamefont {Fang}\ \emph {et~al.}(2012)\citenamefont {Fang}, \citenamefont {Gilbert},\ and\ \citenamefont {Bernevig}}]{PhysRevB.86.115112}%
  \BibitemOpen
  \bibfield  {author} {\bibinfo {author} {\bibfnamefont {C.}~\bibnamefont {Fang}}, \bibinfo {author} {\bibfnamefont {M.~J.}\ \bibnamefont {Gilbert}},\ and\ \bibinfo {author} {\bibfnamefont {B.~A.}\ \bibnamefont {Bernevig}},\ }\bibfield  {title} {\bibinfo {title} {Bulk topological invariants in noninteracting point group symmetric insulators},\ }\href {https://doi.org/10.1103/PhysRevB.86.115112} {\bibfield  {journal} {\bibinfo  {journal} {Phys. Rev. B}\ }\textbf {\bibinfo {volume} {86}},\ \bibinfo {pages} {115112} (\bibinfo {year} {2012})}\BibitemShut {NoStop}%
\bibitem [{\citenamefont {Ahn}\ and\ \citenamefont {Yang}(2019)}]{PhysRevB.99.235125}%
  \BibitemOpen
  \bibfield  {author} {\bibinfo {author} {\bibfnamefont {J.}~\bibnamefont {Ahn}}\ and\ \bibinfo {author} {\bibfnamefont {B.-J.}\ \bibnamefont {Yang}},\ }\bibfield  {title} {\bibinfo {title} {{Symmetry representation approach to topological invariants in ${C}_{2z}T$-symmetric systems}},\ }\href {https://doi.org/10.1103/PhysRevB.99.235125} {\bibfield  {journal} {\bibinfo  {journal} {Phys. Rev. B}\ }\textbf {\bibinfo {volume} {99}},\ \bibinfo {pages} {235125} (\bibinfo {year} {2019})}\BibitemShut {NoStop}%
\bibitem [{\citenamefont {Song}\ \emph {et~al.}(2019)\citenamefont {Song}, \citenamefont {Wang}, \citenamefont {Shi}, \citenamefont {Li}, \citenamefont {Fang},\ and\ \citenamefont {Bernevig}}]{PhysRevLett.123.036401}%
  \BibitemOpen
  \bibfield  {author} {\bibinfo {author} {\bibfnamefont {Z.}~\bibnamefont {Song}}, \bibinfo {author} {\bibfnamefont {Z.}~\bibnamefont {Wang}}, \bibinfo {author} {\bibfnamefont {W.}~\bibnamefont {Shi}}, \bibinfo {author} {\bibfnamefont {G.}~\bibnamefont {Li}}, \bibinfo {author} {\bibfnamefont {C.}~\bibnamefont {Fang}},\ and\ \bibinfo {author} {\bibfnamefont {B.~A.}\ \bibnamefont {Bernevig}},\ }\bibfield  {title} {\bibinfo {title} {{All Magic Angles in Twisted Bilayer Graphene are Topological}},\ }\href {https://doi.org/10.1103/PhysRevLett.123.036401} {\bibfield  {journal} {\bibinfo  {journal} {Phys. Rev. Lett.}\ }\textbf {\bibinfo {volume} {123}},\ \bibinfo {pages} {036401} (\bibinfo {year} {2019})}\BibitemShut {NoStop}%
\bibitem [{\citenamefont {Shiozaki}\ and\ \citenamefont {Sato}(2014)}]{PhysRevB.90.165114}%
  \BibitemOpen
  \bibfield  {author} {\bibinfo {author} {\bibfnamefont {K.}~\bibnamefont {Shiozaki}}\ and\ \bibinfo {author} {\bibfnamefont {M.}~\bibnamefont {Sato}},\ }\bibfield  {title} {\bibinfo {title} {Topology of crystalline insulators and superconductors},\ }\href {https://doi.org/10.1103/PhysRevB.90.165114} {\bibfield  {journal} {\bibinfo  {journal} {Phys. Rev. B}\ }\textbf {\bibinfo {volume} {90}},\ \bibinfo {pages} {165114} (\bibinfo {year} {2014})}\BibitemShut {NoStop}%
\bibitem [{\citenamefont {Bouhon}\ and\ \citenamefont {Slager}(2022)}]{bouhon2022multigaptopologicalconversioneuler}%
  \BibitemOpen
  \bibfield  {author} {\bibinfo {author} {\bibfnamefont {A.}~\bibnamefont {Bouhon}}\ and\ \bibinfo {author} {\bibfnamefont {R.-J.}\ \bibnamefont {Slager}},\ }\href {https://arxiv.org/abs/2203.16741} {\bibinfo {title} {{Multi-gap topological conversion of Euler class via band-node braiding: minimal models, $PT$-linked nodal rings, and chiral heirs}}} (\bibinfo {year} {2022}),\ \Eprint {https://arxiv.org/abs/2203.16741} {arXiv:2203.16741 [cond-mat.mes-hall]} \BibitemShut {NoStop}%
\bibitem [{\citenamefont {Guinea}\ \emph {et~al.}(1983)\citenamefont {Guinea}, \citenamefont {Tejedor}, \citenamefont {Flores},\ and\ \citenamefont {Louis}}]{PhysRevB.28.4397}%
  \BibitemOpen
  \bibfield  {author} {\bibinfo {author} {\bibfnamefont {F.}~\bibnamefont {Guinea}}, \bibinfo {author} {\bibfnamefont {C.}~\bibnamefont {Tejedor}}, \bibinfo {author} {\bibfnamefont {F.}~\bibnamefont {Flores}},\ and\ \bibinfo {author} {\bibfnamefont {E.}~\bibnamefont {Louis}},\ }\bibfield  {title} {\bibinfo {title} {{Effective two-dimensional Hamiltonian at surfaces}},\ }\href {https://doi.org/10.1103/PhysRevB.28.4397} {\bibfield  {journal} {\bibinfo  {journal} {Phys. Rev. B}\ }\textbf {\bibinfo {volume} {28}},\ \bibinfo {pages} {4397} (\bibinfo {year} {1983})}\BibitemShut {NoStop}%
\bibitem [{\citenamefont {Sancho}\ \emph {et~al.}(1984)\citenamefont {Sancho}, \citenamefont {Sancho},\ and\ \citenamefont {Rubio}}]{MPLopezSancho_1984}%
  \BibitemOpen
  \bibfield  {author} {\bibinfo {author} {\bibfnamefont {M.~P.~L.}\ \bibnamefont {Sancho}}, \bibinfo {author} {\bibfnamefont {J.~M.~L.}\ \bibnamefont {Sancho}},\ and\ \bibinfo {author} {\bibfnamefont {J.}~\bibnamefont {Rubio}},\ }\bibfield  {title} {\bibinfo {title} {{Quick iterative scheme for the calculation of transfer matrices: application to Mo (100)}},\ }\href {https://doi.org/10.1088/0305-4608/14/5/016} {\bibfield  {journal} {\bibinfo  {journal} {Journal of Physics F: Metal Physics}\ }\textbf {\bibinfo {volume} {14}},\ \bibinfo {pages} {1205} (\bibinfo {year} {1984})}\BibitemShut {NoStop}%
\bibitem [{\citenamefont {Sancho}\ \emph {et~al.}(1985)\citenamefont {Sancho}, \citenamefont {Sancho}, \citenamefont {Sancho},\ and\ \citenamefont {Rubio}}]{MPLopezSancho_1985}%
  \BibitemOpen
  \bibfield  {author} {\bibinfo {author} {\bibfnamefont {M.~P.~L.}\ \bibnamefont {Sancho}}, \bibinfo {author} {\bibfnamefont {J.~M.~L.}\ \bibnamefont {Sancho}}, \bibinfo {author} {\bibfnamefont {J.~M.~L.}\ \bibnamefont {Sancho}},\ and\ \bibinfo {author} {\bibfnamefont {J.}~\bibnamefont {Rubio}},\ }\bibfield  {title} {\bibinfo {title} {{Highly convergent schemes for the calculation of bulk and surface Green functions}},\ }\href {https://doi.org/10.1088/0305-4608/15/4/009} {\bibfield  {journal} {\bibinfo  {journal} {Journal of Physics F: Metal Physics}\ }\textbf {\bibinfo {volume} {15}},\ \bibinfo {pages} {851} (\bibinfo {year} {1985})}\BibitemShut {NoStop}%
\bibitem [{\citenamefont {Wu}\ \emph {et~al.}(2018)\citenamefont {Wu}, \citenamefont {Zhang}, \citenamefont {Song}, \citenamefont {Troyer},\ and\ \citenamefont {Soluyanov}}]{WU2017}%
  \BibitemOpen
  \bibfield  {author} {\bibinfo {author} {\bibfnamefont {Q.}~\bibnamefont {Wu}}, \bibinfo {author} {\bibfnamefont {S.}~\bibnamefont {Zhang}}, \bibinfo {author} {\bibfnamefont {H.-F.}\ \bibnamefont {Song}}, \bibinfo {author} {\bibfnamefont {M.}~\bibnamefont {Troyer}},\ and\ \bibinfo {author} {\bibfnamefont {A.~A.}\ \bibnamefont {Soluyanov}},\ }\bibfield  {title} {\bibinfo {title} {Wanniertools : An open-source software package for novel topological materials},\ }\href {https://doi.org/https://doi.org/10.1016/j.cpc.2017.09.033} {\bibfield  {journal} {\bibinfo  {journal} {Computer Physics Communications}\ }\textbf {\bibinfo {volume} {224}},\ \bibinfo {pages} {405 } (\bibinfo {year} {2018})}\BibitemShut {NoStop}%
\bibitem [{Note1()}]{Note1}%
  \BibitemOpen
  \bibinfo {note} {We also denote the mirror eigenvalues $\pm 1$ by $\eta $ with a slight abuse of notation.}\BibitemShut {Stop}%
\bibitem [{\citenamefont {Mathai}\ and\ \citenamefont {Thiang}(2017)}]{Mathai2017}%
  \BibitemOpen
  \bibfield  {author} {\bibinfo {author} {\bibfnamefont {V.}~\bibnamefont {Mathai}}\ and\ \bibinfo {author} {\bibfnamefont {G.~C.}\ \bibnamefont {Thiang}},\ }\bibfield  {title} {\bibinfo {title} {{Differential Topology of Semimetals}},\ }\href {https://doi.org/10.1007/s00220-017-2965-z} {\bibfield  {journal} {\bibinfo  {journal} {Communications in Mathematical Physics}\ }\textbf {\bibinfo {volume} {355}},\ \bibinfo {pages} {561} (\bibinfo {year} {2017})}\BibitemShut {NoStop}%
\bibitem [{\citenamefont {Wan}\ \emph {et~al.}(2011)\citenamefont {Wan}, \citenamefont {Turner}, \citenamefont {Vishwanath},\ and\ \citenamefont {Savrasov}}]{PhysRevB.83.205101}%
  \BibitemOpen
  \bibfield  {author} {\bibinfo {author} {\bibfnamefont {X.}~\bibnamefont {Wan}}, \bibinfo {author} {\bibfnamefont {A.~M.}\ \bibnamefont {Turner}}, \bibinfo {author} {\bibfnamefont {A.}~\bibnamefont {Vishwanath}},\ and\ \bibinfo {author} {\bibfnamefont {S.~Y.}\ \bibnamefont {Savrasov}},\ }\bibfield  {title} {\bibinfo {title} {{Topological semimetal and Fermi-arc surface states in the electronic structure of pyrochlore iridates}},\ }\href {https://doi.org/10.1103/PhysRevB.83.205101} {\bibfield  {journal} {\bibinfo  {journal} {Phys. Rev. B}\ }\textbf {\bibinfo {volume} {83}},\ \bibinfo {pages} {205101} (\bibinfo {year} {2011})}\BibitemShut {NoStop}%
\bibitem [{\citenamefont {Okugawa}\ and\ \citenamefont {Murakami}(2014)}]{PhysRevB.89.235315}%
  \BibitemOpen
  \bibfield  {author} {\bibinfo {author} {\bibfnamefont {R.}~\bibnamefont {Okugawa}}\ and\ \bibinfo {author} {\bibfnamefont {S.}~\bibnamefont {Murakami}},\ }\bibfield  {title} {\bibinfo {title} {{Dispersion of Fermi arcs in Weyl semimetals and their evolutions to Dirac cones}},\ }\href {https://doi.org/10.1103/PhysRevB.89.235315} {\bibfield  {journal} {\bibinfo  {journal} {Phys. Rev. B}\ }\textbf {\bibinfo {volume} {89}},\ \bibinfo {pages} {235315} (\bibinfo {year} {2014})}\BibitemShut {NoStop}%
\bibitem [{\citenamefont {Sato}\ \emph {et~al.}(2024)\citenamefont {Sato}, \citenamefont {Bouaziz}, \citenamefont {Sumita}, \citenamefont {Kobayashi}, \citenamefont {Tateishi}, \citenamefont {Bl{\"u}gel}, \citenamefont {Furusaki},\ and\ \citenamefont {Hirayama}}]{Sato2024}%
  \BibitemOpen
  \bibfield  {author} {\bibinfo {author} {\bibfnamefont {M.}~\bibnamefont {Sato}}, \bibinfo {author} {\bibfnamefont {J.}~\bibnamefont {Bouaziz}}, \bibinfo {author} {\bibfnamefont {S.}~\bibnamefont {Sumita}}, \bibinfo {author} {\bibfnamefont {S.}~\bibnamefont {Kobayashi}}, \bibinfo {author} {\bibfnamefont {I.}~\bibnamefont {Tateishi}}, \bibinfo {author} {\bibfnamefont {S.}~\bibnamefont {Bl{\"u}gel}}, \bibinfo {author} {\bibfnamefont {A.}~\bibnamefont {Furusaki}},\ and\ \bibinfo {author} {\bibfnamefont {M.}~\bibnamefont {Hirayama}},\ }\bibfield  {title} {\bibinfo {title} {{Ideal spin-orbit-free Dirac semimetal and diverse topological transitions in Y$_8$CoIn$_3$ family}},\ }\href {https://doi.org/10.1038/s43246-024-00635-9} {\bibfield  {journal} {\bibinfo  {journal} {Communications Materials}\ }\textbf {\bibinfo {volume} {5}},\ \bibinfo {pages} {253} (\bibinfo {year} {2024})}\BibitemShut {NoStop}%
\end{thebibliography}
%

\end{document}